\documentclass{elsart}
\usepackage{amsmath,amssymb,epsfig,bm}

\numberwithin{equation}{subsection}

\journal{Physics Reports}

\begin{document}

\let\Oldsection\section
\renewcommand{\section}[1]{\Oldsection{\bf #1}}

\begin{frontmatter}

\title{Cosmology with\\ inhomogeneous magnetic fields}

\author{John D. Barrow}
\address{DAMTP, Centre for Mathematical Sciences, Wilberforce
Road, Cambridge~CB3~0WA, UK}

\author{Roy Maartens}
\address{Institute of Cosmology \& Gravitation, University of
Portsmouth, Portsmouth~P01~2EG, UK}

\author{Christos G. Tsagas}
\address{Section of Astrophysics, Astronomy and Mechanics,
Department of Physics, Aristotle University of Thessaloniki,
Thessaloniki 54124, Greece}

\vspace{1cm}

\tableofcontents

\newpage

\begin{abstract}
We review spacetime dynamics in the presence of large-scale
electromagnetic fields and then consider the effects of the magnetic
component on perturbations to a spatially homogeneous and isotropic
universe. Using covariant techniques, we refine and extend earlier
work and provide the magnetohydrodynamic equations that describe
inhomogeneous magnetic cosmologies in full general relativity.
Specialising this system to perturbed Friedmann-Robertson-Walker
models, we examine the effects of the field on the expansion
dynamics and on the growth of density inhomogeneities, including
non-adiabatic modes. We look at scalar perturbations and obtain
analytic solutions for their linear evolution in the radiation, dust
and inflationary eras. In the dust case we also calculate the
magnetic analogue of the Jeans length. We then consider the
evolution of vector perturbations and find that the magnetic
presence generally reduces the decay rate of these distortions.
Finally, we examine the implications of magnetic fields for the
evolution of cosmological gravitational waves.\vspace{5mm} \noindent
\end{abstract}


\end{frontmatter}

\newpage

\section{Introduction}\label{sI}
Magnetic fields are a widespread and significant component of the
Universe~\cite{KWM}-\cite{CT}. The Milky Way and many other spiral
galaxies possess magnetic fields with strengths of order a few
micro~Gauss, that are coherent over the plane of the galactic disc.
Magnetic fields are also a common property of galaxy clusters,
extending well beyond their core regions. The strengths of the
ordered magnetic fields in the intracluster medium exceed those
typically associated with the interstellar medium of the Milky Way,
suggesting that galaxy formation, and even cluster dynamics, could
be influenced by magnetic forces. Furthermore, reports of Faraday
rotation in high-redshift, Lyman-$\alpha$ absorption systems suggest
that dynamically significant magnetic fields could be present in
protogalactic condensations.

Despite their widespread presence, however, the precise origin of
any large-scale cosmological magnetic field is still a mystery, and
its likelihood or necessity remains a subject of
debate~\cite{E}-\cite{Gi2}. The alignment of the galactic fields,
especially those seen in spiral galaxies, seems to support the
dynamo idea~\cite{P}-\cite{BS}. However, the galactic dynamo cannot
work without an initial seed field~\cite{Ku}-\cite{DLT} and these
seeds may require $B$-fields of primeval
origin~\cite{H1}-\cite{ITSHO}. Deciding whether galactic and cluster
magnetic fields are primordial relics or post-recombination
artefacts is difficult because the strong amplification of the
fields in these virialized systems has overwhelmed all traces of
their earlier history. In contrast, possible magnetic imprints in
the Cosmic Microwave Background (CMB), or a magnetic field in the
intercluster space, can provide better insight into the early
history because the presence of small-scale magnetic fields leaves
small angular scale features in the CMB undamped~\cite{SB2,SB1}, and
leads to distinctive polarisation signatures~\cite{SSB}. The idea of
primordial magnetism is attractive because it can potentially
explain all the large-scale $B$-fields seen in the universe today,
especially those found in high-redshift protogalaxies. Early
magnetogenesis is not problem-free however. Seed fields generated
during, say, phase-transitions in the radiation era have typical
coherence lengths that are very small and will destabilize the
dynamo. On the other hand, fields which survived an epoch of
inflation are typically too weak to sustain the dynamo. Although the
literature contains several mechanisms of primordial magnetic
amplification~\cite{TW}-\cite{Ts1}, with early reheating looking
like a promising stage~\cite{BTW}, the issue remains open. Any
cosmological magnetic field must also be consistent with a number of
astrophysical constraints. These come from primordial
nucleosynthesis~\cite{GR2,COST}, and the high isotropy of the CMB
radiation~\cite{Z}-\cite{KR}. This constrains the current strength
of a large-scale homogeneous $B$-field below
$\sim10^{-9}$~$G$~\cite{BFS} (see also \S~\ref{ssZeI} here),
although tangled random fields on small scales could reach up to
$\sim10^{-6}$~$G$.

Large-scale magnetic fields of $\mu G$ strength can affect the
evolution of structure in the Universe, and studies of their effects
have a long history. Anisotropic, spatially homogeneous,
relativistic cosmologies permeated by large-scale magnetic fields
have been analysed in~\cite{Th}-\cite{C} and more recently
in~\cite{skew} and~\cite{LbKW}-\cite{KC2}. The bulk of the available
inhomogeneous treatments are Newtonian, with general relativistic
treatments only a recent addition to the literature. The evolution
of linear density and vorticity perturbations in magnetic Newtonian
cosmologies were addressed by~\cite{RR}-\cite{KOR}, and again more
recently by~\cite{BF,VTP}; general relativistic treatments have been
given in~\cite{F}-\cite{BFH-V}. These investigations have
established that magnetic fields are sources of density and
vorticity distortions, but the rather complicated action of the
$B$-field did not allow for analytic solutions to the relativistic
equations. Solutions were provided with the help of covariant and
gauge-invariant techniques, which considerably simplify the
mathematics of cosmological
magnetohydrodynamics~\cite{TB1}-\cite{TM2}. The covariant equations
were also used as the basis for qualitative studies by means of
phase-plane methods~\cite{HD}, while an analogous approach has been
used to analyse two-component charged plasmas~\cite{MDBST,MM}. In
addition to the linear regime, a number of authors have looked into
the non-linear interaction between cosmic magnetic fields and
electrically-charged fluids, primarily on sub-horizon scales. This
includes work on the amplification of the $B$-field by shearing
effects during the anisotropic collapse of a proto-galactic
cloud~\cite{BEO2}-\cite{KC1}. The aforementioned studies combine
analytical and numerical approaches, and apply to the pre- or to the
post-recombination era of the universe. All this has motivated and
facilitated work on the potential effects of magnetism on the CMB
spectrum, including its polarisation~\cite{SSB,ADGR}-\cite{KR}, and
the low-quadrupole moment problem~\cite{CCT}.

In the present article, we use covariant and gauge-invariant
techniques to analyse the effects of cosmological magnetic fields on
the large-scale structure of the universe. We do so by refining and
extending the work of~\cite{TB1}-\cite{TM1}. After a brief
introduction to the 1+3 covariant formalism in \S~\ref{s1+3CD}, we
use \S~\ref{sEMF},~\ref{sGF} and~\ref{sNlCMHD} to provide a detailed
discussion of cosmological electrodynamics and magnetohydrodynamics
in full general relativity. Section~\ref{sPMFRWC} utilises this
information to study the effects of the $B$-field on weakly magnetic
almost-Friedmann-Robertson-Walker (FRW) cosmologies in a
gauge-invariant manner. One of the key results is that linear
perturbations in the magnetic energy density evolve in step with
those in the density of the matter. Focusing on scalar
perturbations, we look at the evolution of linearized matter
perturbations during the radiation and dust eras in \S~\ref{sLDPs}.
During the radiation era, and on super-horizon scales, our solutions
show the magnetic pressure slows down the standard `Jeans' growth
rate of the density contrast. At the same time, the $B$-field
increases the oscillation frequency of small-scale density
perturbations in proportion to its strength relative to the
radiation background density. The same amount of increase is also
observed in the associated Jeans length. After the radiation era
ends, the fluid pressure is effectively zero, and the magnetic field
emerges as the sole source of pressure support. We calculate the
associated `magnetic' Jeans length and find that for $\mu G$ fields
its scale is comparable to the size of a galaxy cluster. Given that
during the dust era the magnetic relative strength decays with time,
we also calculate the late-time magnetic effects on density
perturbations. Here too, we find that the magnetic pressure inhibits
the gravitational clumping of matter. This section closes by looking
at the evolution of magnetic density perturbations within an
inflationary false-vacuum environment. We find that the outcome is
very sensitive to the effective equation of state of the cosmic
medium. Section~\ref{sLIPs} looks at the magnetic effects on
isocurvature density fluctuations, namely on perturbations evolving
on irrotational spatial hypersurfaces of constant curvature. After
establishing the necessary conditions for the existence of such
disturbances in the presence of the $B$-field, we find that they
always decay in time. In \S~\ref{sLEPs}, we treat the magnetic
medium as a two-fluid system and consider the effective entropy
perturbations coming from differences in the dynamical behaviour of
the two components. Our study shows that such disturbances vanish
during the radiation era, while after recombination they either
vanish, or remain constant, depending on the initial conditions. The
magnetic effects on rotational (vector) density perturbations are
considered in \S~\ref{sLVs}. At the linear perturbative level these
distortions are directly related to the amount of vorticity in the
universe. Focusing on the dust era, we show that the magnetic
presence reduces the standard decay rate of rotational
perturbations. This leads to more residual vorticity relative to
magnetic-free universes. The implications of the $B$-field for the
evolution of gravitational waves are examined in \S~\ref{sLGWs}. As
a first step, we introduce the additional constraints needed to
isolate the linear pure-tensor modes in the presence of a magnetic
field. Once these have been imposed, we consider the magnetic
effects on the propagation of gravitational-wave distortions. We
also look at the role of the magnetic anisotropic pressure during
the radiation era and discuss the critical, zero eigenvalue, case
that emerges there.

\section{1+3 covariant description}\label{s1+3CD}
The covariant approach to general relativity and cosmology dates
back to the work of Heckmann, Sch\"{u}cking, and Raychaudhuri in the
1950s~\cite{HS,R} and it has since been employed in numerous
applications by many authors. The formalism uses the kinematic
quantities of the fluid, its energy density and pressure and the
gravito-electromagnetic tensors instead of the metric, which in
itself does not provide a covariant description. The key equations
are the Ricci and Bianchi identities, applied to the fluid
4-velocity vector, while Einstein's equations are incorporated via
algebraic relations between the Ricci and the energy-momentum
tensors. Here, we will only give a brief description of the approach
and direct the reader to a number of review articles for further
details and references~\cite{Eh1}-\cite{TCM}.

\subsection{Local spacetime splitting}\label{ssLSS}
Consider a general spacetime with a Lorentzian metric $g_{ab}$ of
signature ($-,\,+,\,+,\,+$). Introduce a family of fundamental
observers along a timelike congruence of worldlines tangent to the
4-velocity vector\footnote{Latin indices run between~0 and~3 and
Greek vary from~1 to~3. We use geometrized units with $c=1=8\pi G$,
which means that all geometrical variables have physical dimensions
that are integer powers of length.}
\begin{equation}
u^{a}= \frac{\mathrm{d}x^{a}}{\mathrm{d}\tau}\,,  \label{ua}
\end{equation}
where $\tau$ is the associated proper time and
$u_{a}u^{a}=-1$~\cite{EvE}. This fundamental velocity field
introduces a local, 1+3 `threading' of the spacetime into time and
space. The vector $u_{a}$ determines the time direction, and the
tensor $h_{ab}=g_{ab}+u_{a}u_{b}$ projects orthogonal to 4-velocity
field into what is known as the observer's instantaneous rest space.
In the absence of rotation, $u_{a}$ is hypersurface orthogonal and
$h_{ab}$ acts as the metric of the 3-D spatial sections.

Employing $u_{a}$ and $h_{ab},$ we define the covariant time
derivative and the orthogonally projected gradient of any given
tensor field $S_{ab\cdots}{}^{cd\cdots}$ according to
\begin{equation}
\dot{S}_{ab\cdots}{}^{cd\cdots}=
u^{e}\nabla_{e}S_{ab\cdots}{}^{cd\cdots} \hspace{5mm} \mathrm{and}
\hspace{5mm} \mathrm{D}_{e}S_{ab\cdots}{}^{cd\cdots }=
h_{e}{}^{s}h_{a}{}^{f}h_{b}{}^{p}h_{q}{}^{c}h_{r}{}^{d}\cdots
\nabla_{s}S_{fp\cdots}{}^{qr\cdots}\,,  \label{deriv}
\end{equation}
respectively. The former indicates differentiation along the
timelike direction and the latter operates on the observer's rest
space.

\subsection{Matter fields}\label{ssMF}
Relative to the fundamental observers, the energy-momentum tensor of
a general imperfect fluid decomposes into its irreducible parts
as~\cite{EvE}
\begin{equation}
T_{ab}= \rho u_{a}u_{b}+ ph_{ab}+ 2q_{(a}u_{b)}+ \pi_{ab}\,.
\label{Tab}
\end{equation}
Here, $\rho=T_{ab}u^{a}u^{b}$ and $p=T_{ab}h^{ab}/3$ are,
respectively, the energy density and the isotropic pressure of the
medium, $q_{a}=-h_{a}{}^{b}T_{bc}u^{c}$ is the energy-flux vector
relative to $u_{a}$, and $\pi_{ab}=h_{\langle a}{}^{c}
h_{b\rangle}{}^{d}T_{cd}$ is the symmetric and trace-free tensor
that describes the anisotropic pressure of the
fluid.\footnote{Angled brackets denote the symmetric and trace-free
part of second-rank tensors projected orthogonally to $u_{a}$ and
the projected component of vectors (i.e.~$v_{\langle a\rangle}=
h_{a}{}^{b}v_{b}$).} It follows that $q_{a}u^{a}=0=\pi_{ab}u^{a}$.
When the fluid is perfect, both $q_{a}$ and $\pi_{ab}$ are
identically zero, and the remaining degrees of freedom are
determined by the equation of state. For a barotropic medium the
latter reduces to $p=p\,(\rho)$, with $c_{s}^{2}=
\mathrm{d}p/\mathrm{d}\rho$ giving the the square of the associated
adiabatic sound speed.

When dealing with a multi-component medium, or when allowing for
peculiar velocities, one needs to account for the velocity `tilt'
between the matter components and the fundamental observers
(e.g.~see~\cite{KE}-\cite{ET}). Here we will consider a
single-component fluid and we will assume that the fundamental
observers move with it.

\subsection{Covariant kinematics}\label{ssCKs}
The observers' motion is characterized by the irreducible
kinematical quantities of the $u_{a}$-congruence, which emerge from
the covariant decomposition of the 4-velocity gradient
\begin{equation}
\nabla_{b}u_{a}= \sigma_{ab}+ \omega_{ab}+ {\frac{1}{3}}\,\Theta
h_{ab}- A_{a}u_{b}\,,  \label{Nbua}
\end{equation}
where $\sigma_{ab}=\mathrm{D}_{\langle b}u_{a\rangle}$,
$\omega_{ab}=\mathrm{D}_{[b}u_{a]}$,
$\Theta=\nabla^{a}u_{a}=\mathrm{D}^{a}u_{a}$ and
$A_{a}=u^{b}\nabla_{b}u_{a}$ are respectively the shear and the
vorticity tensors, the volume expansion (or contraction) scalar, and
the 4-acceleration vector~\cite{EvE}. Then,
$\sigma_{ab}u^{a}=0=\omega_{ab}u^{a}=A_{a}u^{a}$ by construction.
The volume scalar is used to introduce a representative length scale
(the cosmological scale factor $a$) by means of definition
$\dot{a}/a=\Theta/3$. Also, on using the orthogonally projected
alternating tensor $\varepsilon_{abc}$ (with
$\dot{\varepsilon}_{abc}=3u_{[a}\varepsilon_{bc]d}A^{d}$ --
see~\cite{MGE}), one defines the vorticity vector
$\omega_{a}=\varepsilon_{abc}\omega^{bc}/2$. Note that
$\epsilon_{abc}=\eta_{abcd}u^{d}$, where $\eta_{abcd}$ is the
totally antisymmetric permutation tensor of the spacetime. This is a
covariantly constant quantity, with $\eta_{abcd}\eta^{efpq}=
-4!\delta_{[a}{}^{e}\delta_{b}{}^{f}\delta_{c}{}^{p}
\delta_{d]}{}^{q}$ and $\eta^{0123}=
[-\mathrm{det}(g_{ab})]^{-1/2}$. The tensor
$v_{ab}=\mathrm{D}_{b}u_{a}=\sigma_{ab}+\omega_{ab}+
(\Theta/3)h_{ab}$ describes the relative motion of neighbouring
observers (with the same 4-velocity). In particular,
$v_{a}=v_{ab}\chi^{b}$ is the relative velocity of the associated
worldlines and $\chi_{a}$ is their relative position vector
(see~\cite{El1,El2} for details).

The non-linear covariant kinematics is determined by a set of three
propagation equations complemented by an equal number of
constraints~\cite{EvE}. The former contains Raychaudhuri's formula
\begin{equation}
\dot{\Theta}= -{\frac{1}{3}}\,\Theta^{2}- {\frac{1}{2}}\,(\rho+3p)-
2(\sigma^{2}-\omega^{2})+ \mathrm{D}^{a}A_{a}+ A_{a}A^{a}\,,
\label{Ray}
\end{equation}
for the time evolution of $\Theta$; the shear propagation equation
\begin{equation}
\dot{\sigma}_{\langle ab\rangle}= -{\frac{2}{3}}\,\Theta\sigma_{ab}-
\sigma_{c\langle a}\sigma^{c}{}_{b\rangle}- \omega_{\langle a}
\omega_{b\rangle}+ \mathrm{D}_{\langle a}A_{b\rangle}+ A_{\langle a}
A_{b\rangle}- E_{ab}+ {\frac{1}{2}}\,\pi_{ab}\,,  \label{sigmadot}
\end{equation}
which describes kinematical anisotropies; and the evolution equation
of the vorticity
\begin{equation}
\dot{\omega}_{\langle a\rangle}= -{\frac{2}{3}}\,\Theta\omega_{a}-
{\frac{1}{2}}\,\mathrm{curl\,}A_{a}+ \sigma_{ab}\omega^{b}\,.
\label{omegadot}
\end{equation}
Here $\sigma^{2}=\sigma_{ab}\sigma^{ab}/2$ and
$\omega^{2}=\omega_{ab}\omega^{ab}/2=\omega_{a}\omega^{a}$ are
respectively the scalar magnitudes of the shear and the vorticity,
while $E_{ab}$ is the electric component of the Weyl tensor (see
\S~\ref{ssLrWC} below). Also, $\mathrm{curl\,}v_{a}=
\varepsilon_{abc}\mathrm{D}^{b}v^{c}$ for any orthogonally projected
vector $v_{a}$.

Equations (\ref{Ray}), (\ref{sigmadot}) and (\ref{omegadot}) are
complemented by a set of three non-linear constraints. These are the
shear
\begin{equation}
\mathrm{D}^{b}\sigma_{ab}= {\frac{2}{3}}\,\mathrm{D}_{a}\Theta+
\mathrm{curl\,}\omega_{a}+ 2\varepsilon_{abc}A^{b}\omega^{c}-
q_{a}\,,  \label{shearcon}
\end{equation}
the vorticity
\begin{equation}
\mathrm{D}^{a}\omega_{a}= A_{a}\omega^{a}\,,  \label{omegacon}
\end{equation}
and the magnetic Weyl constraint
\begin{equation}
H_{ab}= \mathrm{curl\,}\sigma_{ab}+ \mathrm{D}_{\langle a}
\omega_{b\rangle}+ 2A_{\langle a}\omega_{b\rangle}\,,  \label{Hcon}
\end{equation}
where $\mathrm{curl\,}T_{ab}\equiv\varepsilon_{cd\langle a}
\mathrm{D}^{c}T_{b\rangle}{}^{d}$ for any symmetric, orthogonally
projected tensor $T_{ab}$. We finally note that the cosmological
constant has been set to zero (i.e.~$\Lambda=0$) throughout this
review.

\section{Electromagnetic fields}\label{sEMF}
Covariant studies of electromagnetic fields date back to the work of
Ehlers~\cite{Eh1} and Ellis~\cite{El2}. In addition to its inherent
mathematical compactness and clarity, the formalism facilitates a
physically intuitive fluid description of the Maxwell field. This is
represented as an imperfect fluid with properties specified by its
electric and magnetic components. For a fully covariant study of
electromagnetic fields in curved spacetimes the reader is referred
to~\cite{Ts}.

\subsection{Electric and magnetic components}\label{ssEMC}
The Maxwell field is covariantly characterized by the antisymmetric
electromagnetic (Faraday) tensor $F_{ab}$. Relative to a fundamental
observer, the latter decomposes as~\cite{El2}
\begin{equation}
F_{ab}= 2u_{[a}E_{b]}+ \varepsilon_{abc}B^{c}\,,  \label{Fab}
\end{equation}
where $E_{a}=F_{ab}u^{b}$ and $B_{a}=\varepsilon_{abc}F^{bc}/2$ are
respectively the electric and magnetic fields measured by the
observer. Note that $E_{a}u^{a}=0=B_{a}u^{a}$, ensuring that both
$E_{a}$ and $B_{a}$ are spacelike vectors. Also,
$B_{a}=\varepsilon_{abc}F^{bc}/2$ guarantees that $B_{a}$ is the
dual of $F_{ab}$.

The Faraday tensor also determines the energy-momentum tensor of the
Maxwell field according to
\begin{equation}
T_{ab}^{(em)}= -F_{ac}F^{c}{}_{b}-
{\frac{1}{4}}\,F_{cd}F^{cd}g_{ab}\,.  \label{Tem1}
\end{equation}
The above combines with (\ref{Fab}) to facilitate the irreducible
decomposition of $T_{ab}^{(em)}$ relative to the
$u_{a}$-frame~\cite{El2},
\begin{equation}
T_{ab}^{(em)}= {\frac{1}{2}}\,\left( E^{2}+B^{2}\right)u_{a}u_{b}+
{\frac{1}{6}}\left(E^{2}+B^{2}\right)h_{ab}+
2{\mathcal{Q}}_{(a}u_{b)}+ {\mathcal{P}}_{ab}\,.  \label{Tem}
\end{equation}
Here $E^{2}=E_{a}E^{a}$ and $B^{2}=B_{a}B^{a}$ are the magnitudes of
the two fields, ${\mathcal{Q}}_{a}=\varepsilon_{abc}E^{b}B^{c}$ is
the electromagnetic Poynting vector and ${\mathcal{P}}_{ab}$ is a
symmetric, trace-free tensor given by
\begin{equation}
{\mathcal{P}}_{ab}= {\mathcal{P}}_{\langle ab\rangle}=
{\frac{1}{3}}\,\left(E^{2}+B^{2}\right)h_{ab}- E_{a}E_{b}-
B_{a}B_{b}\,.  \label{cP}
\end{equation}
Expression (\ref{Tem}) provides a fluid description of the Maxwell
field and manifests its generically anisotropic nature. In
particular, the electromagnetic field corresponds to an imperfect
fluid with energy density $(E^{2}+B^{2})/2$, isotropic pressure
$(E^{2}+B^{2})/6$, anisotropic stresses given by
${\mathcal{P}}_{ab}$ and an energy-flux vector represented by
${\mathcal{Q}}_{a}$. Equation (\ref{Tem}) also ensures that
$T_{a}^{(em)\;a}=0$, in agreement with the trace-free nature of the
radiation stress-energy tensor. Finally, we note that by putting the
isotropic and anisotropic pressure together one arrives at the
familiar Maxwell tensor, which assumes the covariant form
\begin{equation}
{\mathcal{M}}_{ab}= {\frac{1}{2}}\,\left( E^{2}+B^{2}\right)h_{ab}-
E_{a}E_{b}- B_{a}B_{b}\,.  \label{cM}
\end{equation}

\subsection{Maxwell's equations}\label{ssMEs}
We follow the evolution of the electromagnetic field by means of
Maxwell's equations. In their standard tensor form these read
\begin{equation}
\nabla_{[c}F_{ab]}=0 \hspace{15mm} \mathrm{and} \hspace{15mm}
\nabla^{b} F_{ab}=J_{a}\,,  \label{Max}
\end{equation}
where (\ref{Max}a) reflects the existence of a 4-potential and
$J_{a}$ is the 4-current that sources the electromagnetic field.
With respect to the $u_{a}$-congruence, the 4-current splits into
its irreducible parts according to
\begin{equation}
J_{a}= \rho_{e}u_{a}+ {\mathcal{J}}_{a}\,,  \label{Ja}
\end{equation}
with $\rho_{e}=-J_{a}u^{a}$ representing the measurable charge
density and ${\mathcal{J}}_{a}=h_{a}{}^{b}J_{b}$ the orthogonally
projected current (i.e.~${\mathcal{J}}_{a}u^{a}=0$).

Relative to a fundamental observer, each one of Maxwell's equations
decomposes into a timelike and a spacelike component. Projecting
(\ref{Max}a) and (\ref{Max}b) along and orthogonal to the 4-velocity
vector $u_{a}$, one obtains a set of two propagation
equations~\cite{El2}
\begin{eqnarray}
\dot{E}_{\langle a\rangle}&=&
\left(\sigma_{ab}+\varepsilon_{abc}\omega^{c}-{\frac{2}{3}}\,\Theta
h_{ab}\right)E^{b}+ \varepsilon_{abc}A^{b}B^{c}+
\mathrm{curl\,}B_{a}- {\mathcal{J}}_{a}\,,  \label{M1}\\
\dot{B}_{\langle a\rangle}&=&
\left(\sigma_{ab}+\varepsilon_{abc}\omega^{c}-{\frac{2}{3}}\,\Theta
h_{ab}\right)B^{b}- \varepsilon_{abc}A^{b}E^{c}-
\mathrm{curl\,}E_{a}\,,  \label{M2}
\end{eqnarray}
and the following pair of constraints
\begin{eqnarray}
\mathrm{D}^{a}E_{a}&=& \rho_{e}- 2\omega^{a}B_{a}\,,  \label{M3}\\
\mathrm{D}^{a}B_{a}&=& 2\omega^{a}E_{a}\,. \label{M4}
\end{eqnarray}
Note that, in addition to the usual `curl' and `divergence' terms,
Eqs.~(\ref{M1})-(\ref{M4}) also contain effects triggered by the
relative motion of neighbouring observers (with the same 4-velocity
-- see \S~\ref{ssCKs}). These are carried by the kinematic terms in
the right-hand side of the above. Thus,
$\tilde{\rho}_{e}=-2\omega_{a}B^{a}$ is an effective electric charge
caused by the relative motion of the magnetic field, while
$2\omega^{a}E_{a}$ acts as an effective magnetic charge triggered by
the relatively moving $E$-field. The acceleration terms in
(\ref{M1}) and (\ref{M2}), on the other hand, also reflect the fact
that spacetime is treated as a single entity.

\subsection{Conservation laws}\label{ssCLs1}
The twice contracted Bianchi identities guarantee the conservation
of the total energy momentum tensor, namely that
$\nabla^{b}T_{ab}=0$. This constraint splits into a timelike and a
spacelike part, which respectively lead to the energy density and
the momentum-density conservation laws. The energy momentum tensor
of the electromagnetic field satisfies the constraint
$\nabla^{b}T_{ab}^{(em)}=-F_{ab}J^{b}$, with the Faraday tensor
given by (\ref{Fab}) and the quantity in the right-hand side
representing the Lorentz 4-force. Thus, for charged matter the
conservation of the total energy-momentum tensor
$T_{ab}=T_{ab}^{(m)}+T_{am}^{(em)}$ leads to the formulae
\begin{equation}
\dot{\rho}= -\Theta(\rho+p)- \mathrm{D}^{a}q_{a}- 2A^{a}q_{a}-
\sigma^{ab}\pi_{ab}+ E_{a}\mathcal{J}^{a}\,,  \label{edc2}
\end{equation}
for the energy density, and
\begin{eqnarray}
(\rho+p)A_{a}&=& -\mathrm{D}_{a}p- \dot{q}_{\langle a\rangle}-
{\frac{4}{3}}\,\Theta q_{a}- (\sigma_{ab}+\omega_{ab})q^{b}-
\mathrm{D}^{b}\pi_{ab}- \pi_{ab}A^{b} \nonumber\\ &&+\rho_{e}E_{a}+
\varepsilon_{abc}\mathcal{J}^{b}B^{c}\,,  \label{mdc2}
\end{eqnarray}
for the momentum density. The last two terms in the right-hand side
of (\ref{mdc2}) represent the familiar form of the Lorentz force. We
also note that the electromagnetic effects depend on the electrical
properties of the medium (see \S~\ref{ssOL} below).

The antisymmetry of the Faraday tensor (see Eq.~(\ref{Fab})) and the
second of Maxwell's formulae (see Eq.~(\ref{Max}b)) imply
$\nabla^{a}J_{a}=0$ and therefore ensure the conservation of the
4-current density. Using decomposition (\ref{Ja}), we arrive at the
covariant form of the charge-density conservation law~\cite{El2}
\begin{equation}
\dot{\rho}_{e}= -\Theta\rho_{e}- \mathrm{D}^{a}{\mathcal{J}}_{a}-
A^{a}{\mathcal{J}}_{a}\,.  \label{chcon}
\end{equation}
Hence, in the absence of spatial currents, the evolution of the
charge density depends entirely on the volume expansion (or
contraction) of the fluid element.

\subsection{Ohm's law}\label{ssOL}
The relation between the 4-current and the electric field, as
measured by the fundamental observers, is determined by Ohm's law.
Following~\cite{Gr,Ja}, the latter has the covariant form
\begin{equation}
J_{a}= \rho_{e}u_{a}+ \varsigma E_{a}\,,  \label{Ohm}
\end{equation}
where $\varsigma$ is the scalar conductivity of the medium. Thus,
Ohm's law splits the 4-current into a timelike convective component
and a conducting spacelike counterpart. Projecting (\ref{Ohm}) into
the observer's rest space gives
\begin{equation}
{\mathcal{J}}_{a}= \varsigma E_{a}\,.  \label{Ohm1}
\end{equation}
This form of Ohm's law covariantly describes the resistive
magnetohydrodynamic (MHD) approximation in the single-fluid
approach. Note the absence of the induced electric field from the
above, reflecting the fact that the covariant form of Maxwell's
formulae (see expressions (\ref{M1})-(\ref{M4})) already
incorporates the effects of relative motion. According to
(\ref{Ohm1}), non-zero spatial currents are compatible with a
vanishing electric field as long as the conductivity of the medium
is infinite (i.e.~for $\varsigma\rightarrow\infty$). Thus, at the
limit of ideal magnetohydrodynamics, the electric field vanishes in
the frame of the fluid. On the other hand, zero electrical
conductivity implies that the spatial currents vanish, even when the
electric field is non-zero. The electrical conductivity is typically
treated in a phenomenological manner and here we will also assume
that it remains constant throughout the medium.

\section{Gravitational field}\label{sGF}
Covariantly, the local gravitational field is described by a set of
algebraic relations between the Ricci curvature tensor and the
energy-momentum tensor of the matter. The free gravitational field,
on the other hand, is described by the electric and magnetic
components of the conformal curvature (Weyl) tensor.

\subsection{Local Ricci curvature}\label{ssLRC}
In the general-relativistic geometrical interpretation of gravity,
matter determines the spacetime curvature, which in turn dictates
the motion of the matter. This interaction is evident in the
Einstein field equations, which in the absence of a cosmological
constant take the form
\begin{equation}
R_{ab}= T_{ab}- {\frac{1}{2}}\,Tg_{ab}\,,  \label{EFE}
\end{equation}
where $R_{ab}=R^c{}_{acb}$ is the spacetime Ricci tensor, $T_{ab}$
is the energy-momentum tensor of the matter fields, with
$T=T_{a}{}^{a}$ being the trace. For our purposes the total
energy-momentum tensor has the form
$T_{ab}=T_{ab}^{(f)}+T_{ab}^{(em)}$, where $T_{ab}^{(f)}$ is given
by Eq.~(\ref{Tab}) and $T_{ab}^{(em)}$ by (\ref{Tem}). Thus,
\begin{eqnarray}
T_{ab}&=& \left[\rho+{\frac{1}{2}}\,\left(B^{2}+E^{2}\right)\right]
u_{a}u_{b}+ \left[p+{\frac{1}{6}}\,\left(B^{2}+E^{2}\right)\right]
h_{ab}+ 2(q_{(a}+{\mathcal{Q}}_{(a})u_{b)} \nonumber\\ &&+\pi_{ab}+
{\mathcal{P}}_{ab}\,,  \label{tTab}
\end{eqnarray}
ensuring that $\rho+(B^{2}+E^{2})/2$ is the total energy density of
the system, $p+(B^{2}+E^{2})/6$ is the total isotropic pressure,
$q_{a}+{\mathcal{Q}}_{a}$ is the total heat-flux vector and
$\pi_{ab}+{\mathcal{P}}_{ab}$ is the total anisotropic pressure. The
inclusion of electromagnetic terms in the energy-momentum tensor of
the matter guarantees that the contribution of the Maxwell field to
the spacetime geometry is fully accounted for.

The successive contraction of the Einstein field equations, assuming
that $T_{ab}$ is given by Eq.~(\ref{tTab}), leads to the following
algebraic relations:
\begin{eqnarray}
R_{ab}u^{a}u^{b}&=&
{\frac{1}{2}}\,\left(\rho+3p+E^{2}+B^{2}\right)\,,  \label{EFE1}\\
h_{a}{}^{b}R_{bc}u^{c}&=& -\left(q_{a}+{\mathcal{Q}}_{a}\right)\,,
\label{EFE2}\\ h_{a}{}^{c}h_{b}{}^{d}R_{cd}&=&
\left\{{\frac{1}{2}}\,
\left[\rho-p+{\frac{1}{3}}\,\left(E^{2}+B^{2}\right)\right]\right\}
h_{ab}+ \pi_{ab}+ \mathcal{P}_{ab}\,.  \label{EFE3}
\end{eqnarray}
In addition, the trace of (\ref{EFE}) gives $R=-T$, with
$R=R_{a}{}^{a}$ and $T=T_{a}{}^{a}=3p-\rho$, where the latter result
is guaranteed by the trace-free nature of $T_{ab}^{(em)}$. Note that
the above expressions are valid irrespective of the strength of the
electromagnetic components and recall that $q_{a}=0=\pi_{ab}$ when
dealing with a perfect fluid.

\subsection{Long-range Weyl curvature}\label{ssLrWC}
The Ricci tensor describes the local gravitational field of the
nearby matter. The long-range gravitational field, namely
gravitational waves and tidal forces, propagates through the Weyl
conformal curvature tensor. The splitting of the gravitational field
into its local and non-local components is demonstrated in the
following decomposition of the Riemann tensor,
\begin{equation}
R_{abcd}= C_{abcd}+{\frac{1}{2}}\,\left(g_{ac}R_{bd}+g_{bd}R_{ac}
-g_{bc}R_{ad}-g_{ad}R_{bc}\right)-
{\frac{1}{6}}\,R\left(g_{ac}g_{bd}-g_{ad}g_{bc}\right)\,,
\label{Riemann}
\end{equation}
where $C_{abcd}$ is the Weyl tensor. This shares all the symmetries
of the Riemann tensor and is also trace-free
(i.e.~$C^{c}{}_{acb}=0$). Relative to the fundamental observers, the
Weyl tensor decomposes into its irreducible parts according to
\begin{equation}
C_{abcd}= \left(g_{abqp}g_{cdsr}-\eta_{abqp}\eta_{cdsr}\right)
u^{q}u^{s}E^{pr}-
\left(\eta_{abqp}g_{cdsr}+g_{abqp}\eta_{cdsr}\right)
u^{q}u^{s}H^{pr}\,,  \label{Weyl}
\end{equation}
where $\eta_{abcd}=\eta_{[abcd]}$ is the spacetime permutation
tensor defined in \S~\ref{ssCKs} and $g_{abcd}=g_{ac}g_{bd}
-g_{ad}g_{bc}$ (e.g.~see~\cite{HE,M}). The symmetric and trace-free
tensors $E_{ab}$ and $H_{ab}$ are known as the electric and magnetic
Weyl components and they are given by
\begin{equation}
E_{ab}= C_{acbd}u^{c}u^{d} \hspace{15mm} \mathrm{and} \hspace{15mm}
H_{ab}= {\frac{1}{2}}\,\varepsilon_{a}{}^{cd}C_{cdbe}u^{e}\,,
\label{EHab}
\end{equation}
with $E_{ab}u^{b}=0=H_{ab}u^{b}$. Given that $E_{ab}$ has a
Newtonian counterpart, the electric part of the Weyl tensor is
associated with the tidal gravitational field. The magnetic
component, on the other hand, has no Newtonian analogue and is
therefore primarily associated with gravitational waves~and spatial
3-curvature anisotropy~\cite{EvE}. Of course, both tensors are
required if gravitational waves are to exist.

The Weyl tensor represents the part of the curvature that is not
determined locally by matter. However, the dynamics of the Weyl
field are not entirely arbitrary because the Riemann tensor
satisfies the Bianchi identities. When contracted, the latter take
the form~\cite{HE}
\begin{equation}
\nabla^{d}C_{abcd}= \nabla_{[b}R_{a]c}+
{\frac{1}{6}}\,g_{c[b}\nabla_{a]}R\,,  \label{Bianchi}
\end{equation}
by means of decomposition (\ref{Riemann}). In one sense the
contracted Bianchi identities act as the field equations for the
Weyl tensor, determining the part of the spacetime curvature that
depends on the matter distribution at other points~\cite{HE}. The
form of the contracted Bianchi identities guarantees that once the
electromagnetic contribution to the Ricci curvature has been
incorporated, through the Einstein field equations, the effect of
the Maxwell field on the Weyl curvature has also been fully
accounted for.

The 1+3 splitting of (\ref{Bianchi}) provides a set of two
propagation and two constraint equations for the evolution of the
long-range gravitational field, namely of tidal forces and gravity
waves. In particular, on using the decomposition (\ref{Weyl}), the
timelike component of (\ref{Bianchi}) leads to~\cite{EvE,MGE}
\begin{eqnarray}
\dot{E}_{\langle ab\rangle}&=& -\Theta E_{ab}-
{\frac{1}{2}}\,\left[\rho+p+{\frac{2}{3}}\left(B^{2}
+E^{2}\right)\right]\sigma_{ab}+ \mathrm{curl\,}H_{ab}-
{\frac{1}{2}}\left(\dot{\pi}_{ab}+\dot{\mathcal{P}}_{ab}\right)
\nonumber\\
&&-{\frac{1}{6}}\,\Theta\left(\pi_{ab}+\mathcal{P}_{ab}\right)-
{\frac{1}{2}}\,\mathrm{D}_{\langle
a}(q_{b\rangle}+\mathcal{Q}_{b\rangle})- A_{\langle a}
(q_{b\rangle}+\mathcal{Q}_{b\rangle}) \nonumber\\
&&+3\sigma_{\langle a}{}^{c}\left[E_{b\rangle c}
-{\frac{1}{6}}\left(\pi_{b\rangle c}+\mathcal{P}_{b\rangle c}
\right)\right]+2\varepsilon_{cd\langle a}A^{c}H_{b\rangle}{}^{d}
\nonumber\\ &&-\varepsilon_{cd\langle a}
\omega^{c}\left[E_{b\rangle}{}^{d}+
{\frac{1}{2}}\,\left(\pi_{b\rangle}{}^{d}
+\mathcal{P}_{b\rangle}{}^{d}\right)\right]  \label{dotEab}
\end{eqnarray}
and
\begin{eqnarray}
\dot{H}_{\langle ab\rangle}&=& -\Theta H_{ab}-
\mathrm{curl\,}E_{ab}+
{\frac{1}{2}}(\mathrm{curl\,}\pi_{ab}+\mathrm{curl\,}\mathcal{P}_{ab})+
3\sigma_{\langle a}{}^{c}H_{b\rangle c}-
{\frac{3}{2}}\,\omega_{\langle
a}(q_{b\rangle}+\mathcal{Q}_{b\rangle}) \nonumber\\
&&-2\varepsilon_{cd\langle a}A^{c}E_{b\rangle}{}^{d}+
\varepsilon_{cd\langle a}
\left[{\frac{1}{2}}\,\sigma^{c}{}_{b\rangle}(q^{d}+\mathcal{Q}^{d})
-\omega^{c}H_{b\rangle}{}^{d}\right]\,.  \label{dotHab}
\end{eqnarray}
Taking the time derivatives of the above one arrives to a pair of
wavelike equations for the electric and the magnetic parts of the
Weyl tensor, showing how curvature distortions propagate in the form
of gravitational waves like ripples in the spacetime fabric. These
waves are also subjected to a set of constraints, which emerge from
the spacelike component of the decomposed Eq.~(\ref{Bianchi}) and
are given by~\cite{EvE,MGE}
\begin{eqnarray}
\mathrm{D}^{b}E_{ab}&=& {\frac{1}{3}}\,\mathrm{D}_{a}
\left[\rho+{\frac {1}{2}}\left(B^{2}+E^{2}\right)\right]-
{\frac{1}{2}}\,\mathrm{D}^{b}\left(\pi_{ab}+\mathcal{P}_{ab}\right)-
{\frac{1}{3}}\,\Theta(q_a+\mathcal{Q}_a)+
{\frac{1}{2}}\,\sigma_{ab}(q^b+\mathcal{Q}^b) \nonumber\\
&&-3H_{ab}\omega^{b}+ \varepsilon_{abc}\left[\sigma^{b}{}_{d}H^{cd}
-{\frac{3}{2}}\,\omega^{b}(q^c+\mathcal{Q}^c)\right]  \label{Weylc1}
\end{eqnarray}
and
\begin{eqnarray}
\mathrm{D}^{b}H_{ab}&=& \left[\rho+p+{\frac{2}{3}}
\left(B^{2}+E^{2}\right)\right]\omega_{a}-
{\frac{1}{2}}\,\mathrm{curl\,}(q_a+\mathcal{Q}_a)+
3E_{ab}\omega^{b}-
{\frac{1}{2}}\left(\pi_{ab}+\mathcal{P}_{ab}\right)\omega^{b}
\nonumber\\ &&-\varepsilon_{abc}\sigma^{b}{}_{d}\left[E^{cd}
+{\frac{1}{2}}\left(\pi^{cd}+\mathcal{P}^{cd}\right)\right]\,,
\label{Weylc2}
\end{eqnarray}
respectively. The above expressions are similar to Maxwell's
formulae, which explains the names of $E_{ab}$ and $H_{ab}$. In
fact, this Maxwell-like form of the free gravitational field
underlines the rich correspondence between electromagnetism and
general relativity, which has been the subject of theoretical debate
for many decades (see~\cite{Be}-\cite{Da} for a representative
list).

\subsection{Spatial curvature}\label{ssSC}
When the fluid is irrotational, the rest spaces of the fundamental
observers mesh together to form spacelike surfaces orthogonal to
their worldlines. These are normal to the $u_a$-congruence and
define the hypersurfaces of simultaneity of all the comoving
observers. In the presence of vorticity, however, Frobenius' theorem
forbids the existence of such integrable surfaces
(e.g.~see~\cite{Wal,Poi}). The $u_a$-congruence is no longer
hypersurface orthogonal. Then, one can still talk about the
observers' spatial rest-space but only locally. The local 3-Riemann
tensor, defined by
\begin{equation}
{\mathcal{R}}_{abcd}=
h_{a}{}^{q}h_{b}{}^{s}h_{c}{}^{f}h_{d}{}^{p}R_{qsfp}- v_{ac}v_{bd}+
v_{ad}v_{bc}\,,  \label{3Riemann1}
\end{equation}
where $v_{ab}=\mathrm{D}_{b}u_{a}$ is the relative position tensor.
On using Eqs.~(\ref{EFE1})-(\ref{EFE3}) and expressions
(\ref{Riemann}), (\ref{Weyl}), definition (\ref{3Riemann1}) gives
\begin{eqnarray}
{\mathcal{R}}_{abcd}&=& -\varepsilon_{abq}\varepsilon_{cds}E^{qs}+
{\frac{1}{3}}\,\left[\rho+{\frac{1}{2}}\left(E^{2}+B^{2}\right)
-{\frac{1}{3}}\,\Theta^{2}\right](h_{ac}h_{bd}-h_{ad}h_{bc})
\nonumber\\
&&+{\frac{1}{2}}\,\left[h_{ac}(\pi_{bd}+{\mathcal{P}}_{bd})
+(\pi_{ac}+{\mathcal{P}}_{ac})h_{bd}-h_{ad}(\pi_{bc}
+{\mathcal{P}}_{bc})-(\pi_{ad}+{\mathcal{P}}_{ad})h_{bc}\right]
\nonumber\\
&&-{\frac{1}{3}}\,\Theta\left[h_{ac}(\sigma_{bd}+\omega_{bd})
+(\sigma_{ac}+\omega_{ac})h_{bd}-h_{ad}(\sigma_{bc}+\omega_{bc})
-(\sigma_{ad}+\omega_{ad})h_{bc}\right] \nonumber\\
&&-(\sigma_{ac}+\omega_{ac})(\sigma_{bd}+\omega_{bd})+(\sigma_{ad}
+\omega_{ad})(\sigma_{bc}+\omega_{bc})\,,  \label{3Riemann2}
\end{eqnarray}
which offers an irreducible decomposition of the local 3-Riemann
tensor orthogonal to $u_{a}$. Then, one can easily show that
\begin{equation}
{\mathcal{R}}_{abcd}= {\mathcal{R}}_{[ab][cd]}\label{3Rsym1}
\end{equation}
and that
\begin{eqnarray}
{\mathcal{R}}_{abcd}-{\mathcal{R}}_{cdab}&=&
-{\frac{2}{3}}\,\Theta\left(h_{ac}\omega_{bd}+\omega_{ac}h_{bd}
-h_{ad}\omega_{bc}-\omega_{ad}h_{bc}\right) \nonumber\\
&&-2\left(\sigma_{ac}\omega_{bd}+\omega_{ac}\sigma_{bd}
-\sigma_{ad}\omega_{bc}-\omega_{ad}\sigma_{bc}\right)\,.
\label{3Rsym2}
\end{eqnarray}
Therefore, in the absence of vorticity,
${\mathcal{R}}_{abcd}={\mathcal{R}}_{cdab}$ and the spatial Riemann
tensor possesses all the symmetries of its 4-D counterpart.

In analogy with $R_{ab}$ and $R$, the local Ricci tensor and Ricci
scalar of the 3-D space orthogonal to $u_{a}$ are defined by
\begin{equation}
{\mathcal{R}}_{ab}= h^{cd}{\mathcal{R}}_{cadb}=
{\mathcal{R}}^{c}{}_{acb} \hspace{15mm} \mathrm{and} \hspace{15mm}
{\mathcal{R}}= h^{ab}{\mathcal{R}}_{ab}\,,  \label{3Ricci}
\end{equation}
respectively. Thus, contracting (\ref{3Riemann2}) along the first
and third indices we arrive at what is usually referred to as the
Gauss-Codacci equation:
\begin{eqnarray}
{\mathcal{R}}_{ab}&=& E_{ab}+
{\frac{2}{3}}\,\left[\rho+{\frac{1}{2}}\left( E^{2}+B^{2}\right)
-{\frac{1}{3}}\,\Theta^{2}+\sigma^{2}-\omega^{2}\right]h_{ab}+
{\frac{1}{2}}\,(\pi_{ab}+{\mathcal{P}}_{ab}) \nonumber\\&&
-{\frac{1}{3}}\,\Theta(\sigma_{ab}+\omega_{ab})+ \sigma_{c\langle a}
\sigma^{c}{}_{b\rangle}- \omega_{c\langle a}
\omega^{c}{}_{b\rangle}+ 2\sigma_{c[a}\omega^{c}{}_{b]}\,.
\label{GC}
\end{eqnarray}
A further contraction leads to the trace of the above and provides
the local Ricci scalar of the 3-dimension space orthogonal to
$u_{a}$
\begin{equation}
{\mathcal{R}}= h^{ab}{\mathcal{R}}_{ab}=
2\left[\rho+{\frac{1}{2}}\left( E^{2}+B^{2}\right)
-{\frac{1}{3}}\,\Theta^{2}+\sigma^{2}-\omega^{2}\right]\,.
\label{3R}
\end{equation}
This expression is of major importance, since it is nothing else but
the generalized Friedmann equation. Indeed, when the electromagnetic
terms and those measuring anisotropy are removed, the above assumes
its familiar FRW form. Note that, on using (\ref{3R}), the
Gauss-Codacci formula (see Eq.~(\ref{GC})) reads
\begin{equation}
{\mathcal{R}}_{ab}= {\frac{1}{3}}\,{\mathcal{R}}h_{ab}+ E_{ab}+
{\frac{1}{2}}\,\left(\pi_{ab}+{\mathcal{P}}_{ab}\right)-
{\frac{1}{3}}\,\Theta(\sigma_{ab}+\omega_{ab})+ \sigma_{c\langle a}
\sigma^{c}{}_{b\rangle}- \omega_{c\langle a}\omega^{c}{}_{b\rangle}+
2\sigma_{c[a}\omega^{c}{}_{b]}\,.  \label{GC1}
\end{equation}
The latter may be used to calculate the curvature of the 3-space
along a chosen direction. For instance, $\mathcal{R}_{ab}B^{a}B^{b}$
gives the 3-curvature distortions along the magnetic field lines.

\section{Non-linear cosmological
magnetohydrodynamics}\label{sNlCMHD}
With the exception of any period of inflation and early reheating,
the universe has been a good conductor throughout its lifetime. As a
result, $B$-fields of cosmological origin must have remained frozen
into the expanding cosmic fluid during most of their evolution. This
allows us to study the magnetic effects on structure formation
within the ideal-MHD limits.

\subsection{Ideal MHD approximation}\label{ssIMHDA}
Consider a general spacetime filled with a single barotropic fluid
of very high conductivity. Ohm's law (see Eq.~(\ref{Ohm1}))
guarantees that in the frame of the fundamental observer the
electric field vanishes despite the presence of non-zero currents.
Therefore, in the ideal MHD limit the energy-momentum tensor of the
residual magnetic field simplifies to~\cite{TB1}
\begin{equation}
T_{ab}^{(B)}= {\frac{1}{2}}\,B^{2}u_{a}u_{b}+
{\frac{1}{6}}\,B^{2}h_{ab}+ \Pi_{ab}\,,  \label{TB}
\end{equation}
with
\begin{equation}
\Pi_{ab}= \Pi_{\langle ab\rangle}= {\frac{1}{3}}\,B^{2}h_{ab}-
B_{a}B_{b}\,.  \label{Pi}
\end{equation}
Accordingly, the $B$-field corresponds to an imperfect fluid with
energy density $\rho_{B}=B^{2}/2$, isotropic pressure
$p_{B}=B^{2}/6$, and anisotropic stresses represented by the
symmetric and trace-free tensor $\Pi_{ab}$.

Similarly, in the absence of an electric field, Maxwell's equations
reduce to a single propagation formula, namely the covariant
magnetic induction equation,
\begin{equation}
\dot{B}_{\langle a\rangle}=
\left(\sigma_{ab}+\varepsilon_{abc}\omega^{c} -{\frac{2}{3}}\,\Theta
h_{ab}\right)B^{b}\,,  \label{Bdot}
\end{equation}
and the following three constraints
\begin{eqnarray}
\mathrm{curl\,}B_{a}&=&{\mathcal{J}}_{a}-
\varepsilon_{abc}A^{b}B^{c}\,,  \label{C1}\\
\omega^{a}B_{a}&=& {\frac{1}{2}}\,\rho_{e}\,,  \label{C2}\\
\mathrm{D}^{a}B_{a}&=&0\,.  \label{C3}
\end{eqnarray}
The right-hand side of (\ref{Bdot}) is due to the relative motion of
the neighbouring observes and guarantees that the magnetic field
lines always connect the same matter particles~\cite{TB1}. This
means that the field remains frozen-in with the highly conducting
fluid. Expression (\ref{C1}) provides a direct relation between the
spatial currents, which are responsible for keeping the field lines
frozen-in with the matter, and the magnetic field itself
(e.g.~see~\cite{ZRS}). Note Eq.~(\ref{C2}) which shows that rotating
neighbouring observers will measure a non-zero charge density,
triggered by their relative motion, unless $\omega^{a}B_{a}=0$.
Finally, (\ref{C3}) demonstrates that in the absence of magnetic
monopoles the field lines remain closed and $B_{a}$ is a solenoidal
vector.

\subsection{Magnetic evolution}\label{ssME}
The magnetic induction equation also provides the non-linear
evolution law for the energy density of the field. More precisely,
contracting Eq.~(\ref{Bdot}) with $B_{a}$ and then using (\ref{Pi})
we arrive at
\begin{equation}
\left( B^{2}\right)^{\displaystyle\cdot}= -{\frac{4}{3}}\,\Theta
B^{2}- 2\sigma_{ab}\Pi^{ab}\,.  \label{B2dot}
\end{equation}
This shows that in a highly conducting cosmic medium we have
$B^{2}\propto a^{-4}$ always unless there is substantial anisotropy,
in which case the $B$-field behaves as an anisotropic radiative
fluid. In fact, in a homogeneous and anisotropic radiation-dominated
universe this latter situation arises even when the anisotropy is
small because close to isotropy $\Pi^{ab}\propto\rho$ and the
evolution of $B^{2}$ is determined at second-order with
$B^{2}/\rho_{rad}\rightarrow\sigma/\Theta\propto1/\log(t)$ during
the radiation era (see~\cite{Z,skew} and also~\cite{BFS,TM2}).

The nonlinear evolution of the anisotropic magnetic stresses comes
from the time derivative of (\ref{Pi}), which by means of
Eqs.~(\ref{Bdot}) and (\ref{B2dot}) leads to
\begin{equation}
\dot{\Pi}_{ab}= -{\frac{4}{3}}\,\Theta\Pi_{ab}-
{\frac{2}{3}}\,B^{2}\sigma_{ab} +2\sigma_{c\langle a}
\Pi^{c}{}_{b\rangle}- 2\omega_{c\langle a}\Pi^{c}{}_{b\rangle}\,.
\label{Pidot}
\end{equation}

\subsection{Conservation laws}\label{ssCL2}
The energy momentum tensor corresponding to a magnetic single
perfect fluid of infinite conductivity is given by
\begin{equation}
T_{ab}= \left(\rho+{\frac{1}{2}}\,B^{2}\right)u_{a}u_{b}+
\left(p+{\frac{1}{6}}\,B^{2}\right)h_{ab}+ \Pi_{ab}\,,
\label{MHDTab}
\end{equation}
Accordingly, the medium corresponds to an imperfect fluid with
effective density equal to $\rho+B^{2}/2$, isotropic pressure given
by $p+B^{2}/6$, zero heat flux and solely magnetic anisotropic
stresses represented by $\Pi_{ab}$ (see (\ref{Pi})).

When applied to the above, and using the MHD form of Maxwell's
equations, the standard conservation law $\nabla^{b}T_{ab}=0$
decomposes into the following expressions that respectively describe
the energy-density
\begin{equation}
\dot{\rho}= -(\rho+p)\Theta  \label{MHDedc}
\end{equation}
and the momentum-density
\begin{equation}
\left(\rho+p+{\frac{2}{3}}\,B^{2}\right)A_{a}= -\mathrm{D}_{a}p-
\varepsilon_{abc}B^{b}\mathrm{curl\,}B^{c}- \Pi_{ab}A^{b}\,.
\label{MHDmdc}
\end{equation}
conservation~\cite{TB1,TM1}. Note the absence of magnetic terms in
Eq.~(\ref{MHDedc}). This is guaranteed by the magnetic induction
equation (\ref{Bdot}) and reflects the fact that the magnetic energy
density is separately conserved.\footnote{Expressions (\ref{MHDedc})
and (\ref{MHDmdc}) can also be obtained from the conservation laws
(\ref{edc2}) and (\ref{mdc2}). This is done by taking the ideal-MHD
limit of the latter, using Eq.~(\ref{C1}) and assuming perfect-fluid
matter.} Also, when there are no pressure gradients, (\ref{MHDmdc})
gives $A_{a}B^{a}=0$ to ensure that the field exerts no forces along
its own direction. Finally, the left-hand side of (\ref{MHDmdc})
shows that the magnetic contribution to the total inertial mass of
the system is $2B^{2}/3$.

Contracting Eq.~(\ref{MHDmdc}) along $B_{a}$ we find that the
contribution of the $B$-field to the momentum density vanishes, thus
guaranteeing that the magnetic Lorentz force is always normal to the
field lines. Also, the second term in the right-hand side of
(\ref{MHDmdc}) decomposes as
\begin{equation}
\varepsilon_{abc}B^{b}\mathrm{curl\,}B^{c}=
{\frac{1}{2}}\,\mathrm{D}_{a}B^{2}- B^{b}\mathrm{D}_{b}B_{a}\,.
\label{Lorentz}
\end{equation}
The last term in the above is the result of the magnetic tension. In
so far as this tension stress is not balanced by the pressure
gradients, the field lines are out of equilibrium and there is a
non-zero Lorentz force acting on the particles of the magnetic
fluid.

\subsection{Magnetic tension}\label{ssMT}
The anisotropic nature of magnetic fields is encoded in the
energy-momentum tensor of the field -- and particularly in the
anisotropic pressure tensor (see (\ref{Pi})). This ensures that the
$B$-fields exert a positive pressure orthogonal to their own
direction, while carrying a tension along $B_{a}$. Both of these
very well known magnetic features are reflected in the eigenvalues
of $\Pi_{ab}$, which are positive ($1/3$) perpendicular to $B_{a}$
and negative ($-2/3$) parallel to it.

The magnetic tension also demonstrates the elasticity of the field
lines and their tendency to remain as straight as possible by
reacting to any effect that distorts them from
equilibrium~\cite{P,Me}. Within the context of general relativity,
we can see this tendency by looking at the effect of the field on
the spatial curvature of a magnetic spacetime. In particular,
contracting (\ref{GC}) along $B_{a}$ twice, and ignoring all sources
but the magnetic field, we find that~\cite{T1}
\begin{equation}
{\mathcal{R}}_{ab}B^{a}B^{b}= {\frac{1}{3}}\,B^{4}+
{\frac{1}{2}}\,\Pi_{ab}B^{a}B^{b}=0\,,
\end{equation}
given that $\Pi_{ab}B^{b}=-(2B^{2}/3)B_{a}$ (see (\ref{Pi})).
Accordingly, despite the magnetic presence and the energy density
contribution of the field, the curvature of the 3-space in the
direction of the magnetic force lines is zero. Mathematically
speaking, it is the contribution of the negative magnetic pressure
along $B_{a}$ which cancels out the positive input of the field's
energy density. More intuitively, however, one could argue that it
is the elasticity of the magnetic lines, and their tendency to
remain straight, that maintains the zero curvature along $B_{a}$.

In the presence of sources, ${\mathcal{R}}_{ab}B^{a}B^{b}$ is
generally non-zero and the magnetic lines are forced out of
equilibrium. When allowing for spatial inhomogeneities, the reaction
of the field's tension, as expressed through the second stress in
the right-hand side of (\ref{Lorentz}), to these geometrically
induced distortions generally leads to counter-intuitive effects.
The potential implications of these tension stresses, for cosmology
as well as astrophysics, have been discussed in~\cite{MT}-\cite{T3}.
In spatially homogeneous spacetimes on the other hand, like the
Bianchi~$I$ models, the magnetic Lorentz force vanishes and the
field's tension is manifested only as a negative energy-density
input, through the anisotropic pressure tensor ($\Pi_{ab}$ -- see
Eq.~(\ref{Pi})). Such cosmologies were recently investigated
in~\cite{KC2}.

\subsection{Kinematical evolution}\label{ssKE}
The magnetic presence affects the kinematics of the highly
conducting medium both directly and indirectly. In particular,
Raychaudhuri's equation reads
\begin{equation}
\dot{\Theta}= -{\frac{1}{3}}\,\Theta^{2}-
{\frac{1}{2}}\,(\rho+3p+B^{2})- 2(\sigma^{2}-\omega^{2})+
\mathrm{D}^{a}A_{a}+ A_{a}A^{a}\,,  \label{MHDRay}
\end{equation}
showing that the field's contribution to the total gravitational
mass is $B^{2}$. There are also indirect magnetic effects
propagating through the rest of the terms in the right-hand side of
(\ref{MHDRay}). Of these effects, probably the most important are
carried by the acceleration terms which follow from
Eq.~(\ref{MHDmdc}). Similarly, there are direct and indirect
magnetic effects on the shear evolution. In the magnetic environment
the propagation equation of the shear tensor takes the form
\begin{equation}
\dot{\sigma}_{\langle ab\rangle}= -{\frac{2}{3}}\,\Theta\sigma_{ab}-
\sigma_{c\langle a}\sigma^{c}{}_{b\rangle}- \omega_{\langle a}
\omega_{b\rangle}+ \mathrm{D}_{\langle a}A_{b\rangle}+ A_{\langle a}
A_{b\rangle}- E_{ab}+ {\frac{1}{2}}\,\Pi_{ab}\,,
\label{MHDsigmadot}
\end{equation}
which reveals how the anisotropic pressure of the field directly
induces and also affects kinematical anisotropies. Rotation, on the
other hand hand, is affected only indirectly through the magnetic
effects on the 4-acceleration of the fluid. To be precise,
\begin{equation}
\dot{\omega}_{\langle a\rangle}= -{\frac{2}{3}}\,\Theta\omega_{a}-
{\frac{1}{2}}\,\mathrm{curl\,}A_{a}+ \sigma_{ab}\omega^{b}\,.
\label{MHDomegadot}
\end{equation}
This means that the $B$-field not only affects the rotational
behaviour of the observers' frame but it can also generate
vorticity.

The kinematical constraints are also only indirectly affected by the
field's presence. This is manifested by the absence of explicit
magnetic terms in Eqs.~(\ref{shearcon})-(\ref{Hcon}). When applied
to a highly conducting barotropic fluid expressions (\ref{omegacon})
and (\ref{Hcon}) remain unchanged, while (\ref{shearcon}) reduces to
\begin{equation}
\mathrm{D}^{b}\sigma_{ab}= {\frac{2}{3}}\,\mathrm{D}_{a}\Theta+
\mathrm{curl\,}\omega_{a}+ 2\varepsilon_{abc}A^{b}\omega^{c}\,.
\label{MHDshearcon}
\end{equation}
When $\mathrm{D}^{a}\sigma_{ab}=0$, the shear is divergence-free and
its pure-tensor part has been isolated. In other words, there is no
way of using $\sigma_{ab}$ to construct a scalar or a vector. This
condition will be imposed later, in \S~\ref{sLGWs}, when studying
the magnetic effects of gravitational-wave perturbations.

\subsection{Spacetime curvature}\label{ssStC}
The energy-momentum tensor of a highly conducting magnetic perfect
fluid is given by Eqs.~(\ref{TB}), (\ref{Pi}). In that case formulae
(\ref{EFE1})-(\ref{EFE3}) reduce to
\begin{equation}
R_{ab}u^{a}u^{b}= {\frac{1}{2}}\,\left(\rho+3p+B^{2}\right)\,,
\hspace{25mm} h_{a}{}^{b}R_{bc}u^{c}=0  \label{MHDEFE1}
\end{equation}
and
\begin{equation}
h_{a}{}^{c}h_{b}{}^{d}R_{cd}= \left[{\frac{1}{2}}\,
\left(\rho-p+{\frac{1}{3}}\,B^{2}\right)\right] h_{ab}+ \Pi_{ab}\,,
\label{MHDEFE2}
\end{equation}
respectively. We note that the above are fully non-linear
expressions and that the former quantifies the total gravitational
mass of the MHD fluid (in agreement with (\ref{MHDRay})).

We evaluate the magnetic contribution to the local 3-Ricci tensor
orthogonal to the $u_a$-congruence, by taking the MHD limit of the
geometrical relations given in \S~\ref{ssSC}. Expressions (\ref{GC})
and (\ref{3R}), in particular, simplify to
\begin{eqnarray}
{\mathcal{R}}_{ab}&=& E_{ab}+
{\frac{2}{3}}\,\left(\rho+{\frac{1}{2}}\,B^{2}
-{\frac{1}{3}}\,\Theta^{2}+\sigma^{2}-\omega^{2}\right)h_{ab}
+{\frac{1}{2}}\,\Pi_{ab}-
{\frac{1}{3}}\,\Theta(\sigma_{ab}+\omega_{ab}) \nonumber\\
&&+\sigma_{c\langle a}\sigma^{c}{}_{b\rangle}- \omega_{c\langle a}
\omega^{c}{}_{b\rangle}+ 2\sigma_{c[a}\omega^{c}{}_{b]}
\label{MHDGC}
\end{eqnarray}
and
\begin{equation}
{\mathcal{R}}= 2\left(\rho+{\frac{1}{2}}\,B^{2}
-{\frac{1}{3}}\,\Theta^{2}+\sigma^{2}-\omega^{2}\right)\,,
\label{MHD3R}
\end{equation}
respectively. This is simply the familiar Friedmann equation
generalized to the magnetic and highly conducting environment.

\subsection{Evolution of inhomogeneities}\label{ssEIs}
\subsubsection{Basic variables}\label{sssBVs}
Spatial inhomogeneities of a given physical quantity are covariantly
described by its orthogonally projected gradients. For our purposes,
the basic variables are the comoving fractional gradients,
\begin{equation}
\Delta_{a}= \frac{a}{\rho}\;\mathrm{D}_{a}\rho \hspace{15mm}
\mathrm{and} \hspace{15mm} {\mathcal{B}}_{a}=
\frac{a}{B^{2}}\;\mathrm{D}_{a}B^{2}\,,  \label{cDacBa}
\end{equation}
describing spatial variations in the fluid and the magnetic energy
density distributions (see~\cite{EB}-\cite{EBH}) and~\cite{TB1,TB2}
respectively). Both variables are dimensionless by construction, and
vanish identically when the spacetime is spatially homogeneous. As
we will show later, $\Delta_{a}$ and ${\mathcal{B}}_{a}$
respectively describe spatial variations in the fluid and the
magnetic energy densities, as measured by a pair of neighbouring
fundamental observers, in a gauge-invariant way (see
\S~\ref{sssLIIVs}).

The above are supplemented by a pair of auxiliary variables, which
describe spatial inhomogeneities in the average expansion and the
isotropic pressure of the fluid. These are~\cite{EB}
\begin{equation}
{\mathcal{Z}}_{a}= a\mathrm{D}_{a}\Theta\hspace{15mm} \mathrm{and}
\hspace{15mm} Y_{a}=\mathrm{D}_{a}p\,,  \label{cZaYa}
\end{equation}
respectively. It should be noted that, for barotropic fluids, the
variable $Y_{a}$ is directly related to $\Delta_{a}$ and therefore
is redundant for all practical purposes. Finally, one may also
monitor spatial inhomogeneities in the distribution of the magnetic
field vector by means of the orthogonally projected
tensor~\cite{TB1}
\begin{equation}
{\mathcal{M}}_{ab}= \mathrm{D}_{b}B_{a}\,.  \label{Mab}
\end{equation}
This is trace-free in the ideal MHD limit (see constraint
(\ref{C3})) but, contrary to the gradients (\ref{cDacBa}) and
(\ref{cZaYa}), it does not vanish unless the FRW metric is spatially
flat (see Appendix~A in~\cite{TB2} and also \S~\ref{sssGIIVs} here).

\subsubsection{Evolution equations}\label{sssEEs}
In the presence of magnetic fields, the non-linear evolution of
spatial inhomogeneities in the density distribution of a single,
highly conducting perfect fluid is described by the expression
\begin{eqnarray}
\dot{\Delta}_{\langle a\rangle} &=& w\Theta\Delta_{a}-
(1+w)\mathcal{Z}_{a}+
\frac{a\Theta}{\rho}\,\varepsilon_{abc}B^{b}\mathrm{curl\,}B^{c}+
{\frac{2}{3}}\,c_{\mathrm{a}}^{2}(1+w)a\Theta A_{a}\nonumber\\
&&-\left( \sigma_{ba}+\omega_{ba}\right) \Delta^{b}+
\frac{a\Theta}{\rho }\,\Pi_{ab}A^{b}\,,  \label{dotcDa}
\end{eqnarray}
with $w=p/\rho$ and $c_{\rm a}^2=B^2/\rho(1+w)$ representing the
Alfv\'{e}n speed (see \S~\ref{ssMSAS} below). This equation has been
obtained by taking the proper-time derivative of (\ref{cDacBa}a) and
then projecting it onto the observer's 3-dimensional instantaneous
rest space. In the process, we have also employed the conservation
laws (\ref{MHDedc}) and (\ref{MHDmdc}).

Similarly, starting from definition (\ref{cZaYa}a) and using
(\ref{MHDmdc}), (\ref{MHDRay}), we arrive at the following
non-linear evolution equation for the expansion gradients
\begin{eqnarray}
\dot{{\mathcal{Z}}}_{\langle a\rangle}&=&
-{\frac{2}{3}}\,\Theta{\mathcal{Z}}_{a}-
{\frac{1}{2}}\,\rho\Delta_{a}-
{\frac{1}{2}}\,B^{2}{\mathcal{B}}_{a}+
{\frac{3}{2}}\,a\varepsilon_{abc}B^{b}\mathrm{curl\,}B^{c}+
a\mathrm{D}_{a}A \nonumber\\ &&+\left[{\frac{1}{2}}\,{\mathcal{R}}-
3(\sigma^{2}-\omega^{2})+A +A_{b}A^{b}\right]aA_{a}+
2aA^{b}\mathrm{D}_{a}A_{b}\nonumber\\
&&-(\sigma_{ba}+\omega_{ba}){\mathcal{Z}}^{b}+
{\frac{3}{2}}\,a\Pi_{ab} A^{b}-
2a\mathrm{D}_{a}(\sigma^{2}-\omega^{2})\,,  \label{dotcZa}
\end{eqnarray}
where $A=\mathrm{D}^{a}A_{a}$ and ${\mathcal{R}}$ is the Ricci
scalar of the observer's local 3-D rest space.

Finally, the orthogonally-projected time derivative of
(\ref{cDacBa}b) leads to the non-linear propagation formula
monitoring spatial inhomogeneities in the magnetic energy density
\begin{eqnarray}
\dot{\mathcal{B}}_{\langle a\rangle}&=&
\frac{4}{3(1+w)}\,\dot{\Delta}_{\langle a\rangle}-
\frac{4w\Theta}{3(1+w)}\,\Delta_{a}-
\frac{4a\Theta}{3\rho(1+w)}\,\varepsilon_{abc}B^{b}\mathrm{curl\,}B^{c}-
{\frac{4}{3}}\,a\Theta\left(1+{\frac{2}{3}}\,c_{\mathrm{a}}^{2}\right)A_{a}
\nonumber\\
&&-\left(\sigma_{ba}+\omega_{ba}\right){\mathcal{B}}^{b}+
\frac{4}{3(1+w)}\,\left(\sigma_{ba}+\omega_{ba}\right)\Delta^{b}-
\frac{4a\Theta}{3\rho(1+w)}\,\Pi_{ab}A^{b}-
\frac{2a}{B^{2}}\,\Pi^{bc}\mathrm{D}_{a}\sigma_{bc} \nonumber\\
&&-\frac{2a}{B^{2}}\,\sigma^{bc}\mathrm{D}_{a}\Pi_{bc}+
\frac{2}{B^{2}}\,\sigma_{bc}\Pi^{bc}{\mathcal{B}}_{a}-
\frac{2a}{B^{2}}\,\sigma_{bc}\Pi^{bc}A_{a}\,.  \label{dotcBa}
\end{eqnarray}
In deriving the above, we have employed the non-linear relations
(\ref{Bdot}) and (\ref{B2dot}), as well as (\ref{dotcDa}). This
helped to eliminate the expansion gradients from the right-hand
sideof Eq.~(\ref{dotcBa}). We emphasize that (\ref{dotcBa}) cannot
beused when the equation of state of the cosmic medium has the
vacuumenergy form contributed by a cosmological constant in the
Einsteinequations, namely for $w=p/\rho=-1$. In that case we can no
longeruse (\ref{dotcDa}) to eliminate ${\mathcal{Z}}_{a}$ from
(\ref{dotcBa}) and the evolution of density inhomogeneities is
described by a different system of equations (see \S~\ref{ssEFvEs}).

\section{Perturbed magnetic FRW cosmologies}\label{sPMFRWC}
The symmetries of FRW spacetimes cannot naturally accommodate
generic anisotropic sources like magnetic fields. This means that
when based on a Friedmann background, any magnetic study will
require some degree of approximation. Nevertheless, it is
intuitively plausible that sufficiently weak $B$-fields can be
adequately studied within perturbed almost-FRW models. This belief,
which has been at the core of almost every study of cosmological
magnetic fields, has been confirmed by the analysis of
exact~\cite{skew} and perturbed magnetic Bianchi~$I$
universes~\cite{TM2}.

\subsection{Background evolution}\label{ssBE}
\subsubsection{Zero-order approach}\label{sssZoA}
We consider a spatially flat Friedmann background and allow for the
presence of a weak magnetic field. This is effectively a test field
that does not disrupt isotropy and has $B^{2}\ll\rho$. The
background $B$-field can be either homogeneous or sufficiently
random, so that $\langle B_{a}\rangle=0\neq\langle B^{2}\rangle$
(with $\langle B^{2}\rangle$ homogeneous on all the scales of
interest). In both cases the magnetic anisotropic stress $\Pi_{ab}$
is treated as a first-order perturbation and the only zero-order
magnetic variable is $B^{2}$. First-order perturbations arise
through $\Pi_{ab} $, the spatial gradients $\mathrm{D}_{b}B^{2}$ and
the spatial derivatives $\mathrm{D}_{b}B_{a}$. If the background
test field is homogeneous, the scalar $B^{2}$ describes its local
density and pressure. For a stochastic test field, $B^{2}$ describes
the average density and pressure of the background field. In either
case, the form of the linearized MHD equations guarantees that the
magnetic effects propagate only via $B^{2}$-terms (see
\S~\ref{ssLEs1}) and that, at least when studying scalar
(i.e.~density) perturbations, the two approaches lead to the same
linear equations and results.

Alternatively, one can treat the magnetic field as a first-order
perturbation, usually as a stochastic Gaussian field
(e.g.~see~\cite{Gi6,KR}). One can adapt our analysis to this scheme
by removing the magneto-curvature terms from all the linear formulae
(see \S~\ref{ssLEs1}).

\subsubsection{Zero-order equations}\label{sssZoEs}
Locally, the evolution of the background is governed by the standard
Friedmann and Raychaudhuri equations. Introducing $H=\dot{a}/a$ as
our zero-order Hubble parameter, these are
\begin{equation}
3H^{2}= \rho\hspace{15mm} \mathrm{and} \hspace{15mm} \dot{H}=
-H^{2}- {\frac{1}{6}}\,\rho(1+3w)\,,  \label{bFriedbRay}
\end{equation}
and are supplemented by the conservation laws for the matter and the
magnetic energy densities, namely by
\begin{equation}
\dot{\rho}= -3H\rho(1+w) \hspace{15mm} \mathrm{and} \hspace{15mm}
\left(B^{2}\right)^{\displaystyle\cdot}= -4HB^{2}\,,
\label{bedcbmdc}
\end{equation}
respectively.\footnote{The spatially averaged magnetic energy
density depends only on time and evolves in tune with its local
counterpart. Indeed, using the local result (\ref{bedcbmdc}b), it is
straightforward to show that $\langle
B^{2}\rangle^{\displaystyle\cdot}= -4H\langle B^{2}\rangle$ on any
background domain because $\langle H\rangle=H$ in FRW models
(e.g.~see~\cite{B1} for details).} This is obtained from the
magnetic induction equation (see (\ref{Bdot})), which guarantees
that
\begin{equation}
\dot{B}_{a}= -2HB_{a}\,.  \label{bBdot}
\end{equation}
Note that the weakness of the $B$-field allows us to neglect the
magnetic contribution to the total energy density and gravitational
mass of the system. This explains the absence of magnetic terms from
the right-hand sides of (\ref{bFriedbRay}a) and (\ref{bFriedbRay}b),
provided $w\neq-1/3$ in the latter case. Also, ~(\ref{bedcbmdc}b)
guarantees the radiation-like evolution $B^{2}\propto a^{-4}$ for
the zero-order field, which means that the background magnetic flux
is conserved.

\subsubsection{Specifying the magnetic medium}\label{ssSMM}
The nature of the cosmic medium is determined by its equation of
state. Here, we will be dealing with a highly conducting barotropic
fluid with $p=w\rho$, where $w$ is the barotropic index. This
vanishes in the case of dust, takes the value $w=1/3$ for radiation
and indicates a stiff fluid when $w=1$. Also, for $w<-1/3$ the total
gravitational mass of the matter is negative and the model enters a
period of accelerated expansion, which corresponds to exponential de
Sitter-type inflation when $w=-1$. We find that
\begin{equation}
\dot{w}= -(1+w)(c_{s}^{2}-w)\Theta\,,  \label{ldotw}
\end{equation}
where $c_{s}^{2}=\mathrm{d}p/\mathrm{d}\rho$ is the square of the
adiabatic sound speed. Therefore, a time-invariant barotropic index
in a non-static universe (i.e.~$\dot{w}=0$ and $\Theta\neq0$) means
that $c_{s}^{2}=w=$~constant, and vice versa.

\subsubsection{Magnetic strength and Alfv\'en speed}\label{ssMSAS}
The magnetic strength, relative to the matter component, is measured
by the dimensionless energy-density ratio $\beta=B^{2}/\rho$. Then,
given the weakness of the field (i.e.~since $\beta\ll1$), the square
of the Alfv\'{e}n speed is conveniently defined as
\begin{equation}
c_{\mathrm{a}}^{2}= \frac{\beta}{1+w}\,,  \label{Alfven1}
\end{equation}
ensuring that $c_{\mathrm{a}}^{2}\ll1$ at all times. The only
exception is when $w\rightarrow-1$, in which case the proper
definition for the Alfv\'{e}n speed is
$c_{\mathrm{a}}^{2}=\beta/(1+w+\beta)$.\footnote{In the standard
literature the square of the Alfv\'{e}n speed is $c_{\mathrm{a}}^{2}
=B^{2}/\rho=\beta$ (e.g.~see~\cite{Ja}).} Thus, from now on,
$w\neq-1$ unless stated otherwise. On using (\ref{ldotw}) and
Eqs.~(\ref{bedcbmdc}), the time derivative of the above gives
\begin{equation}
(c_{\mathrm{a}}^{2})^{\displaystyle\cdot}=
-(1-3c_{s}^{2})Hc_{\mathrm{a}}^{2}\,,  \label{dotca2}
\end{equation}
to zero order. Accordingly, the Alfv\'{e}n speed does not change
with time when $c_{s}^{2}=1/3$, which agrees with the radiation-like
linear evolution of the magnetic energy density in almost-FRW
environments.\footnote{When shear anisotropy is included, the
parameter $\beta$ and therefore $c_{\mathrm{a}}^{2}$ are no longer
constant during the radiation-dominated epoch but display a slow,
`quasi static', logarithmic decay (see~\cite{Z,skew} and
also~\cite{BFS,TM2}).}

\subsection{Linear evolution}\label{ssLE1}
We linearize the full equations by treating quantities with non-zero
background values as zero-order variables in perturbative terms. On
the other hand, quantities that vanish in the background are of
order one and higher-order terms are ignored. Therefore, the only
zero-order quantities in our case are the fluid energy density
($\rho$) and pressure ($p$), the magnetic energy density ($B^{2}$),
and the Hubble parameter ($H$).

The linear conservation law of the magnetic energy density maintains
its background form, given by Eq.~(\ref{bedcbmdc}b). Similarly, the
linear propagation of the magnetic vector is still given by
(\ref{bBdot}), which implies that $B_{a}\propto a^{-2}$ and that the
magnetic flux remains conserved. On the other hand, the scalar
$B^{a}\mathrm{curl\,}B_{a}$, which is used to measure the helicity
of the field (e.g.~see~\cite{CDK}), obeys the linear evolution law
\begin{equation}
\left(B^{a}\mathrm{curl\,}B_{a}\right)^{\displaystyle\cdot}=
-5HB^{a}\mathrm{curl\,}B_{a}\,.  \label{lmh}
\end{equation}
The above means that the magnetic helicity decays as $a^{-5}$ and
therefore it decreases faster than the energy density of the
$B$-field (recall that $B^{2}\propto a^{-4}$ throughout the linear
regime). Note that in deriving (\ref{lmh}) we have assumed overall
charge neutrality and used Eq.~(\ref{bBdot}), together with the
first-order expression
$B^{a}(\mathrm{curl\,}B_{a})^{\displaystyle\cdot}=
-3HB^{a}\mathrm{curl\,}B_{a}$, obtained by means of commutator
(\ref{A5}) in \S~\ref{AssTDs}.

To first order, the conservation law of the fluid energy-density
retains its the background form (see Eq.~(\ref{bedcbmdc}a)).
However, when linearized around a FRW background, the
momentum-density conservation equation (\ref{MHDmdc}) reduces to
\begin{equation}
(1+w)\left(1+{\frac{2}{3}}\,c_{\mathrm{a}}^{2}\right)\rho A_{a}=
-c_{s}^{2}\mathrm{D}_{a}\rho-
\varepsilon_{abc}B^{b}\mathrm{curl\,}B^{c}\,,  \label{lmdc1}
\end{equation}
with $\mathrm{D}_{a}p=c_{s}^{2}\mathrm{D}_{a}\rho$ due to the
barotropic nature of the medium.

The magnetic field's presence affects the linear kinematics in a
number of ways. The average volume expansion, for example, is
determined by the following version of the Raychaudhuri equation
\begin{equation}
\dot{\Theta}= -{\frac{1}{3}}\,\Theta^{2}-
{\frac{1}{2}}\,\rho(1+3w+\beta)- \frac{c_{s}^{2}}{a(1+w)}
\left(1-{\frac{2}{3}}\,c_{\mathrm{a}}^{2}\right)
\mathrm{D}^{a}\Delta_{a}-
{\frac{c_{\mathrm{a}}^{2}}{2a}}\;\mathrm{D}^{a}\mathcal{B}_{a}+
{\frac{1}{3}}\,c_{\mathrm{a}}^{2}{\mathcal{R}}\,,  \label{lRay}
\end{equation}
where ${\mathcal{R}}=h^{ab}{\mathcal{R}}_{ab}$ is the 3-Ricci scalar
and we have kept up to $c_{\mathrm{a}}^{2}$-order terms. As
expected, any local increase in the magnetic energy density will
decelerate the linear expansion of the model in the same way that
matter perturbations do. However, the $B$-field has an additional
effect due to the magneto-curvature term in the right-hand side of
(\ref{lRay}). This further decelerates the expansion if
${\mathcal{R}}<0$, but tends to accelerate it when ${\mathcal{R}}$
is positive. This counter-intuitive behaviour, whichresults from the
tension of the field, can have nontrivial and veryunexpected
implications for spatially curved magnetic
spacetimes~\cite{MT}-\cite{T3}.

The rotational behaviour of the weakly magnetic fluid is described
by the vorticity propagation equation, which to linear order takes
the form
\begin{equation}
\dot{\omega}_{a}= -2\left[1-{\frac{3}{2}}\,c_{s}^{2}
\left(1-{\frac{2}{3}}\,c_{\mathrm{a}}^{2}\right)
-{\frac{1}{6}}\,c_{\mathrm{a}}^{2}\right]H\omega_{a}-
\frac{1}{2\rho(1+w)}\left(1-{\frac{2}{3}}\,c_{\mathrm{a}}^{2}\right)
B^{b}\mathrm{D}_{b}\mathrm{curl\,}B_{a}\,.  \label{lomegadot}
\end{equation}
This means that the magnetic presence will generate vorticity unless
the field's distribution is curl-free, or more specifically, unless
$\mathrm{curl\,}B_{a}$ remains unchanged along the direction of the
magnetic force lines. Thus, $B$-fields can act as sources of
rotation. The magnetic effect on pre-existing vorticesis more
difficult to quantify. As we will see later, in
section\S~\ref{ssEDE3}, the presence of the field can slow down the
standard decay rate of vorticity and therefore increase the
residualamount of rotation relative to magnetic-free cosmologies.
Recallthat in the absence of the $B$-field, vorticity always decays
unless$c_{s}^{2}>2/3$~\cite{Ba2}.

The evolution of linear kinematical anisotropies in the magnetic
fluid is governed by the following expression of the shear
propagation equation
\begin{eqnarray}
\dot{\sigma}_{ab}&=& -2H\sigma_{ab}+ {\frac{1}{2}}\,\Pi_{ab}-
E_{ab}- \frac{c_{s}^{2}}{a(1+w)}
\left(1-{\frac{2}{3}}\,c_{\mathrm{a}}^{2}\right)\;
\mathrm{D}_{\langle a}\Delta_{b\rangle}-
{\frac{c_{\mathrm{a}}^{2}}{2a}}\;\mathrm{D}_{\langle a}
\mathcal{B}_{b\rangle} \nonumber\\
&&+{\frac{1}{\rho(1+w)}}
\left(1-{\frac{2}{3}}\,c_{\mathrm{a}}^{2}\right)
B^{c}\mathrm{D}_{c}\mathrm{D}_{\langle a}B_{b\rangle}+
{\frac{1}{3}}\,c_{\mathrm{a}}^2\,\mathcal{R}_{\langle ab\rangle}\,.
\label{lsigmadot}
\end{eqnarray}
The magnetic effects are diverse. The anisotropic pressure of the
field is supplemented by anisotropies in the distribution of the
magnetic vector and by those in the energy density of the $B$-field.
In addition, there is a purely geometrical magneto-curvature effect,
provided that ${\mathcal{R}}_{\langle ab\rangle}$ is non-zero at the
linear level.

In the ideal MHD limit the kinematical variables also satisfy
constraints (\ref{MHDshearcon}), (\ref{omegacon}) and (\ref{Hcon}),
which then linearize to
\begin{equation}
\mathrm{D}^{b}\sigma_{ab}= {\frac{2}{3}}\,\mathrm{D}_{a}\Theta+
\mathrm{curl\,}\omega_{a}\,, \hspace{15mm} \mathrm{D}^{a}\omega_{a}=
0\label{lcon1}
\end{equation}
and
\begin{equation}
H_{ab}= \mathrm{curl\,}\sigma_{ab}+ \mathrm{D}_{\langle a}
\omega_{b\rangle}\,,  \label{lHcon}
\end{equation}
respectively. According to the above, $\omega_{a}$ is a linear
solenoidal vector and in the absence of rotation the magnetic Weyl
component is fully determined by the shear tensor. Additional
(geometrical) constraints between the kinematical and the dynamical
quantities in the highly conducting medium are provided by
(\ref{MHDGC}) and (\ref{MHD3R}). The respective linearized
counterparts of these expressions are
\begin{equation}
{\mathcal{R}}_{ab}= E_{ab}+ {\frac{2}{3}}\,
\left(\rho+{\frac{1}{2}}\,B^{2} -{\frac{1}{3}}\,\Theta^{2}\right)
h_{ab}+ {\frac{1}{2}}\,\Pi_{ab}- H(\sigma_{ab}+\omega_{ab})
\label{lGC}
\end{equation}
and
\begin{equation}
{\mathcal{R}}= 2\left(\rho+{\frac{1}{2}}\,B^{2}
-{\frac{1}{3}}\,\Theta^{2}\right)\,. \label{l3R}
\end{equation}

\subsection{Spatial inhomogeneities}\label{ssMSIs}
\subsubsection{Local interpretation of the inhomogeneity
variables}\label{sssLIIVs}
To linear order, the inhomogeneity variables defined in
section~\ref{sssBVs} describe measurable differences in the spatial
distribution of physical quantities. In order to verify this,
consider the relative position vector $\chi_{a}$ connecting the same
two points on the worldlines of neighbouring fundamental observers.
Following~\cite{EvE,EB}, we have $\chi_{a}u^{a}=0$ and
\begin{equation}
\dot{\chi}_{a}= H\chi_{a}+ (\sigma_{ab}+\omega_{ab})\chi^{b}\,,
\label{deltaxdot}
\end{equation}
which means that in a FRW universe $\chi_{a}=a\chi_{a}^{0}$ (with
$\chi_{a}^{0}=$ constant and $a_{0}=1$). If $P$ and $\tilde{P}$ are
points on the two worldlines with coordinates ($x_{a}$) and
($x_{a}+\chi_{a}$) respectively and $\rho$, $\tilde{\rho}$ are the
associated values of the energy density, a Taylor expansion around
$P$ gives
\begin{equation}
\tilde{\rho}-\rho= \chi^{a}\mathrm{D}_{a}\rho\,,  \label{deltarho}
\end{equation}
to leading order. Since $\chi_{a}\propto a$ in the FRW background
and using definition (\ref{cDacBa}a), the above translates into
\begin{equation}
\delta\rho=\chi_0^a\Delta_{a}\,,  \label{cDa}
\end{equation}
where $\delta\rho=(\tilde{\rho}-\rho)/\rho$. In other words, the
comoving fractional gradient $\Delta_{a}$ describes the measurable
local density variation between two neighbouring fundamental
observers. Moreover, $\Delta_{a}$ closely corresponds to the
familiar energy-density contrast of the non-covariant studies. The
same analysis also applies to the rest of the variables defined
in~\ref{sssBVs}.

\subsubsection{Gauge invariance of the inhomogeneity
variables}\label{sssGIIVs}
Cosmological perturbations have long been known to suffer from
gauge-related ambiguities, reflecting the fact that in perturbation
theory one deals with two spacetime manifolds. The first is the
physical spacetime ($\mathcal{W}$) that corresponds to the real
universe, while the second ($\mathcal{\overline{W}}$) is a
fictitious background described by an idealised mathematical model.
To proceed we need to establish an one-to-one correspondence, namely
a gauge $\phi:\mathcal{\overline{W}}\rightarrow\mathcal{W}$, between
these two manifolds. Such point identification maps are generally
arbitrary and the gauge problem stems from our inherent freedom to
make gauge transformations. The latter differ from ordinary
coordinate transformations because they also change the point
identification between $\mathcal{\overline{W}}$ and $\mathcal{W}$.

We define perturbations as the difference between the value of a
given quantity at some event in the realistic universe and its value
at the corresponding (through the gauge) event in the background
spacetime. Gauge transformations, however, generally change this
correspondence and therefore the value of the perturbation itself.
This makes perturbations gauge-dependent and arbitrary. For
instance, we can select the gauge in such a way that a given
perturbation vanishes~\cite{EB}. Following the Stewart \& Walker
lema, the simplest quantities that remain invariant under
gauge-transformations are scalars that are constant in the
background spacetime and tensors that vanish
there~\cite{SW}.\footnote{The only alternative way of constructing
gauge-independent variables is by using tensors which are linear
combinations of products of the Kronecker deldas with constant
coefficients~\cite{SW}.} Note that the same general criteria also
apply to second-order perturbations, but this time the Stewart \&
Walker requirements must be satisfied by the first-order
variables~\cite{BMMS}.

In a Friedmann universe all physical variables are functions of
cosmic time only, though they acquire additional spatial dependence
in perturbed `nearly-FRW' universes. As a result of the spatial
homogeneity of the unperturbed Friedmann background, the variables
(\ref{cDacBa}) and (\ref{cZaYa}) vanish to zero order. Consider the
density gradient for example. Then,
\begin{equation}
\mathrm{D}_{a}\rho= h_{a}{}^{b}\nabla_{b}\rho=
h_{a}{}^{0}\nabla_{0}\rho+ h_{a}{}^{\alpha}\nabla_{\alpha}\rho= 0\,,
\label{gi}
\end{equation}
because $\nabla_{\alpha}\rho=0$ and in a comoving frame
$h_{a}{}^{0}=0$ (with $\alpha=1,2,3$ -- see also~\cite{EB}). This
result guarantees the vanishing of the inhomogeneity variables
(\ref{cDacBa}), (\ref{cZaYa}), in the background, and therefore
ensures their gauge invariance at the linear perturbative
level~\cite{SW}. Note, however, that the gauge invariance of
$\mathrm{D}_{b}B_{a}$, and therefore of ${\mathcal{M}}_{ab}$, are
guaranteed only in spatially flat FRW backgrounds~\cite{TB2}. One
can see that by applying the 3-Ricci identity to a zero-order
3-vector $v_{a}$ (see Eq.~(\ref{A2}) in \S~\ref{AssOPGs}). In the
absence of rotation the latter reads
\begin{equation}
2\mathrm{D}_{[a}\mathrm{D}_{b]}v_{c}= {\mathcal{R}}_{abcd}v^{d}\,,
\label{3Ricci1}
\end{equation}
which means that $\mathrm{D}_{b}v_{a}\neq0$ unless
${\mathcal{R}}_{abcd}=0$~\cite{TB2}.

\subsubsection{Irreducible inhomogeneity variables}\label{sssIIVs}
The orthogonally projected density gradient, $\Delta_{a}$, contains
collective information about three types of inhomogeneities, namely
density perturbations, vortices and shape-distortions. This
information is encoded in the dimensionless comoving gradient
\begin{equation}
\Delta_{ab}= a\mathrm{D}_{b}\Delta_{a}  \label{Deltaab}
\end{equation}
and can be decoded by splitting the latter into its irreducible
components as follows~\cite{EBH,MaT}
\begin{equation}
\Delta_{ab}= {\frac{1}{3}}\,a\mathrm{D}^{c}\Delta_{c}h_{ab}+
a\mathrm{D}_{[b}\Delta_{a]}+ a\mathrm{D}_{\langle b}
\Delta_{a\rangle}\,.  \label{dDeltaab}
\end{equation}
The quantities on the right-hand side are associated with the
aforementioned three different types of inhomogeneity. The scalar
$\mathrm{D}^{a}\Delta_{a}$ describes spatial variations in the
matter density (i.e.~overdensities or underdensities), the skew part
is related to magnitude-preserving changes of $\Delta_{a}$
(i.e.~rotations), and the symmetric and trace-free tensor
$\mathrm{D}_{\langle b}\Delta_{a\rangle}$ describes shape
distortions in the anisotropy pattern of the gradient field
(e.g.~pancakes or cigar-like structures).

In an exactly analogous way, all the information regarding
perturbations in the volume expansion and in the magnetic energy
density is stored in the dimensionless second-rank tensors
\begin{equation}
{\mathcal{Z}}_{ab}= a\mathrm{D}_{b}{\mathcal{Z}}_{a} \hspace{15mm}
\mathrm{and} \hspace{15mm} {\mathcal{B}}_{ab}=
a\mathrm{D}_{b}{\mathcal{B}}_{a}\,,  \label{cZabcBab}
\end{equation}
respectively. These decompose into their irreducible parts according
to~\cite{EBH,TM1}
\begin{equation}
{\mathcal{Z}}_{ab}=
{\frac{1}{3}}\,a\mathrm{D}^{c}{\mathcal{Z}}_{c}h_{ab}+
a\mathrm{D}_{[b}{\mathcal{Z}}_{a]}+ a\mathrm{D}_{\langle b}
{\mathcal{Z}}_{a\rangle}  \label{dcZab}
\end{equation}
and
\begin{equation}
{\mathcal{B}}_{ab}=
{\frac{1}{3}}\,a\mathrm{D}^{c}{\mathcal{B}}_{c}h_{ab}+
a\mathrm{D}_{[b}{\mathcal{B}}_{a]}+ a\mathrm{D}_{\langle b}
{\mathcal{B}}_{a\rangle}\,.  \label{dcBab}
\end{equation}

\subsubsection{Linear equations}\label{sssLEs}
When linearized around a FRW background permeated by a weak magnetic
field, the non-linear propagation equations of section~\ref{sssEEs}
reduce to
\begin{eqnarray}
\dot{\Delta}_{a}&=& 3wH\Delta_{a}- (1+w){\mathcal{Z}}_{a}+
\frac{3aH}{\rho}\,\varepsilon_{abc}B^{b}\mathrm{curl\,}B^{c}+
2c_{\mathrm{a}}^{2}(1+w)aHA_{a}\,,  \label{ldotcDa1}\\
\dot{\mathcal{Z}}_{a}&=& -2H{\mathcal{Z}}_{a}-
{\frac{1}{2}}\,\rho\Delta_{a}-
{\frac{1}{2}}\,B^{2}{\mathcal{B}}_{a}+
{\frac{3}{2}}\,a\varepsilon_{abc} B^{b}\mathrm{curl\,}B^{c}+
a\mathrm{D}_{a}A \label{ldotcZa}
\end{eqnarray}
and
\begin{equation}
\dot{\mathcal{B}}_{a}= \frac{4}{3(1+w)}\,\dot{\Delta}_{a}-
\frac{4wH}{1+w}\,\Delta_{a}-
\frac{4aH}{\rho(1+w)}\,\varepsilon_{abc}B^{b}\mathrm{curl\,}B^{c}-
4aH\left(1+{\frac{2}{3}}\,c_{\mathrm{a}}^{2}\right)A_{a}\,,
\label{ldotcBa}
\end{equation}
respectively. Therefore, the linear evolution of the inhomogeneities
depends on the 4-acceleration of the highly conducting matter. To
first order, the latter satisfies the momentum-density conservation
law (see Eq.~(\ref{lmdc1}))
\begin{equation}
(1+w)\left(1+{\frac{2}{3}}\,c_{\mathrm{a}}^{2}\right)a\rho A_{a}=
-c_{s}^{2}\rho\Delta_{a}-
a\varepsilon_{abc}B^{b}\mathrm{curl\,}B^{c}\,.  \label{lmdc3}
\end{equation}
Using the commutation law
$\mathrm{D}_{[a}\mathrm{D}_{b]}v_{c}=-\omega_{ab}\dot{v}_{\langle
c\rangle}+{\mathcal{R}}_{dcba}v^{d}/2$ between the spatial gradients
of $v_{a}$ (with $v_{a}u^{a}=0$ -- see Eq.~(\ref{A2}) in
\S~\ref{AssOPGs})), and the constraint (\ref{C3}), the projected
divergence of the above leads to
\begin{equation}
A= -\frac{c_{s}^{2}}{a(1+w)}
\left(1-{\frac{2}{3}}\,c_{\mathrm{a}}^{2}\right)\mathrm{D}^{a}\Delta_{a}-
\frac{c_{\mathrm{a}}^{2}}{2a}\,\mathrm{D}^{a}{\mathcal{B}}_{a}+
{\frac{1}{3}}\,c_{\mathrm{a}}^{2}{\mathcal{R}}\,,  \label{lA}
\end{equation}
where
\begin{equation}
{\mathcal{R}}= 2\left(\rho+{\frac{1}{2}}\,B^{2}-
{\frac{1}{3}}\,\Theta^{2}\right)\,,  \label{lcR}
\end{equation}
is the linearized 3-Ricci scalar (see Eq.~(\ref{l3R})). Substituting
(\ref{lmdc3}) into the right-hand side of (\ref{ldotcBa}) we arrive
at
\begin{equation}
\dot{\mathcal{B}}_{a}= \frac{4}{3(1+w)}\,\dot{\Delta}_{a}+
\frac{4(c_{s}^{2}-w)H}{1+w}\,\Delta_{a}\,,  \label{ldotcBa1}
\end{equation}
which means that linear inhomogeneities in the magnetic energy
density are only affected by those of the matter. To close the
system of (\ref{ldotcDa1}), (\ref{ldotcZa}) and (\ref{ldotcBa1}) we
require the propagation equation of the 3-Ricci scalar. Taking the
time derivative of (\ref{l3R}) and linearising we obtain
\begin{equation}
\dot{\mathcal{R}}=
-2\left(1+{\frac{2}{3}}\,c_{\mathrm{a}}^{2}\right)H{\mathcal{R}}+
\frac{4c_{s}^{2}}{a(1+w)}
\left(1-{\frac{2}{3}}\,c_{\mathrm{a}}^{2}\right)H{\rm D}^a\Delta_a+
\frac{2c_{\mathrm{a}}^{2}H}{a}\,{\rm D}^a\mathcal{B}_a\,.
\label{ldotcR}
\end{equation}

We note that for $c_{s}^{2}=w$, as it happens during the radiation
and the dust epochs when the equation of state of the cosmic medium
remains unchanged (i.e.~when $\dot{w}=0$ - see Eq.~(\ref{ldotw})),
the above expression reduces to
\begin{equation}
\dot{\mathcal{B}}_{a}= \frac{4}{3(1+w)}\dot{\Delta}_{a}\,.
\label{ldotcBa2}
\end{equation}
This guarantees that linear perturbations in the magnetic energy
density evolve in step with those in the density of the highly
conducting matter. When dealing with a radiative fluid the above
gives $\dot{\mathcal{B}}_{a}=\dot{\Delta}_{a}$, while
$\dot{\mathcal{B}}_{a}=4\dot{\Delta}_{a}/3$ for non-relativistic
dust.\footnote{The covariantly derived result (\ref{ldotcBa2}) is in
complete agreement with relation $B\propto\rho^{2/3(1+w)}$, familiar
from conventional studies of weakly magnetized FRW models. Indeed,
taking the square of the above, and then its variation, we find that
$\delta B^{2}/B^{2}=4\delta\rho/3\rho(1+w)$ with $\delta
B^{2}/B^{2}$ and $\delta\rho/\rho$ corresponding to
${\mathcal{B}}_{a}$ and $\Delta_{a}$ respectively.} The implication
of (\ref{ldotcBa2}) is that the effective entropy perturbations
coming from the different dynamical behaviour of the two components
(i.e.~the fluid and the magnetic field) are constant, or zero (see
section~\ref{sLEPs}).

\section{Density perturbations}\label{sLDPs}
\subsection{Basic variables}\label{ssBVs1}
Decomposition (\ref{dDeltaab}) allows for a fully covariant and
gauge-invariant study of linear density perturbations, vortices and
shape distortions. Here, we will focus on the scalar component of
$\Delta_{ab}$. This is defined by
\begin{equation}
\Delta= a\mathrm{D}^{a}\Delta_{a}\,,  \label{Delta}
\end{equation}
and when positive (negative) it describes overdensities
(underdensities) in the distribution of the perturbed fluid. In
other words, $\Delta$ corresponds to the relative increase or
decrease in the matter density as measured by a pair of neighbouring
fundamental observers. Similarly,
${\mathcal{Z}}=a\mathrm{D}^{a}{\mathcal{Z}}_{a}$ and
${\mathcal{B}}=a\mathrm{D}^{a}{\mathcal{B}}_{a}$ describe scalar
perturbations in the volume expansion and the magnetic energy
density respectively.

To linear order definition (\ref{Delta}) reduces to
$\Delta=(a^{2}/\rho)\mathrm{D}^{2}\rho$, while analogous expressions
hold for ${\mathcal{B}}$ and ${\mathcal{Z}}$. In addition, it helps
to introduce the following rescaling
\begin{equation}
{\mathcal{K}}=a^{2}{\mathcal{R}}\,,  \label{cK}
\end{equation}
of the perturbed 3-Ricci scalar ${\mathcal{R}}$. This is also gauge
invariant, given the spatial flatness of the FRW background.

\subsection{Linear equations}\label{ssLEs1}
Taking the orthogonally projected divergence of
Eq.~(\ref{ldotcDa1}), using (\ref{lmdc3}) and keeping up to
$c_{\mathrm{a}}^{2}$-order terms, given the weakness of the magnetic
field, we arrive at the following equation
\begin{equation}
\dot{\Delta}= 3w\left(1-{\frac{2}{3}}\,c_{\mathrm{a}}^{2}\right)
H\Delta-(1+w){\mathcal{Z}}+ {\frac{3}{2}}\,c_{\mathrm{a}}^{2}
(1+w)H{\mathcal{B}}- c_{\mathrm{a}}^{2}(1+w)H{\mathcal{K}}\,,
\label{ldotDelta}
\end{equation}
for the linear evolution of $\Delta$. Note that in deriving the
above we have also set $\dot{w}=0$ to zero order, which means that
$w=c_{s}^{2}=$ constant in the background, and employed the linear
relation $a\mathrm{D}^{a}\dot{\Delta}_{a}=\dot{\Delta}$. Following
(\ref{ldotDelta}), the field will generally act as a source for
density perturbations even when there are no such distortions
present initially. Also, the magnetic field's presence has a direct
and an indirect effect on $\Delta$. The former results from the
pressure part of the Lorentz force (see decomposition
(\ref{Lorentz}) in \S~\ref{ssCL2}) and carries the effects of the
isotropic magnetic pressure. The latter comes from the tension
component of the Lorentz force and it is triggered by the magnetic
coupling to the spatial curvature of the perturbed model.
Surprisingly, a positive 3-curvature perturbation causes $\Delta$ to
decrease, while a negative ${\mathcal{K}}$ has the opposite effect.
This rather counter-intuitive behaviour of the magneto-curvature
term in (\ref{ldotDelta}) is a direct consequence of the elasticity
of the field lines (see also Eq.~(\ref{ldotcZ}) below).

Similarly, using (\ref{lA}), together with the linear results
$a\mathrm{D}^{a}\dot{\mathcal{Z}}_{a}=\dot{\mathcal{Z}}$ and
$a\mathrm{D}^{a}\dot{\mathcal{B}}_{a}=\dot{\mathcal{B}}$, the
linearized orthogonally-projected divergences of (\ref{ldotcZa}) and
(\ref{ldotcBa1}) lead to
\begin{eqnarray}
\dot{\mathcal{Z}}&=&
-2\left(1+{\frac{2}{3}}\,c_{\mathrm{a}}^{2}\right)H{\mathcal{Z}}-
{\frac{1}{2}}\,\rho\left(1-{\frac{4}{3}}\,c_{\mathrm{a}}^{2}\right)
\Delta+ {\frac{1}{4}}\,c_{\mathrm{a}}^{2}(1+w)\rho{\mathcal{B}}-
{\frac{1}{2}}\,c_{\mathrm{a}}^{2}(1+w)\rho{\mathcal{K}} \nonumber\\
&&-\frac{c_{s}^{2}}{1+w}\left(1-{\frac{2}{3}}\,c_{\mathrm{a}}^{2}\right)
\mathrm{D}^{2}\Delta-
{\frac{1}{2}}\,c_{\mathrm{a}}^{2}\mathrm{D}^{2}{\mathcal{B}}
\label{ldotcZ}
\end{eqnarray}
and
\begin{equation}
\dot{\mathcal{B}}= \frac{4}{3(1+w)}\dot{\Delta}+
\frac{4(c_{s}^{2}-w)H}{1+w}\Delta\,,  \label{ldotcB}
\end{equation}
respectively. Finally, starting from the linear propagation equation
of the 3-Ricci scalar we obtain
\begin{equation}
\dot{\mathcal{K}}= -{\frac{4}{3}}\,c_{\mathrm{a}}^{2}H{\mathcal{K}}+
\frac{4c_{s}^{2}}{1+w}\left(1-{\frac{2}{3}}\,c_{\mathrm{a}}^{2}\right)
H\Delta+ 2c_{\mathrm{a}}^{2}H{\mathcal{B}}\,.  \label{ldotcK}
\end{equation}
The system (\ref{ldotDelta})-(\ref{ldotcK}) describes the linear
evolution of scalar inhomogeneities in the density distribution of
the matter in a weakly magnetic and spatially flat almost-FRW
universe.\footnote{When $B^{2}$ is treated as a first order
perturbation, the 3-Ricci terms in the right-hand side of
Eqs.~(\ref{ldotDelta}), (\ref{ldotcZ}) are second order and
therefore vanish at the linear level. In that case linearized
magnetized matter fluctuations evolve free of spatial curvature
effects and (\ref{ldotcK}) is no longer required. Note that
$c_{\mathrm{a}}^{2}{\mathcal{B}}=a^{2}\mathrm{D}^{2}B^{2}/\rho$ by
definition and therefore the variable
$c_{\mathrm{a}}^{2}{\mathcal{B}}$ is still linear.} When the cosmic
medium has a time-independent equation of state (i.e.~for
$\dot{w}=0$) we have $c_{s}^{2}=w=$~constant and Eq.~(\ref{ldotcB})
reduces to
\begin{equation}
\dot{\mathcal{B}}=\frac{4}{3(1+w)}\dot{\Delta}\,.  \label{lcBDelta}
\end{equation}

It should be noted that in~\cite{TB1}-\cite{TM1} factors of the form
$1\pm c_{\mathrm{a}}^{2}$ in the coefficients of the perturbed
variables were set to unity, because of the overall weakness of the
$B$-field. Here, retaining the factors $1\pm c_{\mathrm{a}}^{2}$ in
the system of (\ref{ldotDelta})-(\ref{ldotcK}) has improved the
accuracy of our linear equations and will help us refine the results
of~\cite{TB1}-\cite{TM1}. When $1\pm c_{\mathrm{a}}^{2}\simeq1$,
expressions (\ref{ldotDelta})-(\ref{ldotcK}) immediately reduce to
their corresponding formulae of~\cite{TB1}-\cite{TM1}. The reduction
is obvious and straightforward for
Eqs.~(\ref{ldotDelta})-(\ref{ldotcB}) but not for (\ref{ldotcK}).
There, one needs to use definition (\ref{cK}) to show that
$\dot{\mathcal{K}}+(4c_{\mathrm{a}}^{2}/3)H{\mathcal{K}}\simeq\dot
{\mathcal{K}}$ when $1\pm c_{\mathrm{a}}^{2}\simeq1$. This
requirement was overlooked in~\cite{HD}, resulting in the
inconsistent linearisation of their equations.

\subsection{Evolution in the radiation era}\label{ssERE}
During the radiation epoch the background dynamics is determined by
the parameters $w=1/3=c_{s}^{2}$, $H=1/2t$ and $\rho=3/4t^{2}$. In
addition, given the weakness of the magnetic field we have
$c_{\mathrm{a}}^{2}=3\beta/4$, where $\beta=B^{2}/\rho=$~constant
$\ll1$. To proceed further we harmonically decompose the perturbed
variables. In particular, we set
$\Delta=\sum_{k}\Delta^{(k)}{\mathcal{Q}}_{(k)}$, where
${\mathcal{Q}}_{(k)}$ are the standard scalar harmonics (with $k$
representing the associated wavenumber) and
$\mathrm{D}_{a}\Delta^{(k)}=0$. The harmonics are time-independent
functions (i.e.~$\dot{\mathcal{Q}}_{(k)}=0$) and satisfy the scalar
Laplace-Beltrami equation
\begin{equation}
\mathrm{D}^{2}{\mathcal{Q}}_{(k)}=
-\left(\frac{k}{a}\right)^{2}{\mathcal{Q}}_{(k)}\,.  \label{LB}
\end{equation}
Similarly, the rest of the perturbed variables decompose as
${\mathcal{Z}}={\mathcal{Z}}^{(k)}{\mathcal{Q}}_{(k)}$,
${\mathcal{B}}={\mathcal{B}}^{(k)}{\mathcal{Q}}_{(k)}$ and
${\mathcal{K}}={\mathcal{K}}^{(k)}{\mathcal{Q}}_{(k)}$, with
$\mathrm{D}_{a}{\mathcal{Z}}^{(k)}=0=\mathrm{D}_{a}
{\mathcal{B}}^{(k)}=\mathrm{D}_{a}{\mathcal{K}}^{(k)}$. Implementing
this harmonic decomposition, Eqs.~(\ref{ldotDelta})-(\ref{ldotcK})
now read
\begin{eqnarray}
\dot{\Delta}^{(k)}&=&
{\frac{1}{2}}\left(1-{\frac{1}{2}}\,\beta\right)t^{-1}\Delta^{(k)}-
{\frac{4}{3}}\,{\mathcal{Z}}^{(k)}- {\frac{1}{2}}\,\beta
t^{-1}{\mathcal{K}}^{(k)}+ {\frac{3}{4}}\,\beta
t^{-1}{\mathcal{B}}^{(k)}\,, \label{rdotDel}\\
\dot{\mathcal{Z}}^{(k)} &=&
-\left(1+{\frac{1}{2}}\,\beta\right)t^{-1}{\mathcal{Z}}^{(k)}-
{\frac{3}{8}}\,(1-\beta)t^{-2}\Delta^{(k)}- {\frac{3}{8}}\,\beta
t^{-2}{\mathcal{K}}^{(k)}+ {\frac{3}{16}}\,\beta
t^{-2}{\mathcal{B}}^{(k)} \nonumber\\
&&+{\frac{1}{4}}\,\left(\frac{k}{a}\right)^{2}
\left(1-{\frac{1}{2}}\,\beta\right)\Delta^{(k)}+
{\frac{3}{8}}\left(\frac{k}{a}\right)^{2}\beta{\mathcal{B}}^{(k)}\,,
\label{rdotZeta}\\ \dot{\mathcal{K}}^{(k)}&=& -{\frac{1}{2}}\,\beta
t^{-1}{\mathcal{K}}^{(k)}+
{\frac{1}{2}}\left(1-{\frac{1}{2}}\,\beta\right)t^{-1}\Delta^{(k)}+
{\frac{3}{4}}\,\beta t^{-1}{\mathcal{B}}^{(k)}\,,  \label{rdotcK}\\
\dot{\mathcal{B}}^{(k)} &=& \dot{\Delta}^{(k)}\,.  \label{rdotcB}
\end{eqnarray}

During the radiation era the Hubble radius, in comoving proper time
$t$, is given by $\lambda_{H}\equiv1/H=2t$. Also, physical
wavelengths and comoving wave numbers are related by
$\lambda_k=a/k$. Substituting these expressions into
Eq.~(\ref{rdotZeta}), and keeping up to $\beta$-order terms (recall
that $\beta\ll1$), the system (\ref{rdotDel})-(\ref{rdotcB}) becomes
\begin{eqnarray}
\dot{\Delta}^{(k)}&=&
{\frac{1}{2}}\left(1-{\frac{1}{2}}\,\beta\right)t^{-1}\Delta^{(k)}-
{\frac{4}{3}}\,{\mathcal{Z}}^{(k)}- {\frac{1}{2}}\,\beta
t^{-1}{\mathcal{K}}^{(k)}+ {\frac{3}{4}}\,\beta
t^{-1}{\mathcal{B}}^{(k)}\,,  \label{rdotDel1}\\
\dot{\mathcal{Z}}^{(k)}&=&
-\left(1+{\frac{1}{2}}\,\beta\right)t^{-1}{\mathcal{Z}}^{(k)}-
{\frac{3}{8}}\,(1-\beta)t^{-2}\left[1-{\frac{1}{6}}
\left(\frac{\lambda_{H}}{\lambda_k}\right)^{2}
\left(1+{\frac{1}{2}}\,\beta\right)\right]\Delta^{(k)}  \nonumber\\
&&-{\frac{3}{8}}\,\beta t^{-2}{\mathcal{K}}^{(k)}+
{\frac{3}{16}}\,\beta t^{-2}\left[1+{\frac{1}{2}}
\left(\frac{\lambda_{H}}{\lambda_k}\right)^{2}\right]
{\mathcal{B}}^{(k)}\,,  \label{rdotZeta1}\\
\dot{\mathcal{K}}^{(k)}&=& -{\frac{1}{2}}\,\beta
t^{-1}{\mathcal{K}}^{(k)}+
{\frac{1}{2}}\left(1-{\frac{1}{2}}\,\beta\right)t^{-1}\Delta^{(k)}+
{\frac{3}{4}}\,\beta t^{-1}{\mathcal{B}}^{(k)}\,,  \label{rdotcK1}\\
\dot{\mathcal{B}}^{(k)}&=& \dot{\Delta}^{(k)}\,,  \label{rdotcB1}
\end{eqnarray}
where $\lambda_H/\lambda_k=k/aH$. The above describe the evolution
of linear matter perturbations in a weakly magnetic almost-FRW
universe filled with a single highly conducting perfect fluid.

\subsubsection{Super-horizon scales}\label{sssSphSs}
When dealing with perturbations on super-Hubble lengths we have
$\lambda_k\gg\lambda_{H}$. Then, given also that $\beta\ll1$,
Eqs.~(\ref{rdotDel1})-(\ref{rdotcB1}) reduce to the
scale-independent system
\begin{eqnarray}
\dot{\Delta}&=& {\frac{1}{2}}\left(1-{\frac{1}{2}}\,\beta\right)
t^{-1}\Delta- {\frac{4}{3}}\,{\mathcal{Z}}- {\frac{1}{2}}\,\beta
t^{-1}{\mathcal{K}}+ {\frac{3}{4}}\,\beta t^{-1}{\mathcal{B}}\,,
\label{rldotDel}\\ \dot{\mathcal{Z}}&=&
-\left(1+{\frac{1}{2}}\,\beta\right)t^{-1}{\mathcal{Z}}-
{\frac{3}{8}}\,(1-\beta)t^{-2}\Delta-{\frac{3}{8}}\,\beta
t^{-2}{\mathcal{K}}+ {\frac{3}{16}}\,\beta t^{-2}{\mathcal{B}}\,,
\label{rldotZeta}\\ \dot{\mathcal{K}}&=& -{\frac{1}{2}}\,\beta
t^{-1}{\mathcal{K}}+
{\frac{1}{2}}\,\left(1-{\frac{1}{2}}\,\beta\right)t^{-1}\Delta+
{\frac{3}{4}}\,\beta t^{-1}{\mathcal{B}}\,,  \label{rldotcK}\\
\dot{\mathcal{B}}&=& \dot{\Delta}\,.  \label{rldotcB}
\end{eqnarray}
These can be solved analytically and lead to the following simple
power-law evolution for the density contrast
\begin{equation}
\Delta= \Delta(t)= {\mathcal{C}}_{0}+ {\mathcal{C}}_{i}t^{z_{i}}\,,
\label{rlDelta}
\end{equation}
where ${\mathcal{C}}_{0}$, ${\mathcal{C}}_{i}$ are constants (with
$i=1,\,2,\,3$) and the $z_{i}$ are roots of a cubic algebraic
equation. Ignoring terms of order ${\mathcal{O}}(\beta^{2})$ and
higher, the latter reads
\begin{equation}
8z^{3}- 4(1-\beta)z^{2}- 4(1-\beta)z= 4\beta\,.  \label{rlz}
\end{equation}
At the weak-field limit the above is solved perturbatively giving
$z_{1}=-1/2+5\beta/6$, $z_{2}=-\beta$ and $z_{3}=1-\beta/3$, which
are all physically allowed solutions. Consequently, at this level of
approximation, the linear evolution of $\Delta$ proceeds as a sum of
power-laws:
\begin{equation}
\Delta= {\mathcal{C}}_{0}+
{\mathcal{C}}_{1}t^{-{\textstyle{\frac{1}{2}}}+
{\textstyle{\frac{5}{6}}}\beta}+ {\mathcal{C}}_{2}t^{-\beta}+
{\mathcal{C}}_{3}t^{1-{\textstyle{\frac{1}{3}}}\beta}\,.
\label{rlDelta1}
\end{equation}
Note that in the absence of the magnetic field (i.e.~for $\beta=0$),
we recover the standard evolution for the density contrast, familiar
from the linear study of perturbed magnetic-free FRW models
(e.g.~see~\cite{Pee}-\cite{T4}). Thus, in the weak-field limit the
main magnetic effect is to reduce the growth rate of the dominant
density mode. In addition, the field also decreases the rate of the
standard decay mode and introduces a new `non-adiabatic' decay
mode.\footnote{The magnetic suppression of the growing $\Delta$-mode
was first observed in~\cite{TB2}, although there the 3-curvature
effects were switched off. Allowing for the curvature effects, the
analysis of~\cite{TM1} showed no magnetic effect on the growing mode
to lowest order in $\beta$, but a slight magnetic enhancement at
higher order (see solution (54), (55) there). Here, the refined
solutions showed a suppression proportional to $\beta$ in line with
the results of~\cite{TB2}.}

\subsubsection{Sub-horizon scales}\label{sssSbhSs}
Well below the horizon scale, $\lambda_{H}/\lambda_k\gg1$ and the
scale-dependent terms inside the brackets of Eq.~(\ref{rdotZeta1})
become important. Then, on sub-horizon scales the system
(\ref{rdotDel1})-(\ref{rdotcB1}) reads
\begin{eqnarray}
\dot{\Delta}^{(k)}&=&
{\frac{1}{2}}\left(1-{\frac{1}{2}}\,\beta\right)t^{-1}\Delta^{(k)}-
{\frac{4}{3}}\,{\mathcal{Z}}^{(k)}- {\frac{1}{2}}\,\beta
t^{-1}{\mathcal{K}}^{(k)}+ {\frac{3}{4}}\,\beta
t^{-1}{\mathcal{B}}^{(k)}\,,  \label{rsdotDel}\\
\dot{\mathcal{Z}}^{(k)}&=&
-\left(1+{\frac{1}{2}}\,\beta\right)t^{-1}{\mathcal{Z}}^{(k)}+
{\frac{1}{16}}\left(\frac{\lambda_{H}}{\lambda_k}\right)_{0}^{2}
\left(1-{\frac{1}{2}}\,\beta\right)t_{0}^{-1}t^{-1}\Delta^{(k)}
\nonumber\\
&&-{\frac{3}{8}}\,\beta t^{-2}{\mathcal{K}}^{(k)}+
{\frac{3}{32}}\,\beta
\left(\frac{\lambda_{H}}{\lambda_k}\right)_{0}^{2}t_{0}^{-1}t^{-1}
{\mathcal{B}}^{(k)}\,,  \label{rsdotZeta}\\
\dot{\mathcal{K}}^{(k)}&=& -{\frac{1}{2}}\,\beta
t^{-1}{\mathcal{K}}^{(k)}+
{\frac{1}{2}}\left(1-{\frac{1}{2}}\,\beta\right)t^{-1}\Delta^{(k)}+
{\frac{3}{4}}\,\beta t^{-1}{\mathcal{B}}^{(k)}\,,  \label{rsdotcK}\\
\dot{\mathcal{B}}^{(k)}&=& \dot{\Delta}^{(k)}\,,  \label{rsdotcB}
\end{eqnarray}
given that during the radiation era
$\lambda_{H}/\lambda_k=(\lambda_{H}/\lambda_k)_{0}(t/t_{0})^{1/2}$.

When the 3-curvature effects are switched off, as is physically
appropriate during the early universe, the system can be solved
analytically. In particular, the solution for the linear matter
perturbations reads
\begin{equation}
\Delta^{(k)}= \Delta^{(k)}(t)= {\mathcal{C}}_{1}+
{\mathcal{C}}_{2}t^{1/4}\mathrm{J}_{1/2+\beta}(\chi)+
{\mathcal{C}}_{3}t^{1/4}\mathrm{Y}_{1/2+\beta}(\chi)\,,
\label{rsDelta1}
\end{equation}
where ${\mathcal{C}}_{i}$ are the integration constants (with
$i=1,\,2,\,3$) and $\mathrm{J}_{1/2+\beta}(\chi)$ and
$\mathrm{Y}_{1/2+\beta}(\chi)$ are Bessel functions of the first and
second kind respectively, with arguments determined by
\begin{equation}
\chi= c_{s}\left(\frac{\lambda_{H}}{\lambda_k}\right)
\left(1+{\frac{1}{2}}\,\beta\right)=
c_{s}\left(\frac{\lambda_{H}}{\lambda_k}\right)_{0}
\left(\frac{t}{t_{0}}\right)^{1/2}
\left(1+{\frac{1}{2}}\,\beta\right)\,,  \label{chi}
\end{equation}
and $c_{s}=1/\sqrt{3}$. Also, recalling that $\beta\ll1$ and keeping
up to $\beta$-order terms, we arrive at the following large-scale
solution for ${\mathcal{B}}$:
\begin{equation}
{\mathcal{B}}^{(k)}= {\mathcal{B}}^{(k)}(t)=
-{\frac{2}{3}}\,{\mathcal{C}}_{1}
\left(1-{\frac{1}{2}}\,\beta\right)\beta^{-1}+
{\mathcal{C}}_{2}t^{1/4}\mathrm{J}_{1/2+\beta}(\chi)+
{\mathcal{C}}_{3}t^{1/4}\mathrm{Y}_{1/2+\beta}(\chi)\,.
\label{rscB1}
\end{equation}
As expected (see Eq.~(\ref{rsdotcB})), the above is identical to
(\ref{rsDelta1}) up to a constant. Setting
$\Delta^{(k)}=\Delta_{0}^{(k)}$ and
${\mathcal{B}}^{(k)}={\mathcal{B}}_{0}^{(k)}$ initially, solutions
(\ref{rsDelta1}) and (\ref{rscB1}) combine to give
${\mathcal{C}}_{1}\simeq3\beta(\Delta_{0}^{(k)}
-{\mathcal{B}}_{0}^{(k)})/2$. Then,
\begin{equation}
\Delta^{(k)}= {\frac{3}{2}}\,\beta\left(\Delta_{0}^{(k)}
-{\mathcal{B}}_{0}^{(k)}\right)+
{\mathcal{C}}_{2}t^{1/4}\mathrm{J}_{1/2+\beta}(\chi)+
{\mathcal{C}}_{3}t^{1/4}\mathrm{Y}_{1/2+\beta}(\chi)\,.
\label{rsDelta2}
\end{equation}

Based on the weakness of the magnetic field, we may approximate the
two Bessel functions in the above given solutions by
$\mathrm{J}_{1/2+\beta}\simeq\mathrm{J}_{1/2}$ and
$\mathrm{Y}_{1/2+\beta}\simeq\mathrm{Y}_{1/2}$ respectively. In this
case (\ref{rsDelta2}) reduces to
\begin{equation}
\Delta^{(k)}= {\frac{3}{2}}\,\beta
\left(\Delta_{0}^{(k)}-{\mathcal{B}}_{0}^{(k)}\right)+
{\mathcal{C}}_{2}\sqrt{\frac{2}{\pi\alpha}} \sin\left(\alpha
t^{1/2}\right)+ {\mathcal{C}}_{3}\sqrt{\frac{2}{\pi\alpha}}
\cos\left(\alpha t^{1/2}\right)\,,  \label{rsDelta3}
\end{equation}
where
$\alpha=c_{s}(\lambda_{H}/\lambda_k)_{0}t_{0}^{-1/2}(1+\beta/2)$.
Therefore, we find that small-scale matter perturbations oscillate
like magneto-sonic waves. The magnetic presence tends to reduce the
amplitude of the oscillation and increase its frequency (in
agreement with~\cite{TB2,TM1}). In both cases the effect of the
field is proportional to its relative strength (i.e.~to the ratio
$\beta=B^{2}/\rho$). As pointed out in~\cite{ADGR}, the increased
frequency of $\Delta^{(k)}$ should bring the peaks of
short-wavelength oscillations in the density of the radiation
component closer. This in turn could produce a potentially
observable signature in the CMB.

An additional magnetic effect arises from the presence of a constant
mode in solution (\ref{rsDelta3}). This suggests that, unlike the
magnetic-free case (e.g.~see~\cite{Pa}), magnetic matter
perturbations in the pre-equality universe oscillate around a
generally non-zero average value. This depends on the relation
between $\Delta$ and ${\mathcal{B}}$ initially. For example,
isocurvature initial conditions typically correspond to zero
perturbation in the total energy density~\cite{MFB,LL}. In our case
$\rho_{t}=\rho+B^{2}/2$ and the isocurvature requirement translates
into $\Delta_{0}=(\beta/2){\mathcal{B}}_{0}$ (recall that
$\beta=$~constant throughout this epoch). Following
(\ref{rsDelta3}), the latter implies that by the end of the
radiation era the associated density contrast oscillates around the
average value $\langle\Delta\rangle\simeq3\Delta_{0}$. We finally
note that, strictly speaking, the analysis of \S~\ref{ssERE} holds
as long as the electrons are still relativistic (i.e.~up to $T\sim
T_e$). Subsequently, and until the end of the radiation era, one
should examine the magnetic effects on perturbed non-relativistic
matter on an expanding radiative background.

\subsection{Evolution in the dust era}\label{ssEDE}
After the end of the radiation era, the unperturbed background is
well approximated by $w=0=c_{s}^{2}$, $H=2/3t$, $\rho=4/3t^{2}$.
Also, $c_{\mathrm{a}}^{2}=\beta$ and it is no longer constant but
decreases with time according to $\beta\propto a^{-1}\propto
t^{-2/3}$. Applying the usual harmonic decomposition to the
perturbation variables, and keeping up to $\beta$-order terms, the
system (\ref{ldotDelta})-(\ref{ldotcK}) reads
\begin{eqnarray}
\dot{\Delta}^{(k)}&=& -{\mathcal{Z}}^{(k)}- {\frac{2}{3}}\,\beta
t^{-1}{\mathcal{K}}^{(k)}+ \beta t^{-1}{\mathcal{B}}^{(k)}\,,
\label{ddotDelta}\\ \dot{{\mathcal{Z}}}^{(k)}&=&
-{\frac{4}{3}}\left(1+{\frac{2}{3}}\,\beta\right)
t^{-1}{\mathcal{Z}}^{(k)}-
{\frac{2}{3}}\left(1-{\frac{4}{3}}\,\beta\right)t^{-2}\Delta^{(k)}-
{\frac{2}{3}}\,\beta t^{-2}{\mathcal{K}}^{(k)} \nonumber\\
&&+{\frac{1}{3}}\,\beta\left[1+{\frac{2}{3}}
\left(\frac{\lambda_{H}}{\lambda_k}\right)^{2}\right]
t^{-2}{\mathcal{B}}^{(k)}\,,  \label{ddotZeta}\\
\dot{\mathcal{K}}^{(k)} &=& -{\frac{8}{9}}\,\beta
t^{-1}{\mathcal{K}}^{(k)}+ {\frac{4}{3}}\,\beta
t^{-1}{\mathcal{B}}^{(k)}\,,  \label{ddotcK}\\
\dot{\mathcal{B}}^{(k)}&=& {\frac{4}{3}}\,\dot{\Delta}^{(k)}\,,
\label{ddotcB}
\end{eqnarray}
with $\lambda_{H}=3t/2$. The above system has no simple analytical
solution because during the dust era the dimensionless parameter
$\beta$ is no longer constant. Nevertheless, as we shall see next,
one can still extract useful qualitative and quantitative
information about the magnetic effects on density perturbations in
the post-recombination universe.

\subsubsection{Magnetically-induced Jeans' length}\label{sssMiJL}
The wave equation of the density contrast reveals the role of the
magnetic pressure against gravitational collapse and helps to
establish the scale of the magnetic Jeans' length. In particular,
taking the time derivative of (\ref{ddotDelta}) and using the rest
of the formulae we arrive at
\begin{eqnarray}
\ddot{\Delta}^{(k)}&=&
-{\frac{4}{3}}\left(1-{\frac{1}{3}}\,\beta\right)
t^{-1}\dot{\Delta}^{(k)}+
{\frac{2}{3}}\left\{1-{\frac{8}{3}}\,\beta\left[1+{\frac{1}{6}}
\left(\frac{\lambda_{H}}{\lambda_k}\right)^{2}\right]
\right\}t^{-2}\Delta^{(k)} \nonumber\\
&&+{\frac{8}{9}}\,\beta t^{-2}\mathcal{K}^{(k)}\,.
\label{dddotDelta}
\end{eqnarray}
Note that, in deriving the above, we have set
${\mathcal{B}}_{0}^{(k)}=4\Delta_{0}^{(k)}/3$. This guarantees that
${\mathcal{B}}^{(k)}=4\Delta^{(k)}/3$ always (see
Eq.~(\ref{ddotcB})).

The first term on the right-hand side of (\ref{dddotDelta}) shows
that the magnetic presence reduces the diluting effect of the
expansion by a small amount proportional to the field's relative
strength. The second term gauges the opposing effects of gravity and
magnetic pressure. When the curvature term is removed form
(\ref{dddotDelta}), and the quantity in the angled brackets is
positive, gravity prevails and the density contrast grows. When the
brackets take a negative value, the pressure of the magnetic field
dominates and forces the inhomogeneity into an oscillatory phase. In
the latter case the magnetic effect is relatively weak and also
decays in time. Gravitational attraction and magnetic pressure
balance each other out at the threshold
$1-(8\beta/3)[1+(\lambda_{H}/\lambda_k)^{2}/6]=0$. This also
determines an associated Jeans length, defined as the scale below
which the magnetic pressure gradients dominate and the
inhomogeneities oscillate. This `magnetic' Jeans length is given by
\begin{equation}
\lambda_{J}= \sqrt{\frac{4\beta}{9-24\beta}}\;\lambda_{H}
\label{mJeans}
\end{equation}
and it is considerably smaller than the corresponding Hubble radius
since $\beta\ll1$ (see also~\cite{Pe,KOR} for analogous
expressions). The anisotropy of the CMB constrains the current
magnitude of a (homogeneous) cosmological magnetic field below
$10^{-9}$~$G$~\cite{ADGR,BFS}. In that case
$B_{0}^{2}\sim10^{-58}$~GeV$^{4}$, $\beta_{0}\sim10^{-11}$ (recall
that $\rho=\rho_{\mathrm{cr}}\sim10^{-47}$~GeV$^{4}$ today) and the
associated magnetic Jeans length is
\begin{equation}
\lambda_{J}\sim10^{-5}\lambda_{H}\sim10~{\rm Kpc}\,,
\label{lambdaJ1}
\end{equation}
today. Random $B $-fields, however, can have magnitudes up to
$10^{-6}$~$G$ in today's values. If $B_{0}\sim10^{-7}$~$G$, a
strength supported by observations in galaxy clusters and
high-redshift protogalactic structures, we find that
\begin{equation}
\lambda_{J}\sim10^{-3}\lambda_{H}\sim1~{\rm Mpc}\,,
\label{lambdaJ2}
\end{equation}
at present. This number, which can also be obtained through a fully
Newtonian treatment~\cite{VTP}, determines the dimensions of the
region that is gravitationally supported by the magnetic pressure
gradients, and lies intriguingly close to the size of a galaxy
cluster.

\subsubsection{Late-time evolution}\label{sssLtE}
One can obtain an analytical solution for the evolution of magnetic
density perturbations in the dust era by taking the time derivative
of (\ref{dddotDelta}). Then, recalling that
$\lambda_{H}/\lambda_k=(\lambda_{H}/\lambda_k)_{0}(t/t_{0})^{1/3}$,
using (\ref{ddotcK}) and (\ref{ddotcB}), and keeping up to
$\beta$-order terms we arrive at
\begin{eqnarray}
\dddot{\Delta}^{(k)}&=& -4\left[1-{\frac{1}{9}}\,\beta_{0}
\left(\frac{t_{0}}{t}\right)^{2/3}\right]t^{-1}\ddot{\Delta}^{(k)}-
{\frac{14}{9}}\left[1+{\frac{4}{21}}\,\beta_{0}
\left(\frac{\lambda_{H}}{\lambda_k}\right)_{0}^{2}
+{\frac{6}{7}}\,\beta_{0}\left(\frac{t_{0}}{t}\right)^{2/3}\right]
t^{-2}\dot{\Delta}^{(k)} \nonumber\\
&&+{\frac{4}{9}}\left[1-{\frac{4}{9}}\,\beta_{0}
\left(\frac{\lambda_{H}}{\lambda_k}\right)_{0}^{2}\right]
t^{-3}\Delta^{(k)}+{\mathcal{C}}_{\mathcal{B}}t^{-3}\,.
\label{ddddotDelta}
\end{eqnarray}
The constant ${\mathcal{C}}_{\mathcal{B}}$ is inversely proportional
to the scale in question and vanishes when
${\mathcal{B}}_{0}=4\Delta_{0}/3$. As we will see in \S~\ref{sLEPs},
the aforementioned initial conditions guarantee effective
adiabaticity. Ignoring the weak and time-decaying terms from the
right-hand side of (\ref{ddddotDelta}), we obtain the following
late-time solution (see also Eq.~(67) in~\cite{TM1})
\begin{equation}
\Delta^{(k)}= {\mathcal{C}}_{1}t^{\alpha_{1}}+
{\mathcal{C}}_{2}t^{\alpha_{2}}+ {\mathcal{C}}_{3}t^{-2/3}+
{\mathcal{C}}_{4}\,,  \label{dltDelta}
\end{equation}
with
\begin{equation}
\alpha_{1,2}= -{\frac{1}{6}}\,
\left[1\pm5\sqrt{1-{\frac{32}{75}}\,\beta_{0}
\left(\frac{\lambda_{H}}{\lambda_k}\right)_{0}^{2}}\right]\,.
\label{alphas}
\end{equation}
In the absence of the $B$-field (i.e.~when $\beta_{0}=0$), we
immediately recover the standard magnetic-free solution with
$\alpha_{1}=2/3$ and $\alpha_{2}=-1$
(e.g.~see~\cite{Pee}-\cite{T4}). Therefore, the main magnetic effect
is to reduce the growth rate of density perturbations by an amount
proportional to its relative strength (i.e.~to the ratio
$\beta_{0}=(B^{2}/\rho)_{0}$). It should be noted that the
inhibiting role of the field was first observed in the Newtonian
treatment of~\cite{RR} and later in the relativistic studies
of~\cite{TB2,TM1}. According to (\ref{dltDelta}) and (\ref{alphas}),
the magnetic impact is inversely proportional to the scale in
question. On super-horizon lengths, in particular, the above given
solution reads
\begin{equation}
\Delta^{(k)}= {\mathcal{C}}_{1}t^{2/3}+ {\mathcal{C}}_{2}t^{-1}+
{\mathcal{C}}_{3}t^{-2/3}\,,  \label{dltlsDelta}
\end{equation}
since
${\mathcal{C}}_{4}\simeq(\beta_{0}/3)(\lambda_{H}/\lambda_k)_{0}^{2}
[{\mathcal{B}}_{0}-(4/3)\Delta_{0}]\ll1$. Hence, on large scales the
introduction of the $B$-field simply adds the decaying $t^{-2/3}$
mode to the standard magnetic-free result. Note also that $\Delta$
describes the directionally-averaged gravitational clumping of the
matter. Generally, the perturbations will grow at different rates
parallel and perpendicular to the magnetic field and so there will
also be non-spherical evolution in the shapes of these distortions.

\subsection{Evolution in false-vacuum environments}\label{ssEFvEs}
False vacuum cosmological environments have been largely associated
with inflation, since they typically lead to exponential, de
Sitter-type expansion. Dynamically, such environments can be
achieved by introducing a fluid with a non-conventional equation of
state. Here, we will consider linear inhomogeneities in the energy
density of a barotropic and of a non-barotropic magnetized medium
with a false-vacuum equation of state.

Nonlinear spatial inhomogeneities in the density distribution of a
perfect, though not necessarily barotropic, fluid propagate
according to\footnote{One can derive (\ref{fvdotDeltaa}) from first
principle, namely by taking the proper-time derivative of definition
(\ref{cDacBa}a), or directly from (\ref{dotcDa}) by substituting the
4-acceleration from Eq.~(\ref{mdc2}).}
\begin{equation}
\dot{\Delta}_{\langle a\rangle}= {\frac{3p}{\rho}}\,H\Delta_{a}-
{\frac{3aH}{\rho}}\,\mathrm{D}_{a}p- (\rho+p){\mathcal{Z}}_{a}-
3a(\rho+p)HA_{a}- (\sigma_{ba}+\omega_{ba})\Delta^{b}\,,
\label{fvdotDeltaa}
\end{equation}
The 4-acceleration vector seen above satisfies the momentum-density
conservation law. In the absence of a magnetic field the latter
reads
\begin{equation}
(\rho+p)A_{a}= -\mathrm{D}_{a}p\,,  \label{fvmdc1}
\end{equation}
with $\rho+p$ representing the inertial mass of the fluid. When
dealing with a magnetized medium, however, (\ref{fvmdc1}) is
replaced by
\begin{equation}
\left(\rho+p+{\frac{2}{3}}\,B^{2}\right)A_{a}= -\mathrm{D}_{a}p-
\varepsilon_{abc}B^{b}\mathrm{curl\,}B^{c}- \Pi_{ab}A^{b}\,,
\label{fvmdc2}
\end{equation}
which incorporates the magnetic contribution to the total inertial
mass of the system.

One way of achieving the false vacuum environment is by introducing
an effective (barotropic) perfect fluid with $p=-\rho$. In that case
Eq.~(\ref{fvmdc1}) acts as a constraint demanding that
$\mathrm{D}_{a}p=0$. This immediately implies $\mathrm{D}_{a}\rho=0$
and therefore zero density inhomogeneities. This strict constraint
is a direct and unavoidable consequence of having a medium with zero
inertial mass. When a magnetic field is present, however, the
4-acceleration is given by (\ref{fvmdc2}) and this expression allows
for nonzero pressure gradients even if $\rho+p=0$. In that case we
have $\mathrm{D}_{a}p=-\mathrm{D}_{a}\rho\neq0$ and
Eq.~(\ref{fvdotDeltaa}) reduces to
\begin{equation}
\dot{\Delta}_{a}= 0\,,  \label{lfvdotDeltaa1}
\end{equation}
at the linear perturbative level. Accordingly, all three types of
linear density inhomogeneities (i.e.~matter perturbations, vortices
and shape distortions) remain constant as long as $\rho+p=0$. Note
that we always assume that $B^{2}\ll\rho$, which ensures that the
magnetic field has no effect of the exponential expansion of the
model (i.e.~$\rho+3p+B^{2}\simeq\rho+3p=-2\rho<0$).

Standard de Sitter inflation is typically achieved by means of a
single minimally coupled scalar field ($\varphi$), which generally
does not behave like a barotropic medium. Instead, the
$\varphi$-field acts as an effective perfect fluid with energy
density and pressure given by
\begin{equation}
\rho= \rho^{(\varphi)}= {\frac{1}{2}}\,\dot{\varphi}^{2}+ V(\varphi)
\hspace{15mm} \mathrm{and} \hspace{15mm} p= p^{(\varphi)}=
{\frac{1}{2}}\,\dot{\varphi}^{2}- V(\varphi)\,,  \label{sf}
\end{equation}
respectively (e.g.~see~\cite{EM,BED}). Therefore, the scalar field
corresponds to a barotropic fluid only when its kinetic or potential
energies vanishes.\footnote{Non-minimally coupled scalar fields
generally correspond to imperfect fluids with more complicated
expressions for their associated effective density and pressure
(e.g.~see~\cite{Mad}).} In the first instance we have
$\dot{\varphi}=0$ and $p^{(\varphi)}=-\rho^{(\varphi)}=-V(\varphi)$,
while in the second Eqs.~(\ref{sf}) give
$p^{(\varphi)}=\rho^{(\varphi)}=\dot{\varphi}^{2}/2$. In any other
case $p=\rho-2V(\varphi)=\dot{\varphi}^{2}-\rho$. The perfect-fluid
description of a minimally coupled scalar field refers to a family
of observers moving with the (timelike) 4-velocity
$u_{a}=-\nabla_{a}\varphi/\dot{\varphi}$, where
$\nabla_{a}\varphi\neq0$ and $\dot{\varphi}=u^{a}\nabla_{a}\varphi$.
Relative to the same observers, $\mathrm{D}_{a}\varphi=0$ and
therefore $\mathrm{D}_{a}p^{(\varphi)}=
\mathrm{D}_{a}\rho^{(\varphi)}=
\dot{\varphi}\mathrm{D}_{a}\dot{\varphi}$~\cite{BED}. This last
result indicates a major change relative to the barotropic-fluid
case and it will have a profound effect on the evolution of the
inhomogeneities.

The de Sitter phase occurs when the scalar field rolls slowly down
its potential. Throughout this slow-rolling regime,
$\dot{\varphi}^{2}\ll V(\varphi)$ and expressions (\ref{sf}) imply
that $p^{(\varphi)}\simeq-\rho^{(\varphi)}\simeq-V(\varphi)$. In
this environment we may, to first approximation, ignore the last
three terms in the right-hand side of (\ref{fvdotDeltaa}), which
then linearizes to
\begin{equation}
\dot{\Delta}_{a}= -6H\Delta_{a}\,,  \label{lfvdotDeltaa2}
\end{equation}
where $H=\dot{a}/a\simeq$~constant and $a\propto\exp(Ht)$. This
result suggests an exponential decay for all three types of density
inhomogeneities irrespective of the magnetic presence and in line
with the cosmic no-hair theorems~\cite{JB}-\cite{Wald}.

We note that here we have only considered the dynamical effect of
the exponential expansion on the evolution of perturbations in the
presence of the $B$-field. Changes in the electrical properties of
the cosmic medium, which can occur during the de Sitter regime, have
not been accounted for. Also, typical linear studies, in
magnetic-free universes, consider the second-time derivative of the
density gradients instead of the first. This means that one
decouples the expansion gradients from (\ref{fvdotDeltaa}) before
setting $\rho+p\simeq0$. One could do the same in the magnetic case
as well. However, the calculation is rather involved and goes beyond
the scope of this section. We expect that the inhomogeneities will
still be found to decay exponentially, though probably at a slightly
slower rate.

\section{Isocurvature perturbations}\label{sLIPs}
One can define as isocurvature perturbations those occurring on
hypersurfaces of uniform curvature, namely fluctuations which
maintain $\mathrm{D}_{a}{\mathcal{R}}=0$ at all times~\cite{EB}.
This should be distinguished from the definition typically found in
the literature, where the term isocurvature means distortions in
multi-component systems with zero perturbation in the total
energy-density initially (e.g.~see~\cite{MFB,LL}). Here we will
impose the $\mathrm{D}_{a}{\mathcal{R}}=0$ condition throughout the
evolution of the perturbed mode, noting that one could use the
equations given in section \S ~\ref{sLDPs} to study fluctuations
with an isocurvature initial condition by setting
$\Delta_{0}+(\beta_{0}/2){\mathcal{B}}_{0}=0$ (see
\S~\ref{sssSbhSs}).

\subsection{Consistency condition}\label{ssCC}
Isocurvature perturbations also require zero vorticity to guarantee
the integrability of the 3-D hypersurfaces orthogonal to $u_{a}$. In
the magnetic presence, linear vortices are switched off by imposing
the (self-consistent) constraint
$B^{b}\mathrm{D}_{b}\mathrm{curl\,}B_{a}=0$ (see
Eq.~(\ref{lomegadot})). The zero-curvature condition for
isocurvature magnetic perturbations is obtained through the
orthogonally projected gradient of (\ref{l3R}). To be precise, using
definitions (\ref{cDacBa}), (\ref{cZaYa}a) and linearising around
our FRW background we arrive at
\begin{equation}
a\mathrm{D}_{a}{\mathcal{R}}= 2\rho\Delta_{a}+ c_{\mathrm{a}}^{2}
\rho(1+w){\mathcal{B}}_{a}- 4H{\mathcal{Z}}_{a}\,.  \label{lDacR}
\end{equation}
When $\mathrm{D}_{a}{\mathcal{R}}=0,$ the right-hand side of the
above ensures that linear expansion gradients and those in the fluid
and the magnetic energy densities are connected by a simple
algebraic relation. The projected comoving divergence of the latter
translates into the following linear constraint between the
associated scalar variables
\begin{equation}
2H{\mathcal{Z}}= \rho\left[\Delta
+{\frac{1}{2}}\,c_{\mathrm{a}}^{2}(1+w){\mathcal{B}}\right]\,.
\label{licon}
\end{equation}
In the absence of the magnetic field, this condition is
automatically satisfied for pressure-free dust, but holds on large
scales only when the matter has non-zero pressure~\cite{EB}. If a
magnetic field is present the vanishing of the matter pressure is
not enough to guarantee that $\mathrm{D}_{a}{\mathcal{R}}=0$ at all
times, because of the Lorentz-force contribution to the
4-acceleration. In that case the consistency condition for linear
isocurvature perturbations is satisfied on large scales only.
Indeed, on using the linearized propagation equation of the spatial
Ricci scalar, we arrive at
\begin{equation}
(\mathrm{D}_{a}{\mathcal{R}})^{\displaystyle\cdot}=
-3\left(1+{\frac{4}{9}}\,c_{\mathrm{a}}^{2}\right)
H\mathrm{D}_{a}{\mathcal{R}}+ {\frac{4c_{s}^{2}H}{a(1+w)}}
\left(1-{\frac{2}{3}}\,c_{\mathrm{a}}^{2}\right)
\mathrm{D}^{2}\Delta_{a}+
{\frac{2c_{\mathrm{a}}^{2}H}{a}}\,\mathrm{D}^{2}{\mathcal{B}}_{a}\,,
\label{dotlicon}
\end{equation}
since the vorticity has already been switched off. This result shows
that the linear isocurvature condition is self-maintained only
asymptotically (i.e.~at infinity) where the Laplacians of the
right-hand side vanish. Nevertheless,
following~\cite{EB}-\cite{EBH}, we will assume that on sufficiently
long wavelengths the source terms in the right-hand side of
(\ref{dotlicon}) are negligible. Note the last term in the
right-hand side of the above, which demonstrates why zero fluid
pressure does not automatically guarantee the consistency of
$\mathrm{D}_{a}{\mathcal{R}}=0$ in the presence of the $B$-field.

Using the isocurvature condition (\ref{licon}), we can eliminate
${\mathcal{Z}}$ from Eq.~(\ref{ldotDelta}), which reduces to
\begin{equation}
\dot{\Delta}=
-{\frac{3}{2}}\left(1-w+{\frac{4}{3}}\,c_{\mathrm{a}}^{2}w\right)
H\Delta+ {\frac{3}{4}}\,c_{\mathrm{a}}^{2}(1-w^{2})H{\mathcal{B}}-
c_{\mathrm{a}}^{2}(1+w)H{\mathcal{K}}\,.  \label{ildotDelta}
\end{equation}
The above, together with (\ref{ldotcB}) and (\ref{ldotcK}),
describes the linear evolution of isocurvature scalar perturbations
on a weakly magnetic flat FRW background filled with a single highly
conducting perfect fluid.

\subsection{Evolution in the radiation era}\label{ssERE2}
Since isocurvature perturbations apply to super-horizon scales only,
all gradients in the fluid and the magnetic pressure have been
switched off. This means that the only support against the
gravitational pull of the matter comes from the expansion of the
universe. When relativistic matter dominates the energy density of
the latter, the system (\ref{ildotDelta}), (\ref{ldotcB}) and
(\ref{ldotcK}) reduces to
\begin{eqnarray}
\dot{\Delta}&=& -{\frac{1}{2}}\left(1+{\frac{1}{2}}\,\beta\right)
t^{-1}\Delta- {\frac{1}{2}}\,\beta t^{-1}{\mathcal{K}}+
{\frac{1}{4}}\,\beta t^{-1}{\mathcal{B}}\,,  \label{irdotDel}\\
\dot{\mathcal{K}}&=& -{\frac{1}{2}}\,\beta t^{-1}{\mathcal{K}}+
{\frac{1}{2}}\left(1-{\frac{1}{2}}\,\beta\right)t^{-1}\Delta+
{\frac{3}{4}}\,\beta t^{-1}{\mathcal{B}}\,,  \label{irdotcK}\\
\dot{\mathcal{B}}&=& \dot{\Delta}\,.  \label{irdotcB}
\end{eqnarray}
Taking into account that $\beta=\mathrm{constant}\ll1$ and keeping
up to $\beta$-order terms, the solution for magnetised isocurvature
perturbations reads
\begin{equation}
\Delta= {\frac{1}{2}}\,\beta(\Delta_{0}- {\mathcal{B}}_{0})+
{\mathcal{C}}_{1}t^{-\beta}+ {\mathcal{C}}_{2}t^{-(1-\beta)/2}\,,
\label{irDelta}
\end{equation}
where $\Delta_{0}$, ${\mathcal{B}}_{0}$ are the perturbed matter and
magnetic energy densities initially and ${\mathcal{C}}_{1,2}$ are
constants (see also~\cite{TM1}). When
$\Delta_{0}\neq{\mathcal{B}}_{0}$, the isocurvature density contrast
approaches a non-zero value which depends on the initial conditions.

\subsection{Evolution in the dust era}\label{ssEDE2}
After matter-radiation equality, the system (\ref{ildotDelta}),
(\ref{ldotcB}) and (\ref{ldotcK}) becomes
\begin{eqnarray}
\dot{\Delta}&=& -t^{-1}\Delta- {\frac{2}{3}}\,\beta
t^{-1}{\mathcal{K}}+ {\frac{1}{2}}\,\beta t^{-1}{\mathcal{B}}\,,
\label{iddotDelta}\\ \dot{\mathcal{K}}&=& -{\frac{8}{9}}\,\beta
t^{-1}{\mathcal{K}}+ {\frac{4}{3}}\,\beta t^{-1}{\mathcal{B}}\,,
\label{iddotcK}\\ \dot{\mathcal{B}}&=&
{\frac{4}{3}}\,\dot{\Delta}\,.  \label{iddotcB}
\end{eqnarray}
with $\beta=\beta_{0}(t_{0}/t)^{2/3}$ throughout the dust era.
Because of this time-variation in the effective sound speed, the
above system has no simple analytical solution. Nevertheless, taking
the time derivative of (\ref{iddotDelta}), using (\ref{iddotcK}),
(\ref{iddotcB}) and keeping up to $\beta_{0}$-order terms we arrive
at
\begin{equation}
\ddot{\Delta}= -{\frac{8}{3}}\left[1-{\frac{1}{4}}\,\beta_{0}
\left(\frac{t_{0}}{t}\right)^{2/3}\right]t^{-1}\dot{\Delta}-
{\frac{2}{3}}\,t^{-2}\Delta\,.  \label{idddotDelta}
\end{equation}
At late times, we may ignore the weak and decaying
$\beta_{0}(t_{0}/t)^{2/3}$ term in the right-hand side of the above,
which then admits the power-law solution~\cite{TM1}
\begin{equation}
\Delta= {\mathcal{C}}_{1}t^{-2/3}+ {\mathcal{C}}_{2}t^{-1}\,.
\label{idDelta}
\end{equation}
Comparing results (\ref{irDelta}) and (\ref{idDelta}) to their
magnetic-free counterparts, shows that the presence of the
$B$-fieldd has only added new decaying isocurvature modes. Recall
that in single-fluid almost-FRW cosmologies, isocurvature modes
evolve as $\Delta\propto t^{-1/2}$ during the radiation era and like
$\Delta\propto t^{-1}$ after equality~\cite{EB}.

\section{Entropy perturbations}\label{sLEPs}
\subsection{Entropy perturbations in multi-component
systems}\label{ssEPMcSs}
Entropy perturbations can and will generically arise in all
multi-component systems. The magnetized, highly conductive
barotropic medium may be treated as a mixture of two comoving
fluids, of which one (the matter component) is perfect and the other
(the $B$-field) is imperfect (see Eq.~(\ref{MHDTab})). Then, it
would be interesting to see whether some of the magnetic effects can
be interpreted as coming from an effective entropy perturbation
caused by differences in the dynamical behaviour of the two fluids.

Assuming that ${\mathcal{E}}_{a}^{(i)}$ is the intrinsic entropy
perturbation associated with the $i$-th component of a multi-fluid
system, the effective entropy perturbation of the total fluid is
given by the expression~\cite{KS}-\cite{Gi7}
\begin{equation}
{\mathcal{E}}_{a}=
\frac{1}{\bar{p}}\sum_{i}p^{(i)}{\mathcal{E}}_{a}^{(i)}+
\frac{1}{2\bar{p}\,\bar{h}}\sum_{ij}h^{(i)}h^{(j)}
\left(c_{s}^{2(i)}-c_{s}^{2(j)}\right)\mathcal{S}_{a}^{(ij)}\,.
\label{cEai}
\end{equation}
Here, $h^{(i)}=\rho^{(i)}+p^{(i)}$ represents the inertial mass of
the species, while $\bar{p}=\sum_{i}p^{(i)}$ and
$\bar{h}=\sum_{i}h^{(i)}$ are respectively the isotropic pressure
and the inertial mass of the mixture. Also, the orthogonally
projected vector
\begin{equation}
\mathcal{S}_{a}^{(ij)}=
{\frac{\rho^{(i)}}{h^{(i)}}}\,\Delta_{a}^{(i)}-
{\frac{\rho^{(j)}}{h^{(j)}}}\,\Delta_{a}^{(j)}\,,  \label{Saij}
\end{equation}
with $\Delta_{a}^{(i)}=(a/\rho^{(i)})\mathrm{D}_{a}\rho^{(i)}$ and
$\mathcal{S}_{a}^{(ij)}=-\mathcal{S}_{a}^{(ji)}$, describes the
effective entropy perturbation triggered by the different dynamical
behaviour of the individual components. Therefore, in a multi-fluid
system, the total entropy fluctuation has one part coming from the
intrinsic entropy perturbation of each species and one coming from
differences in their dynamics. In this sense, a state of effective
overall adiabaticity corresponds to ${\mathcal{E}}_{a}=0$. Note
that, according to (\ref{cEai}), there is no entropy perturbation
coming from differences in the dynamics of the member species when
these share the same effective sound speed.

\subsection{Magnetically-induced entropy
perturbations}\label{ssMiEPs}
Treating the highly conducting magnetized medium as a system of two
comoving fluids, we may only consider entropy perturbations due to
the different dynamical evolution of the two components. These are
determined by
\begin{equation}
\mathcal{S}_{a}^{(12)}= {\frac{1}{1+w}}\,\Delta_{a}-
{\frac{3}{4}}\,{\mathcal{B}}_{a}\,,  \label{Sa1}
\end{equation}
given that $\rho^{(1)}=\rho$, $\rho^{(2)}=B^{2}/2$,
$h^{(1)}=\rho+p$, $h^{(2)}=2B^{2}/3$, $c_{s}^{2(1)}=c_{s}^{2}$,
$c_{s}^{2(2)}=1/3$ (since $p_{B}=\rho_{B}/3=B^{2}/6$),
$\Delta_{a}^{(1)}=\Delta_{a}$, $\Delta_{a}^{(2)}={\mathcal{B}}_{a}$,
$\bar{p}=p+B^{2}/6$ and $\bar{h}=\rho+p+2B^{2}/3$. However, linear
perturbations in the magnetic energy density evolve in step with
their matter counterparts (see Eq.~(\ref{ldotcBa2})). This
guarantees that
\begin{equation}
\mathcal{S}_{a}^{(12)}= {\frac{3}{4}}\;{\mathcal{C}}_{a}\,,
\label{Sa2}
\end{equation}
where
${\mathcal{C}}_{a}={\mathcal{B}}_{a}^{0}-4\Delta_{a}^{0}/3(1+w)$ is
a constant that depends on the initial conditions. Substituted into
definition (\ref{cEai}), the above gives
\begin{equation}
{\mathcal{E}}_{a}= -\frac{3\rho(1+w)(3c_{s}^{2}-1)B^{2}}
{(6p+B^{2})[3\rho(1+w)+2B^{2}]}\,{\mathcal{C}}_{a}\,.  \label{cEa}
\end{equation}
This vanishes during the radiation era -- a result guaranteed by the
radiation-like behaviour of the $B$-field (i.e.~by the fact that
$p_{B}=\rho_{B}/3$ -- see also~\cite{Gi8}). In the dust epoch, on
the other hand, we find that
${\mathcal{E}}_{a}\simeq{\mathcal{C}}_{a}=$~constant (since
$B^{2}\ll\rho$ always). These mean that, within the two-fluid
description, part of the magnetic effect on the linear evolution of
matter inhomogeneities can be interpreted as an effective entropy
perturbation, due to the different dynamics of the two perturbed
fluids, but only after equality. Then, the value of the effective
entropy fluctuation depends on the initial relation between the
magnetic and the fluid energy-density perturbations. In particular,
when ${\mathcal{B}}_{a}=4\Delta_{a}/3$ initially the dynamically
induced entropy perturbation vanishes and we have effective
adiabaticity.

\section{Vector perturbations}\label{sLVs}
A general inhomogeneous perturbation is characterized by its
rotational and deformation properties, in addition to the amount of
matter aggregation (or dilution). Following~\cite{TM1}, we will
consider here the effect of magnetic fields on the linear evolution
of vortex-like distortions in the density distribution of the cosmic
medium.

\subsection{Basic variables}\label{ssBVs2}
Rotational instabilities in the fluid density are described by means
of the antisymmetric, comoving, orthogonally projected tensor
$\mathcal{W}_{ab}=a\mathrm{D}_{[b}\Delta_{a]}$ (see decomposition
(\ref{dDeltaab}) in \S~\ref{sssIIVs}). The antisymmetry of the
latter means that we can use it to define the vector
\begin{equation}
\mathcal{W}_{a}= {1\over2}\,\varepsilon_{abc}\mathcal{W}^{bc}=
-{\frac {a}{2}}\,\mathrm{curl\,}\Delta_{a}\,,  \label{cWa1}
\end{equation}
with $\mathrm{D}^{a}\mathcal{W}_{a}=0$ to linear order. Recalling
that $\Delta_{a}=(a/\rho)\mathrm{D}_{a}\rho$ and using the general
relativistic commutation law
$\mathrm{D}_{[a}\mathrm{D}_{b]}\phi=-\dot{\phi}\omega_{ab}$ between
the projected gradients of scalars (see Eq.~(\ref{A1}) in
\S~\ref{AssOPGs}), we find that
\begin{equation}
\mathcal{W}_{a}= -3a^{2}(1+w)H\omega_{a}\,.  \label{cWa2}
\end{equation}
This result reflects a fundamental property of vortex-like
distortions in general relativity, namely that they are proportional
to the amount of vorticity. Also, (\ref{cWa2}) and the linearized
version of constraint (\ref{omegacon}) guarantee that
$\mathcal{W}_{a}$ is a solenoidal vector to first order. Similarly,
linear vortices in the volume expansion and the magnetic energy
density are also directly related to the vorticity vector and
therefore to those in the density of the matter. In particular,
\begin{equation}
{\frac{a}{2}}\,\mathrm{curl\,}\mathcal{Z}_{a}=
{\frac{\dot{H}}{(1+w)H}}\,\mathcal{W}_{a} \hspace{15mm} \mathrm{and}
\hspace{15mm} {\frac{a}{2}}\,\mathrm{curl\,}\mathcal{B}_{a}=
-{\frac{4}{3(1+w)}}\,\mathcal{W}_{a}\,,  \label{curlcZcBac}
\end{equation}
to first approximation. Finally, combining the antisymmetric part of
the linearized Gauss-Codacci equation (see (\ref{MHDGC}) in
\S~\ref{ssStC}) with Eq.~(\ref{cWa2}), we obtain
\begin{equation}
\mathcal{R}_{a}= {\frac{1}{3a^{2}(1+w)}}\,\mathcal{W}_{a}\,,
\label{cRa}
\end{equation}
where $\mathcal{R}_{a}=\varepsilon_{abc}\mathcal{R}^{bc}/2$ is the
vector component of the local spatial curvature.

\subsection{Linear equations}\label{ssLEs2}
Magnetic fields are unique in the sense that they are the only
large-scale vector source that has ever been observed in the
universe. Given that, the study of vortex-like, vector perturbations
in the presence of the $B$-field acquires particular significance.
In our analysis, effects directly related to the vector nature of
magnetic fields propagate through the tension part of the Lorentz
force (see decomposition (\ref{Lorentz})). Thus, as far as the
scalar/density perturbations are concerned, these effects are
encoded in the magneto-curvature terms seen in
Eqs.~(\ref{ldotDelta}), (\ref{ldotcZ}).

Spatial inhomogeneities in the density distribution of a magnetized
and highly conductive barotropic perfect fluid propagate according
to the linear expression (\ref{ldotcDa1}) in \S ~\ref{sssLEs}. When
combined with (\ref{lmdc3}), the latter takes the form
\begin{eqnarray}
\dot{\Delta}_{a}&=&
3w\left(1-{\frac{2}{3}}\,c_{\mathrm{a}}^{2}\right) H\Delta_{a}-
(1+w)\mathcal{Z}_{a}+
{\frac{3}{2}}\,c_{\mathrm{a}}^{2}(1+w)H\mathcal{B}_{a} \nonumber\\
&&-{\frac{3aH}{\rho}}\left(1-{\frac{2}{3}}\,c_{\mathrm{a}}^{2}\right)
B^{b}\mathrm{D}_{b}B_{a}\,,  \label{ldotcDa2}
\end{eqnarray}
where we have decomposed the magnetic Lorentz force into its
isotropic-pressure and tension parts (see (\ref{Lorentz}) in
\S~\ref{ssCL2}) and kept up to $c_{\mathrm{a}}^{2}$-order terms.
Taking the comoving, orthogonally projected gradient of the above
and isolating its antisymmetric part we arrive at
\begin{eqnarray}
\dot{\mathcal{W}}_{a}&=&
-{\frac{3}{2}}\left\{1-w-{\frac{1}{18}}\,c_{\mathrm{a}}^{2}
\left[(1-3w)^{2}-12w\right]\right\}H\mathcal{W}_{a} \nonumber\\
&&+{\frac{3a^{2}H}{2\rho}}
\left(1-{\frac{2}{3}}\,c_{\mathrm{a}}^{2}\right)B^{b}
\mathrm{D}_{b}\mathrm{curl\,}B_{a}\,,  \label{ldotcWa}
\end{eqnarray}
to linear order. Note that in deriving this expression we have also
used the auxiliary relations (\ref{curlcZcBac}), (\ref{cRa}), the
commutation laws between the spatial gradients of scalars and
orthogonally projected vectors and the zero-order versions of
Eqs.~(\ref{MHDRay}), (\ref{MHD3R}). The above shows that the
$B$-field acts as a source of density vortices when
$B^{b}\mathrm{D}_{b}\mathrm{curl\,}B_{a}\neq0$. Not surprisingly,
the same requirement can also lead to kinematic vorticity (see
Eq.~(\ref{lomegadot}) in \S~\ref{ssLE1}). In fact, recalling that
$w=c_{s}^{2}=$~constant, one can recover (\ref{lomegadot}) from
(\ref{ldotcWa}) by means of relation (\ref{cWa2}). An additional
magnetic effect is the precession of $\mathcal{W}_{a}$, since its no
longer parallel to $\dot{\mathcal{W}}_{a}$ (see also~\cite{TM1}).

For comparison reasons it will help to consider first the solution
of Eq.~(\ref{ldotcWa}) in the absence of the $B$-field. Assuming
conventional matter and dropping the magnetic terms from the
right-hand side of the above, it is straightforward to show that
non-magnetized fluid vortices decay always and on all scales. In
particular, $\mathcal{W}_{a}\propto t^{-1/2}$ during the radiation
era and $\mathcal{W}_{a}\propto t^{-1}$ after equality. In what
follows we will illustrate how the magnetic field changes this
picture by focusing on the dust era, referring the reader
to~\cite{TM1} for further discussion and details.

\subsection{Evolution in the dust era}\label{ssEDE3}
Expression (\ref{ldotcWa}) shows that the presence of the $B$-field
will generally trigger vortices in the density distribution of the
magnetized matter. Nevertheless, the magnetic effect on preexisting
rotational distortions is not clear yet. To quantify the role of the
field we need to go one step further and obtain a decoupled equation
for the evolution of $\mathcal{W}_{a}$. After equality, the
Alfv\'{e}n speed decays with time according to
$c_{\mathrm{a}}^{2}=\beta\propto t^{-2/3}$. This means that, very
soon, the weak Alfv\'{e}n terms in the right-hand side of
(\ref{ldotcWa}) become completely negligible. At this limit, the
time derivative of the latter leads to~\cite{TM1}
\begin{equation}
\ddot{\mathcal{W}}_{a}= -4H\dot{\mathcal{W}}_{a}-
{\frac{1}{2}}\,\rho \mathcal{W}_{a}
+{\frac{1}{3}}\,c_{\mathrm{a}}^{2}\mathrm{D}^{2}\mathcal{W}_{a}\,.
\label{lddotcWa}
\end{equation}
Thus, during the dust era, magnetized vortices propagate like
Alfv\'{e}n waves with signal speed
$v_{\mathrm{a}}=c_{\mathrm{a}}/\sqrt{3}$. Using standard vector
harmonics, we may decompose the solenoidal $\mathcal{W}_{a}$ as
$\mathcal{W}_{a}=\sum_{k}\mathcal{W}_{(k)}\mathcal{Q}_{a}^{(k)}$,
where $\mathrm{D}_{a}\mathcal{W}_{(k)}=0=
\dot{\mathcal{Q}}_{a}^{(k)}=\mathrm{D} ^{a}\mathcal{Q}_{a}^{(k)}$
and $\mathrm{D}^{2}\mathcal{Q}_{a}^{(k)}=
-(k/a)^{2}\mathcal{Q}_{a}^{(k)}$. Then, recalling that
$w=0=c_{s}^{2}$, $a\propto t^{2/3}$, $H=2/3t$ and $\rho=4/3t^{2}$,
the $k$-th Fourier mode propagates according to
\begin{equation}
\ddot{\mathcal{W}}^{(k)}=
-{\frac{8}{3}}\,t^{-1}\dot{\mathcal{W}}^{(k)}-
{\frac{2}{3}}\left[1+{\frac{2}{9}}\,
\left({\frac{\lambda_{\mathrm{a}}}{\lambda_k}}\right)_{0}^{2}\right]t^{-2}
\mathcal{W}^{(k)}\,,  \label{dlddotcW}
\end{equation}
with $\lambda_k=a/k$ representing the wavelength of the perturbation
and $\lambda_{\mathrm{a}}=\sqrt{\beta}\lambda_{H}$. This represents
a characteristic scale of the magnetized environment, which may be
termed the `Alfv\'{e}n horizon'~\cite{TM1}. The above admits the
solution
\begin{equation}
\mathcal{W}^{(k)}= \mathcal{C}_{1}t^{\alpha_{1}}+
\mathcal{C}_{2}t^{\alpha_{2}}\,,  \label{dcW}
\end{equation}
where $\mathcal{C}_{1,\,2}$ are constants and
\begin{equation}
\alpha_{1,\,2}= -{\frac{1}{6}}\left[5\pm\sqrt{1-{\frac{48}{9}}
\left({\frac{\lambda_{\mathrm{a}}}{\lambda_k}}\right)_{0}^{2}}\;
\right]\,.  \label{dalphas}
\end{equation}
On scales far beyond the Alfv\'{e}n horizon,
$\lambda_{\mathrm{a}}/\lambda_k\ll1$ and the perturbed mode exhibits
a power-law decay with $\mathcal{W}\propto t^{-2/3}$, a rate
considerably slower than the one associated with dust dominated,
magnetic-free cosmologies. Recall that $\mathcal{W}\propto t^{-1}$
in the absence of the $B$-field. Well inside $\lambda_{\mathrm{a}}$,
on the other hand, perturbations oscillate like Alfv\'{e}n waves
with a time-decreasing amplitude. In particular, for
$\lambda_{\mathrm{a}}/\lambda_k\gg1$, solution (\ref{dcW}),
(\ref{dalphas}) implies
\begin{equation}
\mathcal{W}_{(k)}\propto \cos\left[{2\sqrt{3}\over9}
\left({\lambda_{\mathrm{a}}\over\lambda_k}\right)_0\ln
t\right]t^{-5/6}\,.  \label{dsscW}
\end{equation}
Thus, the overall, the effect of the $B$-field on a given vortex
mode is to decrease its standard depletion rate.

Not surprisingly, a similar magnetic effect is also observed on
vorticity proper. Using the linear relation (\ref{cWa2}), we find
that after equality large-scale rotational distortions decay as
$\omega_{a}\propto t^{-1}$ as opposed to $\omega_{a}\propto
t^{-4/3}$. On these grounds, magnetized cosmologies would contain
more residual vortices than their magnetic-free counterparts.

\section{Gravitational waves}\label{sLGWs}
Gravitational waves are propagating fluctuations in the geometry of
the spacetime fabric, usually described as weak perturbations of the
background metric. Alternatively, one can describe gravity waves
covariantly by means of the electric and magnetic components of the
Weyl tensor~\cite{H}, which encodes the long-range gravitational
field, namely the one determined by the presence of matter
``elsewhere'' in the space (see \S~\ref{ssLrWC}).

\subsection{Isolating magnetised tensor modes}\label{ssIMGWs}
Covariantly, gravitational waves are described by the transverse
degrees of freedom in the electric ($E_{ab}$) and magnetic
($H_{ab}$) parts of the conformal curvature tensor. The
transversality is necessary to ensure that the pure tensor modes of
the locally free gravitational field have been isolated. The same
condition is also imposed on the shear and any other orthogonally
projected, traceless, second-rank tensor that might be present.
Thus, when studying the propagation of gravitational radiation in
perturbed FRW models with perfect fluid matter we require that (see
\S~\ref{ssCKs} and \S~\ref{ssLrWC})
\begin{equation}
\mathrm{D}^{b}E_{ab}=0= \mathrm{D}^{b}H_{ab}=
\mathrm{D}^{b}\sigma_{ab}\,,  \label{lnmtrans}
\end{equation}
to linear order and at all times (e.g.~see~\cite{Ch}). In
magnetic-free universes, this is achieved by switching the vorticity
off and by setting
$\mathrm{D}_{a}\rho=0=\mathrm{D}_{a}p=\mathrm{D}_{a}\Theta$ (for a
barotropic medium it suffices to ensure that
$\mathrm{D}_{a}\rho=0=\mathrm{D}_{a}\Theta$). These constraints,
which are self-consistent (i.e.~preserved in time) at the linear
perturbative level, guarantee that the 4-acceleration also vanishes
to first approximation. When a magnetic field is included there is
an additional constraint on the anisotropic pressure of the field.
In particular, the system (\ref{lnmtrans}) reads~\cite{T1,MTU}
\begin{eqnarray}
\mathrm{D}^{b}E_{ab}&=& {\frac{1}{3}}\,\mathrm{D}_{a}\rho+
{\frac{1}{6}}\,\mathrm{D}_{a}B^{2}-
{\frac{1}{2}}\,\mathrm{D}^{b}\Pi_{ab}= 0\,,  \label{ltrans1}\\
\mathrm{D}^{b}H_{ab}&=& \rho(1+w)\omega_{a}= 0\,,  \label{ltrans2}\\
\mathrm{D}^{b}\sigma_{ab}&=& {\frac{2}{3}}\,\mathrm{D}_{a}\Theta+
\mathrm{curl\,}\omega_{a}= 0\,,  \label{ltrans3}\\
\mathrm{D}^{b}\Pi_{ab}&=&
\varepsilon_{abc}B^{b}\mathrm{curl\,}B^{c}-
{\frac{1}{6}}\,\mathrm{D}_{a}B^{2}=0\,.  \label{ltrans4}
\end{eqnarray}

When studying the magnetic effects on cosmological gravitational
waves, one needs to take into account the anisotropic nature of the
$B$-field. It is therefore not appropriate to assume a fully random
$B$-field, which simply acts as an additional source of energy
density. Thus, following~\cite{DFK} and also~\cite{T1,MTU}, we will
treat both the energy density and the anisotropic pressure of the
field as first order perturbations.\footnote{Formally, treating
$B^{2}=B_{a}B^{a}$ as a first-order variable, means that $B_{a}$ is
of order 1/2 in perturbative terms. Then, because
$\mathrm{D}_{a}B^{2}$ is of order 1 and
$\mathrm{D}_{a}B^{2}=2B^{b}\mathrm{D}_{a}B_{b}$, consistency demands
that the projected magnetic gradients are also half-order
perturbations.} This approach differs from the perturbative scheme
adopted for the scalar and the vector modes, where there was a weak
$B$-field in the background. Nevertheless, adopting the scheme
of~\cite{DFK} simplifies the mathematics of the gravito-magnetic
interaction without compromising the physics.

We assume a spatially flat FRW background. Then one can isolate the
pure-tensor perturbations by adding to the previously mentioned
perfect-fluid constraints the following linear
conditions~\cite{T1,MTU}
\begin{equation}
\mathrm{D}_{a}B^{2}=0= \varepsilon_{abc}B^{b}\mathrm{curl\,}B^{c}\,.
\label{lmtrans}
\end{equation}
In other words, both magnetic energy-density gradients and the
Lorentz force vanish at all times. These guarantee that there is no
magnetic contribution to the 4-acceleration and subsequently that
the field does not trigger any vorticity, density or expansion
perturbations. The consistency of constraints (\ref{lmtrans}) is
straightforward to show. In particular, we find that
\begin{equation}
(\mathrm{D}_{a}B^{2})^{\displaystyle\cdot}= -5H(\mathrm{D}_{a}B^{2})
\hspace{15mm} \mathrm{and} \hspace{15mm}
(\varepsilon_{abc}B^{b}\mathrm{curl\,}B^{c})^{\displaystyle\cdot}=
-5H\varepsilon_{abc}B^{b}\mathrm{curl\,}B^{c}\,,  \label{lmtranscs}
\end{equation}
to lowest order. This means that the constraints (\ref{lmtrans})
continue to hold after they are initially imposed. Then, the only
remaining nontrivial linear constraints are
\begin{equation}
H_{ab}=\mathrm{curl\,}\sigma_{ab} \hspace{15mm} \mathrm{and}
\hspace{15mm} \mathcal{R}_{\langle ab\rangle}= -H\sigma_{ab}+
E_{ab}+ {\frac{1}{2}}\,\Pi_{ab}\,,  \label{lGWcon}
\end{equation}
where $\mathrm{curl\,}\sigma_{ab}=\varepsilon_{cd\langle
a}\mathrm{D}^{c}\sigma_{b\rangle}{}^{d}$ (see Eqs.~(\ref{Hcon}) and
(\ref{GC}) in \S~\ref{ssCKs} and \S~\ref{ssSC} respectively). Note
that, according to (\ref{lGWcon}b), the linear conditions
$\mathrm{D}^{b}E_{ab}=0=\mathrm{D}^{b}\sigma_{ab}=
\mathrm{D}^{b}\Pi_{ab}$ guarantee that
$\mathrm{D}^{b}\mathcal{R}_{\langle ab\rangle}=0$ as well.

\subsection{Covariant gravitational-wave energy
density}\label{ssCGwED}
The energy density of gravitational radiation is determined by the
pure tensor, namely the transverse traceless part
($h_{\alpha\beta}^{TT}$) of the metric perturbation, according to
(e.g.~see~\cite{CD1})
\begin{equation}
\rho_{GW}= {\frac{(h_{\alpha\beta}^{TT})^{\prime}
(h_{TT}^{\alpha\beta})^{\prime}}{2a^{2}}}\,,  \label{GWrho}
\end{equation}
where a prime indicates differentiation with respect to conformal
time (recall that $c=1=8\pi G$ throughout this review). In a
comoving frame, with $u^{a}=\delta_0{}^au^0$, we have
$\sigma^{2}=\sigma_{ab}\sigma^{ab}/2=\sigma_{\alpha\beta}
\sigma^{\alpha\beta}/2$, with the transverse part of the shear
components given by~\cite{BED,G}
\begin{equation}
\sigma_{\alpha\beta}= a(h_{\alpha\beta}^{TT})^{\prime} \hspace{15mm}
\mathrm{and} \hspace{15mm} \sigma^{\alpha\beta}=
a^{-3}(h_{TT}^{\alpha\beta})^{\prime}\,.  \label{TTsigma}
\end{equation}
Solving these relations for the metric perturbations and then
substituting the results into Eq.~(\ref{GWrho}), we arrive
at~\cite{T1}
\begin{equation}
\rho_{GW}=\sigma^{2}\,,  \label{cGWrho}
\end{equation}
which provides a simple covariant expression for the energy density
of gravitational-wave distortions.

\subsection{Evolution equations}\label{ssLEs3}
In an FRW spacetime the Weyl tensor vanishes identically, which
means that $E_{ab}$ and $H_{ab}$ provide a covariant and
gauge-invariant description of perturbations in the
free-gravitational field. Once the pure tensor modes have been
isolated, we can proceed to linearize the propagation equations
(\ref{dotEab}), (\ref{dotHab}) of \S ~\ref{ssLrWC}. On a Friedmann
background with a single perfect fluid, the latter reduce
to~\cite{T1,MTU}
\begin{eqnarray}
\label{ldotEab1}\dot{E}_{ab}&=& -3HE_{ab}-
{\frac{1}{2}}\,\rho(1+w)\sigma_{ab}+ \mathrm{curl\,}H_{ab}-
{\frac{1}{2}}\,\dot{\Pi}_{ab}- {\frac{1}{2}}\,H\Pi_{ab}\,,\\
\dot{H}_{ab}&=& -3HH_{ab}- \mathrm{curl\,}E_{ab}+
{\frac{1}{2}}\,\mathrm{curl\,}\Pi_{ab}\,\,.  \label{ldotHab}
\end{eqnarray}
Since the magnetic part of the Weyl tensor satisfies the constraint
(\ref{lGWcon}a), the linear evolution of $H_{ab}$ is determined by
the shear evolution (\ref{sigmadot})
\begin{equation}
\dot{\sigma}_{ab}= -2H\sigma_{ab}- E_{ab}+
{\frac{1}{2}}\,\Pi_{ab}\,.  \label{ldotsigma}
\end{equation}
Furthermore, on using the commutation law between the orthogonally
projected gradients of spacelike tensors, constraint (\ref{lGWcon}a)
leads to the auxiliary relation $\mathrm{curl\,}H_{ab}=
-\mathrm{D}^{2}\sigma_{ab}$, and Eq.~(\ref{ldotEab1}) becomes
\begin{equation}
\dot{E}_{ab}= -3HE_{ab}- {\frac{1}{2}}\,\rho(1+w)\sigma_{ab}-
{\frac{1}{2}}\,\dot{\Pi}_{ab}- {\frac{1}{2}}\,H\Pi_{ab}-
\mathrm{D}^{2}\sigma_{ab}\,.  \label{ldotEab2}
\end{equation}
This, together with Eq.~(\ref{ldotsigma}) and the linearized part of
(\ref{Pidot}), namely
\begin{equation}
\dot{\Pi}_{ab}= -4H\Pi_{ab}\,,  \label{ldotPi}
\end{equation}
describe the linear evolution of gravitational waves in almost-FRW
universes containing a single highly conductive barotropic fluid and
a magnetic field.

We can describe this evolution in terms of the shear as follows.
Taking the time derivative of (\ref{ldotsigma}), using
Eqs.~(\ref{ldotEab2}), (\ref{ldotPi}), the background Raychaudhuri
and Friedmann equations, and keeping only linear order terms, we
arrive at the following wave equation for the gravitationally
induced shear~\cite{MTU}
\begin{equation}
\ddot{\sigma}_{ab}- \mathrm{D}^{2}\sigma_{ab}= -5H\dot{\sigma}_{ab}-
{\frac {1}{2}}\,\rho(1-3w)\sigma_{ab}- 2H\Pi_{ab}\,.
\label{lddotsigma}
\end{equation}
This is no longer coupled to the electric Weyl tensor, and together
with Eq.~(\ref{ldotPi}), describes the propagation of gravitational
waves. The magnetic anisotropic stress can act as a source for
gravitational waves, and it is possible to place strong constraints
on magnetic fields via this generation of gravitational
waves~\cite{CD1,CD2}. Also, the anisotropic stress of the $B$-field
slows the decay of shear~\cite{BM}.

Following (\ref{lddotsigma}), the solution of Eqs.~(\ref{ldotPi})
and (\ref{lddotsigma}) depends critically on the equation of state
of the matter component, and in particular on whether $w=1/3$ or
not. In the matter dominated era and on superhorizon scales, we find
\begin{equation}
\sigma= P_1\,t^{-1/3}+ P_2\,t^{-2}+ Q\,t^{-5/3}\,,  \label{dlsigma}
\end{equation}
where $\dot{P}_1=0=\dot{P}_2=\dot{Q}$. The $P$-modes are the usual
Bianchi~$I$~type solutions, while the $Q$-mode carries the effect of
the $B$-field. The latter mode means that, although the magnetic
presence does not alter the standard evolution rate of the shear, it
can affect its magnitude (in a rather involved way --
see~\cite{T1}). Note that in deriving the above we have also
introduced the familiar tensor harmonics $\mathcal{Q}_{ab}^{(k)}$,
with $\mathcal{Q}_{ab}^{(k)}=\mathcal{Q}_{\langle ab\rangle}^{(k)}$,
$\dot{\mathcal{Q}}_{ab}^{(k)}=0={\rm D}^b\mathcal{Q}_{ab}^{(k)}$ and
${\rm D}^2\mathcal{Q}_{ab}^{(k)}=-(k/a)^2\mathcal{Q}_{ab}^{(k)}$.
These were used to decompose $\sigma_{ab}$ and $\Pi_{ab}$ according
to $\sigma_{ab}=\sum_k\sigma_{(k)}\mathcal{Q}_{ab}^{(k)}$ and
$\Pi_{ab}=\sum_k\Pi_{(k)}\mathcal{Q}_{ab}^{(k)}$ respectively (with
${\rm D}_a\sigma^{(k)}=0={\rm D}_a\Pi^{(k)}$).

\subsection{The zero-eigenvalue issue}\label{ssZeI}
The evolution equation of trace-free anisotropic stresses during the
radiation era presents a particular mathematical problem, when
linearised around an FRW background, because of the zero eigenvalue
issue (see~\cite{skew} and references there in). In particular,
following Eq.~(\ref{dotca2}), it appears that the Alfv\'{e}n speed
and therefore the ratio $B^2/\rho$ remain constant as long as
$w=1/3$. This is a critical case, however, and a closer analysis
leads to $B^2/\rho\propto(\log t)^{-1}$ because of second-order
pressure effects. When these are taken into account, we find
that~\cite{BM}
\begin{equation}
\sigma_{ab}=P_{ab}t^{-3/2}+Q_{ab}(t\log t)^{-1}\,,  \label{aniso}
\end{equation}
on superhorizon scales. The magnetized mode falls off slowly due to
the anisotropic pressures, with $\sigma_{ab}\propto(t\log t)^{-1},$
and there is an attractor with $\sigma/H\rightarrow B^{2}/\rho$ and
$\Pi_{ab}\rightarrow2E_{ab}\propto (t^{2}\log t)^{-1}$ when the
dynamics of the radiation era are close to isotropy~\cite{skew,BM}.
In this regime the leading-order diagonal scale factors of the
spacetime metric during the radiation epoch take the form
\begin{equation}
a_{i}(t)\propto t^{1/2}(\log t)^{q_{i}}\,,  \label{ai1}
\end{equation}
with the $\{q_{i},\, i=1,2,3\}$ constants satisfying the constraint
$\sum_{i=1}^{3}q_{i}=0$. In the dust-dominated era the anisotropies
fall as power-law in time and the orthogonal scale factors evolve as
\begin{equation}
a_{i}(t)\propto t^{2/3}(1+V_{i}t^{-s_{i}})\,,  \label{ai2}
\end{equation}
where the $\{V_{i},\, i=1,2,3\}$ and the $\{s_{i},\, i=1,2,3\}$ are
sets of constants with $\sum_{i=1}^{3}s_{i}=0$. Since the observed
microwave background temperature anisotropy is determined by the
value of $\sigma/H$ at recombination, and $\sigma/H\propto
B^{2}/\rho_{m}\propto t^{-2/3}$ after the epoch of matter-radiation
equality, we obtain a strong observational bound on the present-day
magnetic field strength from the large-angular scale anisotropy of
the microwave background at last scattering, when
$z\sim1100$,~\cite{BFS}:
\begin{equation}
\left\vert B\right\vert < 4.0\times10^{-9}\Omega_{0}^{1/2}~G\,.
\label{Bbound}
\end{equation}

An investigation of the behaviour of Yang-Mills fields in the early
universe reveals that chaotic behaviour is possible in the evolution
of the scale factors with time, no matter how weak the Yang-Mills
field, and how close the expansion is to isotropy~\cite{jlev,maeda},
if there is no perfect fluid present in addition to the magnetic
field. However, despite the close relationship between the form of
Yang-Mills fields and magnetic fields, the slow logarithmic decay of
$\sigma/H\propto(\log t)^{-1}$ of the shear distortion with time
characteristic of magnetic fields and radiation does not occur if an
isotropic blackbody radiation fluid is added to the Yang-Mills
fields~\cite{jin}, because of the different time-dependence of their
pressures ($\Pi_{ab}/\rho_{YM}$ is no longer constant), which
hastens their decay, and the shear falls off in Bianchi~$I$
universes at the same rate as it does when anisotropic stresses are
absent, with $\sigma/H\propto t^{-1/2}$. This means that the
microwave background anisotropy provides a rather weak bound on the
energy-density in Yang-Mills fields today of $\Omega_{YM0}<
0.11\Omega_{\gamma0}$, where $\Omega_{\gamma0}$ is the present
density of blackbody fluids in units of the critical density. By
contrast, the slow decay of the shear in the presence of magnetic
stresses led to a far stronger bound on the magnetic field density
today, with $\Omega_{mag0}\lesssim10^{-5}\Omega_{\gamma0}$.

\section{Summary}\label{sD}
Magnetic fields are detected everywhere when the appropriate
observations are possible. The origin of cosmic magnetism is still
an open issue but the presence of large-scale $B$-fields with
similar $\mu$G-order strengths in galaxies, galaxy clusters, and
high-redshift protogalactic structures may suggest a common
(primordial) origin for them. Large-scale, micro-Gauss fields can
significantly affect the evolution of the universe and are extremely
important for astrophysics (see~\cite{Gi2} for a cross-disciplinary
review of the subject). The idea of primordial magnetism is
appealing because it can potentially explain all the large-scale
fields seen in the universe today, especially those found in remote
protogalaxies. As a result, the literature contains many studies
examining the role and the implications of magnetic fields for
cosmology.

In this work we have employed covariant techniques to analyse the
effects of magnetic fields on the evolution of inhomogeneous
relativistic cosmologies. After a brief introduction to nonlinear
cosmological electrodynamics and magnetohydrodynamics, we
investigated weakly magnetized almost-FRW universes. Within the
standard ideal MHD limits, we found that linear inhomogeneities in
the magnetic energy density grow in step with those in the matter
throughout the hot big-bang evolution of the universe. This close
relation between the magnetic and the fluid density gradients is a
key property of weakly-magnetized FRW cosmologies.

We found that during the radiation era and on large scales the
magnetic pressure inhibits the growth of density perturbations by an
amount proportional to the field's relative strength. On sub-horizon
lengths the extra magnetic pressure was shown to trigger
magnetosonic waves with a reduced amplitude and an increased
frequency, relative to the magnetic-free case. Moreover, the
magnetic presence means that small-scale density perturbations no
longer oscillate around a zero average value but have a finite
residual average that depends on the initial conditions. After
equality, the $B$-field is effectively the only source of pressure
support. This reduces the growth rate of matter condensations in a
way exactly analogous to that observed on super-Hubble scales during
the radiation era. The same magnetic pressure can also stabilize
against gravitational collapse by bringing the Jeans length up to
the size of a galaxy cluster. Looking at models with a vacuum
equation of state for the matter, we found that the outcome depends
on the effective equation of state of the medium. In particular, the
comoving density gradients remain constant as long as the barotropic
$p=-\rho$ condition holds. When dealing with non-barotropic
minimally coupled scalar fields, on the other hand, all three types
of density inhomogeneities suffer exponential decay.

Defining as isocurvature large-scale perturbations that maintain the
zero 3-curvature requirement at all times, we found that they
contain decaying modes only. Given that, typically, the term
'isocurvature' refers to fluctuations with zero initial total
energy-density perturbation, we have explained how our equations can
be used to study this type of perturbation mode. By treating the
highly conductive medium as a two-component system we found that the
effective entropy perturbation caused by the different dynamical
evolution of the two species is either constant or zero. The
evolution of vector perturbations in the presence of magnetic
fields, showed that the latter increase the amount of residual
vorticity. Thus, universes containing large-scale inhomogeneous
$B$-fields would rotate more than their magnetic-free counterparts.
Finally, we investigated the implications of the field for the
evolution of gravitational-wave perturbations.

\textbf{Acknowledgements:}
CT wishes to thank the Centre for Mathematical Sciences at Cambridge
University and the Institute of Cosmology and Gravitation at
Portsmouth University for their hospitality on a number of occasions
while this work was in progress and PPARC for support in Cambridge.
The work of RM was partly supported by PPARC.

\appendix
\section{Commutation laws}\label{AsCLs}
According to definition (\ref{deriv}a), the orthogonally projected
covariant derivative operator satisfies the condition
$\mathrm{D}_{a}h_{bc}=0$. This means that we can use $h_{ab}$ to
raise and lower indices in equations acted upon by the
aforementioned operator. Following Frobenius' theorem, however,
rotating spaces do not possess integrable 3-D submanifolds
(e.g.~see~\cite{Wal,Poi}). Therefore, the $\mathrm{D}_{a}$-operator
cannot be used as a standard 3-D derivative in such spaces and it
does not always satisfy the usual commutation laws (see below and
also~\cite{EBH}).

\subsection{Orthogonally projected gradients}\label{AssOPGs}
When acting on a scalar quantity the orthogonally projected
covariant derivative operators commute according to
\begin{equation}
\mathrm{D}_{[a}\mathrm{D}_{b]}f= -\omega_{ab}\dot{f}\,.  \label{A1}
\end{equation}
The above is a purely relativistic result and underlines the
different behaviour of rotating spacetimes within Einstein's theory.
Similarly, the commutation law for the orthogonally projected
derivatives of spacelike vectors reads
\begin{equation}
\mathrm{D}_{[a}\mathrm{D}_{b]}v_{c}= -\omega_{ab}\dot{v}_{\langle
c\rangle}+ {\frac{1}{2}}\,\mathcal{R}_{dcba}v^{d}\,.  \label{A2}
\end{equation}
where $v_{a}u^{a}=0$ and $\mathcal{R}_{abcd}$ represents the Riemann
tensor of the observer's local rest-space. Finally, when dealing
with orthogonally projected tensors, we have
\begin{equation}
\mathrm{D}_{[a}\mathrm{D}_{b]}S_{cd}=
-\omega_{ab}h_{c}{}^{e}h_{d}{}^{f}\dot{S}_{ef}+
{\frac{1}{2}}\,(\mathcal{R}_{ecba}S^{e}{}_{d}
+\mathcal{R}_{edba}S_{c}{}^{e})\,,  \label{A3}
\end{equation}
with $S_{ab}u^{a}=0=S_{ab}u^{b}$. Note that in the absence of
rotation, $\mathcal{R}_{abcd}$ is the Riemann tensor of the
(integrable) 3-D hypersurfaces orthogonal to the $u_{a}$-congruence
For details on the definition, the symmetries and the key equations
involving $\mathcal{R}_{abcd}$, the reader is referred to
\S~\ref{ssSC}. We also note that the above equations are fully
nonlinear and hold at all perturbative levels.

\subsection{Time derivatives}\label{AssTDs}
In general relativity, time derivatives do not generally commute
with their spacelike counterparts. For scalars, in particular, we
have
\begin{equation}
\mathrm{D}_{a}\dot{f}-
h_{a}{}^{b}(\mathrm{D}_{b}f)^{\displaystyle\cdot}= -\dot{f}A_{a}+
{\frac{1}{3}}\,\Theta\mathrm{D}_{a}f+
\mathrm{D}_{b}f\left(\sigma^{b}{}_{a}+\omega^{b}{}_{a}\right)\,,
\label{A4}
\end{equation}
at all perturbative levels. Assuming an FRW background, we find that
the orthogonally projected gradient and the time derivative of the
first-order vector $v_{a}$ commute as
\begin{equation}
a\mathrm{D}_{a}\dot{v}_{b}=
\left(a\mathrm{D}_{a}v_{b}\right)^{\displaystyle\cdot}\,,
\label{A5}
\end{equation}
to linear order. Similarly, when dealing with first-order spacelike
tensors, we have the following linear commutation law
\begin{equation}
a\mathrm{D}_{a}\dot{S}_{bc}=
\left(a\mathrm{D}_{a}S_{bc}\right)^{\displaystyle\cdot}\,.
\label{A6}
\end{equation}

\section{Notation}\label{BsN}
\begin{itemize}
\item \textbf{Spacetime Geometry}\newline Line element:
$\mathrm{d}s^{2}=g_{ab}\mathrm{d}x^{a}\mathrm{d}x^{b}=
-\mathrm{d}\tau^{2}$, with $c=1$.\newline 4-velocity:
$u^{a}=\mathrm{d}x^{a}/\mathrm{d}\tau$, 3-D projection tensor:
$h_{ab}=g_{ab}+u_{a}u_{b}$.\newline 4-D permutation tensor:
$\eta_{abcd}$, 3-D permutation tensor:
$\varepsilon_{abc}=\eta_{abcd}u^{d}$.\newline Covariant derivative:
$\nabla_{b}T_{a}=\partial T_{a}/\partial
x^{b}-\Gamma_{ab}^{c}T_{c}$.\newline Time derivative:
$\dot{T}_{a}=u^{b}\nabla_{b}T_{a}$, 3-D covariant derivative:
$\mathrm{D}_{b}T_{a}=h_{b}{}^{d}h_{a}{}^{c}\nabla_{d}T_{c}$.\newline
Riemann tensor: $R_{abcd}$, Ricci tensor: $R_{ab}=R^{c}{}_{acb}$,
Ricci scalar: $R=R^{a}{}_{a}$.\newline 3-Riemann tensor:
$\mathcal{R}_{abcd}$, 3-Ricci tensor:
$\mathcal{R}_{ab}=\mathcal{R}^{c}{}_{acb}$, 3-Ricci scalar:
$\mathcal{R}=\mathcal{R}^{a}{}_{a}$.\newline 3-curvature index:
$K=0,\,\pm1$, with $\mathcal{R}=6K/a^{2}$ (in FRW models).\newline
Weyl Tensor: $C_{abcd}$, electric Weyl: $E_{ab}=C_{acbd}u^{c}u^{d}$,
magnetic Weyl: $H_{ab}=\varepsilon_{a}{} ^{cd}C_{cdbe}u^{e}/2$.

\item \textbf{Kinematics}\newline Expansion scalar:
$\Theta=\nabla^{a}u_{a}=\mathrm{D}^{a}{}u_{a}$, scale factor: $a$,
with $\dot{a}/a=\Theta /3$.\newline Vorticity tensor:
$\omega_{ab}=\mathrm{D}_{[b}u_{a]}$, vorticity vector:
$\omega_{a}=\varepsilon_{abc}\omega^{bc}/2$.\newline Shear tensor:
$\sigma_{ab}=\mathrm{D}_{\langle
b}u_{b\rangle}=\mathrm{D}_{(b}u_{b)}
-(\mathrm{D}^{c}u_{c})h_{ab}/3$, 4-acceleration:
$A_{a}=u^{b}\nabla_{b}u_{a} $.\newline Hubble parameter:
$H=\dot{a}/a$ (in FRW models).

\item \textbf{Matter Fields}\newline Field equations:
$R_{ab}-(R/2)g_{ab}=T_{ab}$, with $\kappa=8\pi G=1$.\newline Matter
energy-momentum tensor: $T_{ab}= \rho u_{a}u_{b}+ph_{ab}+
2u_{(a}q_{b)}+\pi_{ab}$.\newline Matter density:
$\rho=T_{ab}u^{a}u^{b}$, isotropic pressure: $p=T_{ab}h^{ab}
/3$.\newline Barotropic index: $w=p/\rho$, adiabatic sound speed:
$c^{2}_{\mathrm{s}}=\partial p/\partial\rho$.\newline Energy flux:
$q_{a}=h_{a}{}^{b}T_{bc}u^{c}$, anisotropic pressure:
$\pi_{ab}=T_{\langle ab\rangle}=T_{(ab)}-(T/3)h_{ab}$.

\item \textbf{Electromagnetism}\newline Electromagnetic tensor:
$F_{ab}$, magnetic field: $B_{a}=\varepsilon_{abc}F^{bc}/2$,
electric field: $E_{a}=F_{ab}u^{b}$.\newline Magnetic energy
density: $B^{2}/2$, magnetic isotropic pressure: $B^{2}/6$.\newline
Magnetic anisotropic pressure: $\Pi_{ab}=-B_{\langle a}
B_{b\rangle}=(B^{2}/3)h_{ab}-B_{a}B_{b}$.\newline Alfv\'{e}n speed:
$c_{\mathrm{a}}^{2}=B^{2}/(\rho+p+B^{2})$.\newline Electric
4-current: $J_{a}$, electric 3-current: $\mathcal{J}_{a}=J_{\langle
a\rangle}=h_{a}{}^{b}J_{b}$.\newline Charge density:
$\rho_{e}=-J_{a}u^{a}$, electrical conductivity: $\varsigma$.

\item \textbf{Perturbations}\newline Matter density gradients:
$\Delta_{a}=(a/\rho)\mathrm{D}_{a}\rho$, with
$\Delta_{ab}=a\mathrm{D}_{b}\Delta_{a}$ and
$\Delta=\Delta^{a}{}_{a}$.\newline Matter vortices:
$\mathcal{W}_{ab}=\Delta_{[ab]}$, with
$\mathcal{W}_{a}=\varepsilon_{abc}\mathcal{W} ^{bc}/2$.\newline
Volume expansion gradients: $\mathcal{Z}_{a}=a\mathrm{D}_{a}\Theta$,
with $\mathcal{Z}_{ab}=a\mathrm{D}_{b}\mathcal{Z}_{a}$ and
$\mathcal{Z}=\mathcal{Z}^{a}{}_{a}$.\newline Magnetic density
gradients: $\mathcal{B}_{a}=(a/B^{2})\mathrm{D}_{b}B^{2}$ and
$\mathcal{B}_{ab} =a\mathrm{D}_{b}\mathcal{B}_{a}$, with
$\mathcal{B}=\mathcal{B}^{a}{}_{a} $.\newline Effective entropy
perturbations: $\mathcal{E}_{a}$, $\mathcal{S} _{a}^{(ij)}$, with
$\mathcal{S}_{a}^{(ij)}= -\mathcal{S}_{a}^{(ji)}$.
\end{itemize}


\begin{thebibliography}{999}
\bibitem{KWM} U. Klein, R. Wielebinski and H.W. Morsi, Astron.
Astrophys. \textbf{190}, 41 (1988).
\bibitem{KP} P.P. Kronberg, J.J. Perry and E.L.H. Zukowski,
Astrophys. J. \textbf{387}, 528 (1992).
\bibitem{WLO} A.M. Wolfe, K.M. Lanzetta and A.L. Oren, Astrophys. J.
\textbf{388}, 17 (1992).
\bibitem{K} P.P. Kronberg, Rep. Prog. Phys. \textbf{57}, 325 (1994).
\bibitem{BBMSS} R. Beck, A. Brandenburg, D. Moss, A.A. Shukurov and
D. Sokoloff, Ann. Rev, Astron. Astrophys. \textbf{34}, 155 (1996).
\bibitem{Va} J.P. Vall\'{e}e, New Astron. Rev. \textbf{48}, 763
(2004).
\bibitem{CT} C.L. Carilli and G.B. Taylor, Ann. Rev. Astron.
Astrophys. \textbf{40}, 391 (2004).
\bibitem{E} K. Enqvist, Int. J. Mod. Phys. D \textbf{7}, 331 (1998).
\bibitem{GR1} D. Grasso and H. Rubinstein, Phys. Rep. \textbf{348},
163 (2001).
\bibitem{W} L.M. Widrow, Rev. Mod. Phys. \textbf{74}, 775 (2002).
\bibitem{Gi1} M. Giovannini, in \textit{7th Paris Cosmology
Colloquium on High Energy Astrophysics for and from Space}, Eds. N.
Sanchez and H. de Vega, (2002) (hep-ph/0208152).
\bibitem{Gi2} M. Giovannini, Int. J. Mod. Phys. D \textbf{13}, 391
(2004).
\bibitem{P} E.N. Parker, \emph{Cosmical Magnetic Fields} (Oxford,
Clarendon 1979).
\bibitem{ZRS} Y.B. Zeldovich, A.A. Ruzmaikin and D.D. Sokoloff,
\emph{Magnetic Fields in Astrophysics} (New York, Gordon and Breach
1983).
\bibitem{Kr} F. Krause, in \emph{The Cosmic Dynamo}, Eds. F. Krause,
K.-H. R\"adler and G. R\"udiger (Kluwer, Dordrecht, 1993) p.~487.
\bibitem{Me} L. Mestel, \emph{Stellar Magnetism} (Oxford, Oxford
University Press 1999).
\bibitem{BS} A. Brandenburg and K. Subramanian, Phys. Rep.
\textbf{417}, 1 (2005).
\bibitem{Ku} R.M. Kulsrud, in \textit{Galactic and Extragalactic
Magnetic Fields}, Eds. R. Beck, P.P. Kronberg, and R. Wielebinski
(Reidel, Dordrecht, 1990).
\bibitem{KA} R.M. Kulsrud and S.W. Anderson, Astrophys. J.
\textbf{396}, 606 (1992).
\bibitem{DLT} A.-C. Davis, M. Lilley and O. T\"{o}rnkvist, Phys.
Rev. D \textbf{60}, 021301 (1999).
\bibitem{H1} E.R. Harrison, Mon. Not. R. Astron. Soc. \textbf{147},
279 (1970).
\bibitem{PS} R.E. Pudritz and J. Silk, Astrophys. J. \textbf{342},
650 (1989).
\bibitem{KCOR} R.M. Kulsrud, R. Cen, J.P. Ostriker and D. Ryu,
Astrophys. J. \textbf{480}, 481 (1997).
\bibitem{V} T. Vachaspati, Phys. Lett. B \textbf{265}, 258 (1991).
\bibitem{EO} K. Enqvist and P. Olesen, Phys. Lett. B \textbf{319},
178 (1993).
\bibitem{GGV} M. Gasperini, M. Giovannini and G. Veneziano, Phys.
Rev. Lett. \textbf{75}, 3796 (1995).
\bibitem{BEO1} A. Brandenburg, K. Enqvist and P. Olesen, Phys. Rev.
D \textbf{54}, 1291 (1996).
\bibitem{SJO} G. Sigl, K. Jedamzik and A.V. Olinto, Phys. Rev. D
\textbf{55}, 4582 (1997).
\bibitem{DD1} A.-C. Davis and K. Dimopoulos, Phys. Rev. D
\textbf{55}, 7398 (1997).
\bibitem{CKM} E.A. Calzetta, A. Kandus and F.D Mazzitelli, Phys.
Rev. D \textbf{57}, 7139 (1998).
\bibitem{KCMW} A. Kandus, E.A. Calzetta, F.D Mazzitelli and C.E.M.
Wagner, Phys. Lett. B \textbf{472}, 287 (2000).
\bibitem{Ga1} M. Gasperini, Phys. Rev. D \textbf{63}, 047301 (2001).
\bibitem{CK} E.A. Calzetta and A. Kandus, Phys. Rev. D \textbf{65},
063004 (2002).
\bibitem{DC} R. Durrer and C. Caprini, JCAP \textbf{0311}, 010
(2003).
\bibitem{GT} M.R. Garousi, M. Sami and S. Tsujikava, Phys. Lett. B
\textbf{606}, 1 (2005).
\bibitem{TIOH} K. Takahashi, K. Ichiki, H. Ohno and H. Hanayama,
Phys. Rev. Lett. \textbf{95}, 121301 (2005).
\bibitem{DD2} A.-C. Davis and K. Dimopoulos, Phys. Rev. D
\textbf{72}, 043517 (2005).
\bibitem{Kun} K.E. Kunze, Phys. Lett. B \textbf{623}, 1 (2005).
\bibitem{Ga2} M. Gasperini, in \textit{The Origin and Evolution of
Cosmic Magnetism}, Eds. R. Beck, G. Brunetti, L. Feretti and B.
Gaensler, Astron. Nachr \textbf{327}, 399 (2006).
\bibitem{MMNR}  S.~Matarrese, S.~Mollerach, A.~Notari and A.~Riotto,
Phys. Rev. D, {\bf 71}, 043502 (2005).
\bibitem{GS} R.~Gopal and S.~Sethi, Mon. Not. Roy. Astron. Soc.,
{\bf 363}, 529 (2005).
\bibitem{SF} E.~R.~Siegel and J.~N.~Fry, preprint: astro-ph/0604526.
\bibitem{KMST} T.~Kobayashi, R.~Maartens, T.~Shiromizu and
K.~Takahashi, Phys. Rev. D \textbf{75}, 103501 (2007).
\bibitem{ITSHO} K.~Ichiki, K.~Takahashi, N.~Sugiyama, H.~Hanayama
and H.~Ohno, preprint: astro-ph/0701329.
\bibitem{SB2} K. Subramanian and J.D. Barrow, Phys. Rev. D
\textbf{58}, 083502 (1998).
\bibitem{SB1} K. Subramanian and J.D. Barrow, Phys. Rev. Lett.
\textbf{81}, 3575 (1998).
\bibitem{SSB} K. Subramanian, T.R. Seshadri and J.D. Barrow, Mon.
Not. R. Astron. Soc. \textbf{344}, L31 (2003).
\bibitem{TW} M.S. Turner and L. Widrow, Phys. Rev. D \textbf{37},
2743 (1988).
\bibitem{Ra} B. Ratra, Astrophys. J. \textbf{391}, L1 (1992).
\bibitem{Gio} M. Giovannini, Phys. Rev. D \textbf{62}, 123505
(2000).
\bibitem{BPTV} B.A. Bassett, G. Pollifrone, S. Tsujikawa and F.
Viniegra, Phys. Rev. D \textbf{63}, 103515 (2001).
\bibitem{Ma} A.L. Maroto, Phys. Rev D \textbf{64}, 083006 (2001).
\bibitem{DPTD} K. Dimopoulos, T. Prokopec, O. T\"ornkvist and A.-C.
Davis, Phys. Rev. D \textbf{65}, 063505 (2002).
\bibitem{TDM} C.G. Tsagas, P.K.S. Dunsby and M. Marklund, Phys.
Lett. B \textbf{561}, 17 (2003).
\bibitem{AM} A. Ashoorioon and R.B. Mann, Phys. Rev. D \textbf{71},
103509 (2005).
\bibitem{TK} C.G. Tsagas and A. Kandus, Phys. Rev. D \textbf{71},
103509 (2005).
\bibitem{Ts1} C.G. Tsagas, Phys. Rev. D \textbf{72}, 123509 (2005).
\bibitem{BTW} B.A. Bassett, S. Tsujikawa and D. Wands, Pev. Mod.
Phys. \textbf{78}, 537 (2006).
\bibitem{GR2} D. Grasso and H.R. Rubinstein, Phys. Lett. B
\textbf{379}, 79 (1996).
\bibitem{COST} B. Cheng, A.V. Olinto, D.N. Schramm and J.W. Truran,
Phys. Rev. D \textbf{54}, 4714 (1997).
\bibitem{Z} Y.B. Zeldovich, Sov. Astron. \textbf{13}, 608 (1970).
\bibitem{ZN} Y.B. Zeldovich and I.D. Novikov, \emph{Relativistic
Astrophysics: Volume II} (Chicago, University of Chicago Press
1974).
\bibitem{ADGR} J. Adams, U.H. Danielson, D. Grasso and H.R. Rubinstein,
Phys. Lett. B \textbf{388}, 253 (1996).
\bibitem{KL} A. Kosowsky and A. Loeb, Astrophys. J. \textbf{469}, 1
(1996).
\bibitem{skew} J.D. Barrow, Phys. Rev. D \textbf{55}, 7451 (1997).
\bibitem{BFS} J.D. Barrow, P.G. Ferreira and J. Silk, Phys. Rev.
Lett. \textbf{78}, 3610 (1997).
\bibitem{DKY} R. Durrer, T. Kahniasvili and A. Yates, Phys. Rev. D
\textbf{58}, 123004 (1998).
\bibitem{DFK} R. Durrer, P.G. Ferreira and T. Kahniashvili, Phys.
Rev. D \textbf{61}, 043001 (2000).
\bibitem{PVW} L. Pogosian, T. Vachaspati and S. Winitzki, Phys. Rev.
D \textbf{65}, 083502 (2002).
\bibitem{CCMT} C.A. Clarkson, A.A. Coley, R. Maartens and C.G.
Tsagas, Class. Quantum Grav. \textbf{20}, 1519 (2003).
\bibitem{CDK} C. Caprini, R. Durrer and T. Kahniashvili, Phys. Rev.
D \textbf{69}, 063006 (2004).
\bibitem{L} A. Lewis, Phys. Rev. D \textbf{70}, 043011 (2004).
\bibitem{Gi3} M. Giovannini, Phys. Rev. D \textbf{70}, 123507
(2004).
\bibitem{Gi4} M. Giovannini, Phys. Rev. D \textbf{71}, 021301
(2005).
\bibitem{Gi5} M. Giovannini, Class. Quantum Grav. \textbf{23}, R1
(2006).
\bibitem{Gi6} M. Giovannini, Phys. Rev D \textbf{73}, 101302 (2006).
\bibitem{S} K. Subramanian, in \textit{The Origin and Evolution of
Cosmic Magnetism}, Eds. R. Beck, G. Brunetti, L. Feretti and B.
Gaensler, Astron. Nachr \textbf{327}, 403 (2006).
\bibitem{YIKM}D. Yamazaki, K. Ichiki, T. Kajino and G.J. Mathews,
Astrophys. J. \textbf{646}, 719 (2006).
\bibitem{dur} R. Durrer, New Astron. Rev. \textbf{51}, 275 (2007).
\bibitem{KR}T. Kahniashvili and B. Ratra, Phys. Rev. D \textbf{75},
023002 (2007).
\bibitem{Th} K.S. Thorne, Astrophys. J. \textbf{148}, 51 (1967).
\bibitem{J} K. Jacobs, Astrophys. J. \textbf{155}, 379 (1969).
\bibitem{C} C.B. Collins, Commun. Math. Phys. \textbf{27}, 37 (1972).
\bibitem{LbKW} V.G. LeBlanc, D. Kerr and J. Wainwright, Class.
Quantum Grav. \textbf{12}, 513 (1995).
\bibitem{Lb} V.G. LeBlanc, Class. Quantum Grav. \textbf{14}, 2281
(1997).
\bibitem{We} M. Weaver, Class. Quantum Grav. \textbf{17}, 421
(2000).
\bibitem{PP} A. Pradhan and O.P. Pandey, Int. J. Mod. Phys. D
\textbf{12}, 1299 (2003).
\bibitem{HW} J.T. Horwood and J. Wainwright, Gen. Rel. Grav.
\textbf{36}, 799 (2004).
\bibitem{KC2} E.J. King and P. Coles, Class. Quantum Grav.
\textbf{24}, 2061 (2007).
\bibitem{RR} T.V. Ruzmaikina and A.A. Ruzmaikin, Sov. Astron.
\textbf{14}, 963 (1971).
\bibitem{Wa} I. Wasserman, Astrophys. J. \textbf{224}, 337 (1978).
\bibitem{Pe} P.J.E. Peebles, \emph{The Large-Scale Structure of the
Universe} (Princeton, Princeton University Press 1980).
\bibitem{KOR} E. Kim, A.V. Olinto and R. Rosner, Astrophys. J.
\textbf{468}, 28 (1996).
\bibitem{BF} E. Battaner and E. Florido, Astron. Nachr.
\textbf{328}, 92 (2007).
\bibitem{VTP} L. Vlahos, C.G. Tsagas and D. Papadopoulos, Astrophys.
J. L9 \textbf{629}, (2005).
\bibitem{F} A.G. Fennelly, Phys. Rev. Lett. \textbf{55}, 955 (1980).
\bibitem{PE} D. Papadopoulos and F.P. Esposito, Astrophys. J.
\textbf{257}, 10 (1982).
\bibitem{BFH-V} E. Battaner, E. Florido and J. Jimenez-Vic\'ente,
Astron. Astrophys. \textbf{326}, 13 (1997).
\bibitem{TB1} C.G. Tsagas and J.D. Barrow, Class. Quantum Grav.
\textbf{14}, 2539 (1997).
\bibitem{TB2} C.G. Tsagas and J.D. Barrow, Class. Quantum Grav.
\textbf{15}, 3523 (1998).
\bibitem{TM1} C.G. Tsagas and R. Maartens, Phys. Rev. D \textbf{61},
083519 (2000).
\bibitem{TM2} C.G. Tsagas and R. Maartens, Class. Quantum Grav.
\textbf{17}, 2215 (2000).
\bibitem{HD} S. Hobbs and P.K.S. Dunsby, Phys. Rev. D \textbf{62},
124007 (2000).
\bibitem{MDBST} M. Marklund, P.K.S. Dunsby, G. Betschart, M. Servin
and C.G. Tsagas, Class. Quantum Grav. \textbf{20}, 1823 (2003).
\bibitem{MM} J. Moortgat and M. Marklund, Mon. Not. R. Astron. Soc.
\textbf{369}, 1813 (2006).
\bibitem{BEO2} A. Brandenburg, K. Enqvist and P. Olesen, Phys. Lett.
B \textbf{392}, 395 (1997).
\bibitem{JKO} K. Jedamzik, V. Katalinic and A. Olinto, Phys. Rev. D
\textbf{57}, 3264 (1998).
\bibitem{DBL} K. Dolag, M. Bartelmann and H. Lesch, Astron.
Astrophys. \textbf{348}, 351 (1999).
\bibitem{BMT} M. Bruni, R. Maartens and C.G. Tsagas, Mon. Not. R.
Astron. Soc. \textbf{338}, 785 (2003).
\bibitem{S-OW}  G. Siemieniec-Ozieblo and A. Woszczyna, Astron.
Astrophys. {\bf 414}, 1 (2004).
\bibitem{S-OG}  G. Siemieniec-Ozieblo and Z.A. Golda, Astron.
Astrophys. {\bf 422}, 23 (2004).
\bibitem{BJ} R. Banerjee and Jedamzik K., Phys. Rev D \textbf{70},
123004 (2004).
\bibitem{KGST} K. Dolag, D. Grasso, V. Springel and I. Tkachev, JCAP
\textbf{0501}, 009 (2005).
\bibitem{SS} V.B. Semikoz and D.D. Sokoloff, Int. J. Mod. Phys. D
\textbf{14}, 1839 (2005).
\bibitem{OIV} M. Onofri, H. Isliker and L. Vlahos, Phys. Rev. Lett.
\textbf{96}, 151102 (2006).
\bibitem{KC1} E. King and P. Coles, Mon. Not. R. Astron. Soc.
\textbf{365}, 1288 (2006).
\bibitem{CCT} L. Campanelli, P. Cea and L. Tedesco, Phys. Rev. Lett.
\textbf{97}, 131302 (2006).
\bibitem{HS} O. Heckmann and E. Sch\"{u}cking, Zeits. f. Astrophys.
\textbf{38}, 95 (1955).
\bibitem{R} A.K. Raychaudhuri, Zeits. f. Astrophys. \textbf{43}, 161
(1957).
\bibitem{Eh1} G. Ehlers, Abh. Mainz. Akad. Wiss. Lit. \textbf{11}, 1
(1961).
\bibitem{El1} G.F.R. Ellis, in \emph{General Relativity and
Cosmology} Ed. R.K. Sachs (New York, Academic 1971) p.~104.
\bibitem{EvE} G.F.R Ellis and H. van Elst, in \emph{Theoretical and
Observational Cosmology}, Ed. M. Lachi\'{e}ze-Rey (Dordrecht, Kluwer
1999) p.~1.
\bibitem{TCM} C.G. Tsagas, A. Challinor and R. Maartens, preprint:
arXiv:0705.4397.
\bibitem{KE} A.R. King and G.F.R Ellis, Commun. Math. Phys.
\textbf{31}, 209 (1973).
\bibitem{KS} H. Kodama and M. Sasaki, Prog. Theor. Phys. Suppl.
\textbf{78}, 1 (1984).
\bibitem{DBE} P.K.S. Dunsby, M. Bruni and G.F.R. Ellis, Astrophys.
J. \textbf{395}, 54 (1992).
\bibitem{Gi7} M. Giovannini, Class. Quantum Grav. \textbf{22}, 5243
(2005).
\bibitem{ET} G.F.R. Ellis and C.G. Tsagas, Phys. Rev. D
\textbf{66}, 124015 (2002).
\bibitem{El2} G.F.R. Ellis, in \emph{Carg\`{e}se Lectures in
Physics} vol VI, Ed. E. Schatzmann (New York, Gordon and Breach
1973) p.~1.
\bibitem{Ts} C.G. Tsagas, Class. Quantum Grav. \textbf{22}, 393
(2005).
\bibitem{Gr} P.J. Greenberg, Ap. J. \textbf{164}, 589 (1971).
\bibitem{Ja} J.D. Jackson, \emph{Classical Electrodynamics} (New
York, Wiley 1975).
\bibitem{HE} S.W. Hawking and G.F.R. Ellis, \emph{The Large Scale
Structure of Space-time} (Cambridge, Cambridge University Press
1973).
\bibitem{M} R. Maartens, Phys. Rev. D \textbf{55}, 463 (1997).
\bibitem{MGE} R. Maartens, T. Gebbie and G.F.R. Ellis, Phys. Rev. D
\textbf{59}, 083506 (1998).
\bibitem{Be} L. Bel, C.R. Acad. Sci. Paris \textbf{247}, 1094 (1958).
\bibitem{Pen} R. Penrose, Ann. Phys. \textbf{138}, 59 (1969).
\bibitem{MB} R. Maartens and B. Bassettt, Class. Quantum Grav.
\textbf{15}, 705 (1998).
\bibitem{Da} N. Dadhich, Gen. Rel. Grav. \textbf{32}, 1009 (2000).
\bibitem{Wal} R.M. Wald, \emph{General Relativity} (Chicago:
University of Chicago Press, 1984).
\bibitem{Poi} E. Poisson \emph{A Relativist's Toolkit} (Cambridge,
Cambridge University Press, 2004).
\bibitem{T1} C.G. Tsagas, Class. Quantum Grav. \textbf{19}, 3709
(2002).
\bibitem{MT} D.R. Matravers and C.G. Tsagas, Phys. Rev. D
\textbf{62}, 103519 (2000).
\bibitem{T2} C.G. Tsagas, Phys. Rev. Lett. \textbf{86}, 5421 (2001).
\bibitem{dFSZ} F. de Felice, F. Sorge and S. Zilio, Class. Quantum
Grav. \textbf{22}, 47 (2005).
\bibitem{T3} C.G. Tsagas, Class. Quantum Grav. \textbf{23}, 4323
(2006).
\bibitem{EB} G.F.R Ellis and M. Bruni, Phys. Rev. D \textbf{40},
1804 (1989).
\bibitem{EHB} G.F.R. Ellis, J. Hwang and M. Bruni, Phys. Rev D
\textbf{40}, 1819 (1989).
\bibitem{EBH} G.F.R. Ellis, M. Bruni and J. Hwang, Phys. Rev D
\textbf{42}, 1035 (1990).
\bibitem{B1} T. Buchert, Gen. Rel. Grav. \textbf{33}, 1381 (2001).
\bibitem{Ba2} J.D. Barrow, Mon. Not. R. Astron. Soc. \textbf{179},
47P (1977).
\bibitem{SW} J.M. Stewart and M. Walker, Proc. R. Soc. A
\textbf{341}, 49 (1974).
\bibitem{BMMS}  M. Bruni, S. Matarrese, S. Mollerach and S. Sonego,
Class. Quantum Grav. {\bf 14}, 2584 (1997).
\bibitem{MaT} R. Maartens and J. Triginer, Phys. Rev. D \textbf{56},
4640 (1997).
\bibitem{Pee} P.J.E. Peebles, \emph{The Large-Scale Structure of the
Universe} (Princeton, Princeton University Press, 1980).
\bibitem{Pa} T. Padmanabhan, \emph{Structure Formation in the
Universe} (Cambridge, Cambridge University Press, 1993).
\bibitem{T4} C.G. Tsagas, in \emph{Cosmlogical Crossroads}, Eds.
S.~Cotsakis and E.~Papantonopoulos, Lec. Notes Phys. \textbf{592},
223 (Berlin, Springer, 2002).
\bibitem{EM} G.F.R. Ellis and M.S. Madsen, Class. Quantum Grav.
\textbf{8}, 667 (1991).
\bibitem{BED} M. Bruni, G.F.R. Ellis and P.K.S. Dunsby, Class.
Quantum Grav. \textbf{9}, 921 (1992).
\bibitem{Mad} M.S. Madsen, Class. Quantum Grav. \textbf{5}, 627
(1988).
\bibitem{JB} J.D. Barrow, in \textit{The Very Early Universe}, Eds.
G. Gibbons, S.W. Hawking and S.T.C. Siklos, (Cambridge, Cambridge
University Press, 1983) p.~267.
\bibitem{Gibb} W. Boucher and G.W. Gibbons, in \textit{The Very
Early Universe}, Eds. G. Gibbons, S.W. Hawking and S.T.C. Siklos,
(Cambridge, Cambridge University Press, 1983) p.~273.
\bibitem{Starob} A.A. Starobinsky, Sov. Astron. Lett. \textbf{9},
302 (1983).
\bibitem{Wald} R. Wald, Phys. Rev. D \textbf{28}, 2118 (1983).
\bibitem{MFB} V.F. Mukhanov, H.A. Feldman and B.R. Brandenberger,
Phys. Rep. \textbf{215}, 203 (1992).
\bibitem{Gi8} M. Giovannini, Class. Quantum Grav. \textbf{23}, 4991
(2006).
\bibitem{LL} A.R. Liddle and D.H. Lyth, \emph{Cosmological Inflation
and Large-Scale Structure} (Cambridge, Cambridge University Press,
2000).
\bibitem{H} S.W. Hawking, Astrophys. J. \textbf{145}, 544 (1966).
\bibitem{Ch} A. Challinor, Class. Quantum Grav. \textbf{17}, 871
(2000).
\bibitem{MTU} R. Maartens, C.G. Tsagas and C. Ungarelli, Phys. Rev.
D \textbf{63}, 123507 (2001).
\bibitem{CD1} C. Caprini and R. Durrer, Phys. Rev. D \textbf{65},
023517 (2002).
\bibitem{G} S.W. Goode, Phys. Rev. D \textbf{39}, 2882 (1989).
\bibitem{CD2} C.~Caprini and R.~Durrer, Phys. Rev. D {\bf 74},
063521 (2006).
\bibitem{BM}  J.D.~Barrow and R.~Maartens, Phys. Rev. D {\bf 59},
043502 (1999).
\bibitem{jlev}J.D. Barrow and J. Levin, Phys. Rev. Lett.
\textbf{80}, 656 (1998).
\bibitem{maeda} Y. Jin and K-I. Maeda, Phys. Rev. D \textbf{71},
064007 (2005).
\bibitem{jin} J.D. Barrow, Y. Jin and K-I Maeda, Phys. Rev. D
\textbf{72}, 103512 (2005).
\end{thebibliography}

\end{document}